\documentclass[12pt]{article} 
\usepackage{amssymb}
\usepackage{amsmath} 
\usepackage{cite}
\pdfoutput=1   % ensures pdflatex is used at arxiv
\usepackage{hyperref}
\input xy
\xyoption{all}
\usepackage{rotfloat}   % Note that this only seems to work with pdflatex

\def\sqr#1#2{{\vcenter{\vbox{\hrule height.#2pt
            \hbox{\vrule width.#2pt height#1pt \kern#1pt
                  \vrule width.#2pt}\hrule height.#2pt}}}}

\def\square
 {\mathop{\mathchoice{\sqr{12}{15}}{\sqr{9}{12}}
{\sqr{6.3}{9}}{\sqr{4.5}{9}}}}

\def\sqra#1#2#3{{\vcenter{\vbox{\hrule height.#2pt
            \hbox{\vrule width.#2pt height#1pt \kern5pt %\kern#1pt
#3
                  \vrule width.#2pt}\hrule height.#2pt}}}}

\newcommand{\be}{\begin{equation}}
\newcommand{\ee}{\end{equation}}
\newcommand{\bea}{\begin{eqnarray}}
\newcommand{\eea}{\end{eqnarray}}

\newcommand{\e}{\epsilon}

\newcommand{\la}{\lambda}
\newcommand{\m}{\mu}

\newcommand{\Om}{\Omega}
\newcommand{\om}{\omega}

\newcommand{\C}{\mathbb{C}}

\newcommand{\non}{\nonumber}

\newcommand{\rr}{\rightarrow}

\newcommand{\Z}{\mathbb{Z}}

\newcommand{\GL}{\operatorname{GL}}

\newcommand{\U}{\operatorname{U}}

\newcommand{\lp}{\left(}
\newcommand{\rp}{\right)}
\newcommand{\ls}{\left[}
\newcommand{\rs}{\right]}

\numberwithin{equation}{section}

\numberwithin{table}{section}

\begin{document} 

\begin{center}

{\large\bf A generalization of decomposition in orbifolds}

\vspace*{0.2in}

Daniel G. Robbins$^1$, Eric Sharpe$^2$,
Thomas Vandermeulen$^1$

\begin{tabular}{cc}
{\begin{tabular}{l}
$^1$ Department of Physics\\
University at Albany\\
Albany, NY 12222 \end{tabular}} &
{\begin{tabular}{l}
$^2$ Department of Physics MC 0435\\
850 West Campus Drive\\
Virginia Tech\\
Blacksburg, VA  24061 \end{tabular}}
\end{tabular}

{\tt dgrobbins@albany.edu},
{\tt ersharpe@vt.edu},
{\tt tvandermeulen@albany.edu}

\end{center}

This paper describes a generalization of decomposition in orbifolds.
In general terms, decomposition states that two-dimensional orbifolds
and gauge theories whose gauge groups have
trivially-acting subgroups decompose into
disjoint unions of theories.  However, decomposition can be, at least
naively, broken in orbifolds if the orbifold has discrete torsion in
the trivially-acting subgroup.
(Formally, this breaks finite global one-form symmetries.)  Nevertheless,
even in such cases, one still sees rudiments of decomposition.  In this paper,
we generalize decomposition in orbifolds to include such examples of
discrete torsion, which we check in numerous examples.  
Our analysis includes as special cases (and in one
sense generalizes)
quantum symmetries of abelian orbifolds.

\begin{flushleft}
January 2021
\end{flushleft}

\newpage

\tableofcontents

\newpage

\section{Introduction}

This paper is motivated by a revival of interest in two old topics,
namely 
\begin{itemize}
\item orbifolds, see e.g. \cite{Wang:2017loc,Bhardwaj:2017xup,Tachikawa:2017gyf,Chang:2018iay,Robbins:2019zdb,Robbins:2019ayj,yujitasi2019},
utilizing new methods and insights into
anomalies from topological defect lines and related technologies,
and 
\item decomposition, first described in \cite{Hellerman:2006zs},
an equivalence between two-dimensional theories with what are now called
one-form symmetries (and various generalizations) and disjoint unions of
other two-dimensional theories,
see e.g. 
\cite{Sharpe:2014tca,Sharpe:2019ddn,Tanizaki:2019rbk,Komargodski:2020mxz,Eager:2020rra} 
for recent activity in this area.
\end{itemize}

Decomposition was first introduced in \cite{Hellerman:2006zs}
to understand orbifolds and two-dimensional gauge theories in which a 
finite subgroup
of the gauge group acts trivially on the theory.
(If the subgroup is abelian, this means that the theory
has a finite global one-form symmetry, in modern language,
but decomposition is defined more generally.)  
Briefly, decomposition says that such a quantum field theory is
equivalent to ('decomposes' into) a disjoint union of related quantum
field theories, often constructed from orbifolds and gauge theories
by effectively-acting quotients of the original gauge group.
See \cite{Pantev:2005zs,Pantev:2005rh,Pantev:2005wj} for 
detailed discussions of such examples, and see also
\cite{ajt1,ajt2,ajt3,t1,gt1,xt1} for applications to Gromov-Witten
theory, \cite{Caldararu:2007tc,Hori:2011pd,Addington:2012zv,Sharpe:2012ji,Hori:2013gga,
Hori:2016txh,Knapp:2019cih} for applications to phases of gauged
linear sigma models (GLSMs), \cite{Anderson:2013sia} 
for applications in heterotic
string compactifications, \cite{Eager:2020rra} for applications to
elliptic genera of two-dimensional pure supersymmetric gauge theories,
and \cite{Tanizaki:2019rbk} for four-dimensional analogues,
for example.

The original work on decomposition \cite{Hellerman:2006zs}
studied orbifolds and gauge theories,
but did not consider orbifolds in which discrete
torsion was turned on in a way that obstructed the existence of the
one-form symmetry (or its analogues), except to note that the
decomposition story did not apply to such cases.  The purpose of this
paper is to fill that gap, by generalizing decomposition in orbifolds
to include cases in which discrete torsion is turned on.

As one application, we shall see how quantum symmetries are described
in this framework.  Recall \cite{Vafa:1989ih,Ginsparg:1988ui} 
that a ${\mathbb Z}_k$ orbifold
has a ${\mathbb Z}_k$ quantum symmetry that, when gauged, returns the
original orbifold.  We shall see that the composition of orbifolds
can be described as a single ${\mathbb Z}_k \times {\mathbb Z}_k$ orbifold
with discrete torsion, for which our generalization of decomposition predicts
that the orbifold is equivalent to the original space.  In fact, 
our generalization of decomposition will predict analogous results in
more general cases, that often the effect of gauging a trivially-acting subgroup
$K$ of the orbifold group $G$
with nonzero discrete torsion is to partly `undo' the underlying
$G/K$ orbifold.  Indeed, part of our results is a generalization,
in a certain direction, of quantum symmetries.

We begin in section~\ref{sect:rev} by reviewing decomposition for
the special case of orbifolds, specializing a number of very general
(and necessarily abstract) statements in \cite{Hellerman:2006zs} to the
simpler concrete case that the theory is a finite gauge theory,
an orbifold by a finite group.  Briefly, decomposition in this case
reduces to the statement
\begin{equation}
{\rm QFT}\left( [X/G] \right) \: = \:
{\rm QFT}\left( \left[ \frac{X \times \hat{K} }{G/K} \right]_{\hat{\omega}}
\right),
\end{equation}
where $K \subset G$ is a trivially-acting subgroup, $\hat{K}$ the
set of isomorphism classes of irreducible representations of $K$,
and 
$\hat{\omega}$ represents discrete torsion on the components.
Note that the right-hand size of the expression above has multiple
connected components in general.
In the case of ordinary decomposition in orbifolds, 
the $G$ orbifold $[X/G]$ does not
have discrete torsion.

In section~\ref{sect:conj} we then make our conjecture for the
generalization of decomposition to include $G$ orbifolds with
discrete torsion $\omega$, which we denote
$[X/G]_{\omega}$.  Examples naturally fall into
three classes.
Given an orbifold with orbifold group
$G$, with trivially-acting subgroup $K$, where
\begin{equation}
\label{eq:BasicExtension}
1 \: \longrightarrow \: K \: \stackrel{\iota}{\longrightarrow} \: G \:
\stackrel{\pi}{\longrightarrow} \: G/K \: \longrightarrow \: 1,
\end{equation}
with discrete torsion $\omega \in H^2(G,U(1))$,
the three cases are as follows:
\begin{enumerate}
\item $\iota^* \omega \neq 0$ as an element of $H^2(K,U(1))$,
in which case we conjecture that
\begin{equation}
{\rm QFT}\left( [X/G]_{\omega} \right) \: = \:
{\rm QFT}\left( \left[ \frac{X \times \hat{K}_{\iota^* \omega} }{G/K}
 \right]_{\hat{\omega}}
\right),
\end{equation}
where $\hat{K}_{\iota^* \omega}$ is the set of irreducible projective
representations of $K$,
\item $\iota^* \omega = 0$, $\beta(\omega) \neq 0$,
in which case we conjecture that
\begin{equation}
{\rm QFT}\left( [X/G]_{\omega} \right) \: = \:
{\rm QFT}\left( \left[ \frac{ X \times {\rm Coker}\: \beta(\omega) }{
{\rm Ker}\: \beta(\omega) } \right]_{\hat{\omega}} \right),
\end{equation}
\item $\iota^* \omega = 0$, $\omega = \pi^* \overline{\omega},$
in which case we conjecture that 
\begin{equation}
{\rm QFT}\left( [X/G]_{\omega} \right) \: = \:
{\rm QFT}\left( \left[ \frac{X \times \hat{K} }{ G/K } 
\right]_{\hat{\omega} + \overline{\omega}}
\right),
\end{equation}
\end{enumerate}
where the maps $\pi^*$, $\beta$ arise in the exact sequence\footnote{
This sequence, described in e.g. \cite{hochschild},
is part of a seven-term extension of the standard 
inflation-restriction
exact sequence.
Further details and references are
given in section~\ref{sect:conj}.
}
\begin{equation}
H^2(G/K,U(1)) \: \stackrel{\pi^*}{\longrightarrow} \:
{\rm Ker}\, \iota^* \: \stackrel{\beta}{\longrightarrow} \:
H^1(G/K, H^1(K,U(1)) ),
\end{equation}
and the discrete torsion $\hat{\omega}$ varies between cases, in a fashion
that will be described in section~\ref{sect:conj}.

In the remainder of this paper we check the conjecture in a number
of concrete examples.  In section~\ref{sect:exs:iw}, we discuss
examples with $\iota^* \omega \neq 0$; in
section~\ref{sect:exs:b} we discuss examples with
$\iota^* \omega = 0$ and $\beta(\omega) \neq 0$, and relate $\beta$ to
quantum symmetries;
in section~\ref{sect:exs:pi} we discuss examples with
$\iota^* \omega = 0$ and $\omega = \pi^* \overline{\omega}$ for
some $\overline{\omega} \in H^2(G/K,U(1))$.
Finally in section~\ref{sect:exs:mixed} we discuss examples spanning all three
categories, depending upon the value of discrete torsion $\omega$.

For reference, in appendix~\ref{app:GroupCohomologyAndProjReps} 
we briefly review some pertinent aspects of group cohomology and projective representations, 
in appendix~\ref{app:SomeCocycleCalculations} we give some technical
details of our cocycle computations,
in appendix~\ref{sect:beta} we make explicit the map $\beta$ that
plays a prominent role in our analysis,
and in appendix~\ref{app:gpthy} 
we collect a number of group-theoretic
results on the finite groups appearing in the examples, including
explicit representatives of discrete torsion cocycles 
and genus-one twisted sector phases.

Finally, we should comment on additive versus multiplicative
notation in cocycles.  At various points in this paper, it will greatly improve
readability to use one or the other, so we have adopted both notations, and
leave the reader to infer from context which is being used in any one
section.

After this paper appeared on the arXiv, we learned of
\cite{stt} which also discusses decomposition in the presence of
$B$ fields and derives similar conclusions, albeit from a different
computational perspective.

\section{Review of decomposition in orbifolds}  
\label{sect:rev}

Let $G$ be a finite group acting on a space $X$,
with $K \subset G$ a normal subgroup acting trivially.
It was argued in \cite[section 4.1]{Hellerman:2006zs} 
that this orbifold is equivalent
to a disjoint union, specifically
\begin{equation}  \label{eq:decomp1}
{\rm QFT}\left( [X/G] \right) \: = \:
{\rm QFT}\left( \left[ \frac{ X \times \hat{K}}{ G/K } 
\right]_{\hat{\omega}}\right),
\end{equation}
where $\hat{K}$ denotes the set of isomorphism classes of
irreducible representations of $K$,
and $\hat{\omega}$ denotes discrete torsion we shall describe momentarily.
This is known as decomposition, referring to the fact that the
theory on the right-hand side typically has multiple different 
disjoint components.  These different components or summands are sometimes
referred to in the literature as universes, given that in a string 
compactification, they would define low-energy theories with multiple 
independent decoupled gravitons.

The group $G/K$ acts on $\hat{K}$ as follows:  
pick a section $s:G/K\rr G$, so that $\pi(s(q))=q$ for all $q\in G/K$.  If the extension does not split (if $G$ is not isomorphic to a semi-direct product  
$K\rtimes G/K$), then $s$ cannot be chosen to be a group homomorphism, but it always exists as a map.  For any representation $\phi:K\rr\GL(V)$ of $K$ and any $q\in G/K$, define a new representation $L_q\phi:K\rr\GL(V)$ by
\be
(L_q\phi)(k) \: = \: \phi(s(q)^{-1} k s(q))
\ee
(here we are suppressing the map $\iota$ and are simply taking $K$ to be a normal subgroup of $G$).
It's easy to check that $L_q\phi$ is a homomorphism and that if $\phi$ is irreducible then $L_q\phi$ is also irreducible.  By verifying that $L_1\phi$ is isomorphic to $\phi$, that $L_{q_1}(L_{q_2}\phi)$ is isomorphic to $L_{q_1q_2}\phi$, and that if $\phi_1$ is isomorphic to $\phi_2$ then $L_q\phi_1$ is isomorphic to $L_q\phi_2$, one shows that this defines a $G/K$ action on $\hat{K}$ by $q\cdot[\phi]=[L_q\phi]$.  Moreover, this action is independent of the choice of section $s$.

With this understanding we can interpret (\ref{eq:decomp1}) and determine the discrete torsion $\hat{\omega}$.
Let $\{ \rho_a \}$ be a collection of irreducible representations of $K$
chosen so that the equivalence classes $[\rho_a]$ are representatives
of the distinct orbits of the $G/K$ action on $\hat{K}$.  For each $\rho_a$, let
$H_a \subseteq G/K$ be the stabilizer of $[\rho_a]$ in $\hat{K}$.
Then, decomposition becomes the statement that
\begin{equation}
{\rm QFT}\left( [X/G] \right) \: = \:
{\rm QFT}\left( \coprod_a [X/H_a]_{\hat{\omega}_a} \right),
\end{equation}
where $\hat{\omega}_a\in H^2(H_a,\U(1))$ (our notation and conventions for group cohomology are reviewed in appendix~\ref{subapp:GroupCohomology}) denotes the discrete torsion in the
summand $[X/H_a]$.

If $K$ is abelian, then there is a simple way to determine $\hat{\om}_a$.  First, given $q_1,q_2\in G/K$ define
\be
\label{eq:ExtensionClass-e}
e(q_1,q_2)=s(q_1)s(q_2)s(q_1q_2)^{-1}.
\ee
Since this is in the kernel of the map $\pi$, it must lie in the subgroup $K$.  
Indeed, $e$ turns out to be a 2-cocycle valued in $Z^2(G/K,K)$, 
where $K$ is taken to be a $G/K$-module with action $q\cdot k=s(q)ks(q)^{-1}$.  
Different choices of section $s$ will lead to cocycles that differ from this 
by coboundary terms, but the cohomology class of $e$ in $H^2(G/K,K)$ depends 
only on the extension~(\ref{eq:BasicExtension}) and is 
called the extension class \cite[section IV.3]{brown}.  Since $K$ is finite and abelian, its irreducible representations are all one-dimensional and map $K$ into $\U(1)$.  So
given one of our irreducible representations $\rho_a$, we can apply it to $e$ restricted to $H_a$ to get our discrete torsion $\hat{\om}_a$,
\be
\label{eq:HatRhoForAbelianK}
\hat{\om}_a(h_1,h_2)=\rho_a(s(h_1)s(h_2)s(h_1h_2)^{-1}).
\ee

If in addition, $G$ itself is abelian, 
then the extension class (\ref{eq:ExtensionClass-e}) is symmetric in $q_1$ and $q_2$, and hence so is $\hat{\om}_a$.  But this means that the discrete torsion phases $\e(g,h)=\hat{\om}_a(g,h)/\hat{\om}_a(h,g)$ which appear in the partition function are all unity.  

If $K$ is not abelian, 
one must work harder to define the discrete torsion $\hat{\omega}_a$.  
Before describing 
the general case, let us describe
three examples in which $K$ is abelian.
\begin{itemize}
\item For one example, if $G$ is the trivial extension $K \times H$, with $K$
abelian, then the extension class vanishes, and so
\begin{equation}
{\rm QFT}\left( [X/G] \right) \: = \:
{\rm QFT}\left( \coprod_{\hat{K}} [X/H] \right),
\end{equation}
where no copies have any discrete torsion. 
\item For another example,
if $G=D_4$ and $K = {\mathbb Z}_2$, then using the fact that
$G/K = {\mathbb Z}_2 \times {\mathbb Z}_2$,
and that $D_4 \not\cong {\mathbb Z}_2 \times ( {\mathbb Z}_2 \times
{\mathbb Z}_2)$ (so the extension class is nontrivial),
\begin{equation}
{\rm QFT}\left( [X / D_4] \right) \: = \: 
{\rm QFT}\left( [X/{\mathbb Z}_2 \times {\mathbb Z}_2] \, \coprod \,
[X/{\mathbb Z}_2 \times {\mathbb Z}_2]_{\hat{\om}} \right),
\end{equation}
where the second summand has discrete torsion $\hat{\om}$ given by the nontrivial irreducible representation of $\Z_2$ applied to the extension class, which one can check is the nontrivial
element of $H^2({\mathbb Z}_2 \times {\mathbb Z}_2, U(1))$.
See \cite[section 5.2]{Hellerman:2006zs} 
for a detailed verification that physics
does, indeed, obey this equivalence.
\item For our third example,
let $G = {\mathbb H}$, the eight-element group of
quaternions, and $K = \langle i \rangle \cong {\mathbb Z}_4$.
In this case, $K$ has four irreducible representations, two of which
are invariant under the action of $G/K = {\mathbb Z}_2$, and two of
which are exchanged.  In this case, decomposition predicts
\begin{equation}
{\rm QFT}\left( [X / {\mathbb H} ] \right) \: = \:
{\rm QFT}\left( [X / {\mathbb Z}_2] \, \coprod \, [X/{\mathbb Z}_2] \,
\coprod \, X \right).
\end{equation}
See \cite[section 5.4]{Hellerman:2006zs} for a detailed verification that
physics does, indeed, obey this equivalence.
\end{itemize}

In passing, ultimately because of the Cartan-Leray spectral sequence,
if $X/ (G/K)$ is smooth, then the prescription above will determine
a $B$ field on that quotient, which is in the spirit of the original
phrasing of decomposition in \cite{Hellerman:2006zs}.

Next, we shall describe how the discrete torsion $\hat{\omega}_a$
is determined in the general case, when $K$ need not be abelian.

Since $H_a$ fixes the isomorphism class $[\rho_a]$, it means that for each $h\in H_a$ the representations $\rho_a:K\rr\GL(V_a)$ and $L_h\rho_a$ are isomorphic.  Explicitly, this means that for each $h\in H_a$ we can find an element $f_a(h)\in\GL(V_a)$ so that the following diagram commutes for all $k\in K$,
\begin{equation}  \label{eq:diag:intertwiner}
\xymatrixcolsep{6pc}
\xymatrix{
V_a \ar[r]^{\rho_a(k)} \ar[d]_{f_a(h)} &
V_a \ar[d]^{f_a(h)} \\
V_a \ar[r]^{\rho_a( s_a(h)^{-1} k s_a(h)) } & V_a
}
\end{equation}
The map $f_a:H_a\rr\GL(V_a)$ is called the intertwiner.

Let $G_a = \pi^{-1}(H_a)$, so that we have an exact sequence
\begin{equation}
1 \: \longrightarrow \: K \: \stackrel{\iota}{\longrightarrow} \:
G_a \: \stackrel{\pi}{\longrightarrow} \: H_a \:
\longrightarrow \: 1.
\end{equation}
The idea of the construction will be that for each 
$\rho_a$ we will define a projective representation $\widetilde{\rho}_a$ on $G_a$.  Projective representations are reviewed in appendix~\ref{subapp:ProjReps}.  The associated 2-cocycle $\widetilde{\om}_a$ will turn out to be the pullback of a 2-cocycle on $H_a$ which we'll identify with $\hat{\om}_a^{-1}$.  

Indeed, by restricting our choice of section $s$ to $H_a$, we can write each element of $G_a$ uniquely as $g=s(h)k$ for some $h\in H_a$ and $k\in K$.  Now define a map $\widetilde{\rho}_a:G_a\rr\GL(V_a)$ by
\be
\widetilde{\rho}_a(s(h)k)=f_a(h)^{-1}\rho_a(k).
\ee
We claim that $\widetilde{\rho}_a$ is a projective representation on $G_a$.  Indeed, for any two elements $g_1,g_2\in G_a$, define an operator $C_a(g_1,g_2)$ by
\be
C_a(g_1,g_2)=\widetilde{\rho}_a(g_1)\widetilde{\rho}_a(g_2)\widetilde{\rho}_a(g_1g_2)^{-1}.
\ee
To show that $\widetilde{\rho}_a$ is a projective representation, we need to show that $C_a(g_1,g_2)$ is a scalar multiple of the identity operator $\mathbf{1}\in\GL(V_a)$, say $C_a(g_1,g_2)=\widetilde{\om}_a(g_1,g_2)\mathbf{1}$.  It then follows that
\be
\label{eq:HatRhoMult}
\widetilde{\rho}_a(g_1)\widetilde{\rho}_a(g_2)=\widetilde{\om}_a(g_1,g_2)\widetilde{\rho}_a(g_1g_2).
\ee

Indeed, computing we have (abbreviating $s_a(h_1)=s_1$, $s_a(h_1h_2)=s_{12}$, $f_a(h_2)=f_2$, etc.)
\be
\widetilde{\rho}_a(g_1g_2)=\widetilde{\rho}_a(s_1k_1s_2k_2)=\widetilde{\rho}_a(s_{12}(s_{12}^{-1}s_1s_2)s_2^{-1}k_1s_2k_2)=f_{12}^{-1}\rho_a(s_{12}^{-1}s_1s_2)\rho_a(s_2^{-1}k_1s_2k_2),
\ee
and so
\bea
C_a(g_1,g_2) &=& f_1^{-1}\rho_a(k_1)f_2^{-1}\rho_a(k_2)\rho_a(s_2^{-1}k_1s_2k_2)^{-1}\rho_a(s_{12}^{-1}s_1s_2)^{-1}f_{12},
\non\\
&=& f_1^{-1}f_2^{-1}\rho_a(s_2^{-1}k_1s_2)\rho_a(k_2)\rho_a(k_2)^{-1}\rho_a(s_2^{-1}k_1s_2)^{-1}\rho_a(s_2^{-1}s_1^{-1}s_{12})f_{12},
\non\\
&=& f_1^{-1}f_2^{-1}f_{12}\, \rho_a(s_{12}s_2^{-1}s_1^{-1}).
\eea
Note that $C_a(s(h_1)k_1,s(h_2)k_2)$ is in fact independent of $k_1$ and $k_2$.  To see that $C_a(g_1,g_2)$ is a multiple of the identity, take any $k\in K$ and check that $C_a(g_1,g_2)$ commutes with $\rho_a(k)$,
\bea
C_a(g_1,g_2)\rho_a(k) &=& f_1^{-1}f_2^{-1}f_{12}\rho_a(s_{12}s_2^{-1}s_1^{-1}k)\non\\
&=& \rho_a(ks_{12}s_2^{-1}s_1^{-1})f_1^{-1}f_2^{-1}f_{12}\non\\
&=& \rho_a(k)f_1^{-1}f_2^{-1}f_{12}\, \rho_a(s_{12}s_2^{-1}s_1^{-1})
\: = \: \rho_a(k)C_a(g_1,g_2),
\eea
where we repeatedly used the intertwiner property
\be
\rho_a(k)f_a(h)^{-1}=f_a(h)^{-1}\rho_a(s(h)^{-1}ks(h)).
\ee
Since $\rho_a$ was an irreducible representation of $K$, Schur's lemma tells us that anything that commutes with all the operators $\rho_a(k)$ must be a multiple of the identity.  

Note that associativity of $\widetilde{\rho}_a$ implies the coclosure of $\widetilde{\om}_a$.  Specifically, the cocycle condition follows by plugging (\ref{eq:HatRhoMult}) into
\be
\lp\widetilde{\rho}_a(k_1)\widetilde{\rho}_a(k_2)\rp\widetilde{\rho}_a(k_3)=\widetilde{\rho}_a(k_1)\lp\widetilde{\rho}_a(k_2)\widetilde{\rho}_a(k_3)\rp,
\ee
which clearly holds since the $\widetilde{\rho}_a(k_i)$ are simply products of $\GL(V_a)$ matrices whose multiplication is associative.  Alternatively, one can show by direct calculation that 
\be
C_a(g_2,g_3)C_a(g_1,g_2g_3)C_a(g_1g_2,g_3)^{-1}=C_a(g_1,g_2).
\ee

Finally, we define $\hat{\om}_a(h_1,h_2)$ as the inverse of $\widetilde{\om}_a$, i.e.\ 
\be
\label{eq:decomp:cocycle}
\hat{\om}_a(h_1,h_2)\mathbf{1}=C_a(s_1,s_2)^{-1}=\rho_a(s_1s_2s_{12}^{-1})f_{12}^{-1}f_2f_1.
\ee
The fact that $\widetilde{\om}_a$ is a cocycle implies that $\hat{\om}_a$ is also coclosed, and thus $\hat{\om}_a$ defines a class in\footnote{Technically, our construction only ensures that $\widetilde{\om}_a$ and hence $\hat{\om}_a$ are $\C^\times$-valued, not $\U(1)$-valued.  However, we can always multiply the intertwiners by $\C^\times$ scalars to arrange that their determinants have magnitude one, and in that case our cocycles are $\U(1)$-valued.} $H^2(G/K,\U(1))$.

As a special case, suppose again that $K$ is abelian (but not necessarily
in the center of $G$).  Then all irreducible representations $\rho_a$
are one-dimensional, and the intertwiners $f_a(h)$ are all scalars.
In this case, the cocycle $\hat{\omega}$ given in
equation~(\ref{eq:decomp:cocycle}), modulo coboundaries, reduces to
\begin{equation}
\hat{\omega}(h_1,h_2) \: = \:
\rho_a\left( s_a(h_1) s_a(h_2) s_a(h_1 h_2)^{-1} \right).%^{-1}.
\end{equation}
Now, the product
\begin{equation}
 s_a(h_1) \, s_a(h_2) \, s_a(h_1 h_2)^{-1}
\end{equation}
is precisely the cocycle representing the extension class in
$H^2(H_a,K)$ in the case that
$K$ is abelian \cite[section IV.3]{brown},
\cite[exercise VI.10.1]{hs}, so we see that
$\hat{\omega}$ is the 
image of the extension class of $G_a$ in $H^2(H_a,K)$ under $\rho_a$,
as claimed earlier.
If $K$ is a subgroup of the center of $G$, then each $H_a = G/K$,
and then $\hat{\omega}$ is the 
image of the extension class of $G$
in $H^2(G/K,K)$ under $\rho_a$, also as described earlier.

These statements have been checked in many examples in many ways.
In orbifolds, the original work \cite{Pantev:2005zs,Pantev:2005rh,Pantev:2005wj}
first checked that physics `sees' trivially-acting groups,
through studies of multiloop factorization (target-space unitarity),
as well as
massless spectra and partition functions, which confirmed
that not only does one get distinct theories, but one encounters
physical contradictions if one tries to ignore them. 
Decomposition in such theories was tested in
\cite{Hellerman:2006zs} by comparing partition functions at all genus,
construction of projection operators projecting onto the various
universes, and comparisons of massless spectra and correlation functions,
as well as studies of open string sectors.
Decomposition is also defined for gauge theories with finite
trivially-acting subgroups, and there it has been similarly tested
in both supersymmetric and nonsupersymmetric models, via comparisons
of partition functions and elliptic genera (using supersymmetric localization),
quantum cohomology rings, and mirrors, to name a few.
See for example \cite{Sharpe:2014tca,Sharpe:2019ddn,Eager:2020rra} 
for some more recent discussions and reviews cited therein.

Sometimes, for $K$ abelian, decomposition can be understood in terms
of the existence of finite global one-form symmetries in the theory,
see e.g. \cite{Sharpe:2019ddn} for a recent discussion.
A finite global one-form $K$ symmetry, technically denoted $BK$, 
acts in an orbifold by interchanging the twisted sectors (and in
more general gauge theories, the nonperturbative sectors).  If we think of
each twisted sector contribution to the partition function as associated
with some principal bundle, the action of $BK$ is to tensor\footnote{
These are not vector bundles, so the term `tensor product' is not 
completely accurate, but nevertheless there can exist, under suitable
circumstances, a product structure defined in the obvious fashion.
}
that principal bundle with a principal $K$ bundle to get another
twisted sector contribution.  If $K$ acts trivially, then the two
twisted sectors make identical contributions to the total partition
function.

As a quick consistency check when doing computations,
given a decomposition of the form
\begin{equation}
{\rm QFT}\left( [X/G] \right) \: = \: 
{\rm QFT}\left( \coprod_i [X/G_i]_{\omega_i} \right),
\end{equation}
then it should be true that the total number of irreducible representations
of $G$ should match the sum of the number of $\omega_i$-twisted
irreducible representations of each $G_i$.  (This is a consequence
of the special case that $X$ is a point.)  For example, 
as noted earlier, for $G = D_4$,
decomposition predicts
\begin{equation}
{\rm QFT}\left( [X/D_4] \right) \: = \: 
{\rm QFT}\left( [X/{\mathbb Z}_2 \times {\mathbb Z}_2] \, \coprod \,
[X/{\mathbb Z}_2 \times {\mathbb Z}_2]_{\omega} \right).
\end{equation}
Now, $D_4$ has five irreducible representations,
${\mathbb Z}_2 \times {\mathbb Z}_2$ has four ordinary irreducible
representations, and one irreducible projective representation twisted by the
nontrivial element of $H^2({\mathbb Z}_2 \times {\mathbb Z}_2,U(1))$, as described in appendix~\ref{subapp:ProjReps}.
As expected, $5 = 4 + 1$.  This observation can sometimes be handy
when double-checking expressions for decomposition.

Previous work such as \cite{Hellerman:2006zs} focused on orbifolds (and
related theories) in which a subgroup acted trivially, but no discrete
torsion was turned on that interacted with that subgroup in any way,
aside from making the observation that discrete torsion typically
`broke' the decomposition story, yielding theories with fewer components
than decomposition would predict.  (For example, 
discrete torsion typically breaks much of the finite global one-form symmetry.)
The purpose of this paper is to extend to orbifolds with trivially-acting
subgroups in which discrete torsion has been turned on, and understand what
is happening in such cases.

\section{General conjecture}  \label{sect:conj}

\subsection{Complete argument}

Now, consider an orbifold $[X/G]_{\omega}$ by a finite group $G$ as above, where
one has included discrete torsion, given by some cocycle $\omega$, so
$[\omega] \in H^2(G,U(1))$ (with trivial action on the coefficients).
Suppose as above 
that a normal subgroup $K \subseteq G$ acts trivially on $X$, and
describe $G$ as the extension
\begin{equation} \label{eq:ext}
1 \: \longrightarrow \: K \: \stackrel{\iota}{\longrightarrow} \:
G \: \stackrel{\pi}{\longrightarrow} \: G/K \: \longrightarrow \: 1.
\end{equation}
As before we will choose a section $s:G/K\rr G$ which is not a homomorphism in general.  At the end of the day, nothing physical will depend on the choice of section.

In the previous section, when $\om$ was trivial, we found that the $G$ orbifold decomposed into a number of disjoint pieces, one for each orbit of the $G/K$ action on $\hat{K}$, the set of isomorphism classes of irreducible representations of $K$.  Now, the role of $\hat{K}$ will be replaced by the set $\hat{K}_{\iota^\ast\om}$ of isomorphism classes of irreducible projective representations of $K$ with respect to the cocycle $\iota^\ast\om$ (which, if we view $K$ as a normal subgroup of $G$, is simply the restriction of $\om$ to $K$).  Some facts about projective representations of finite groups are reviewed in appendix~\ref{subapp:ProjReps}.  

Our first task is to understand the natural action of $G/K$ on $\hat{K}_{\iota^\ast\om}$ in this case.  
We can't simply make the naive definition $(L_q\phi)(k)=\phi(s(q)^{-1}ks(q))$, since $L_q\phi$ defined in this way turns out to be projective, 
but not with respect to $\iota^\ast\om$.  
Instead, the cocycle which arises is 
\begin{equation}
\om'(k_1,k_2) \: = \: \om(s(q)^{-1}k_1s(q),s(q)^{-1}k_2s(q)),
\end{equation}
which differs from $\iota^\ast\om$ by a coboundary.  Fortunately, we can cancel coboundary terms by multiplying our projective representations by phases, which leads us to instead make the definition
\be
\label{eq:ProjectiveLqDef}
(L_q\phi)(k)
\: = \:
\frac{
\om(s(q)^{-1}k,s(q))
}{
\om(s(q),s(q)^{-1}k)
}
\,
\phi(s(q)^{-1}ks(q)).
\ee

To check that $L_q\phi$ is a projective representation with respect to $\iota^\ast\om$, we compute (abbreviating $g=s(q)$)
\bea
(L_q\phi)(k_1)(L_q\phi)(k_2) &=& \frac{\om(g^{-1}k_1,g)\om(g^{-1}k_2,g)}{\om(g,g^{-1}k_1)\om(g,g^{-1}k_2)}\phi(g^{-1}k_1g)\phi(g^{-1}k_2g),
\non\\
&=& \frac{\om(g^{-1}k_1,g)\om(g^{-1}k_2,g)}{\om(g,g^{-1}k_1)\om(g,g^{-1}k_2)}\om(g^{-1}k_1g,g^{-1}k_2g)\phi(g^{-1}k_1k_2g),
\non\\
&=& \om(k_1,k_2)(L_q\phi)(k_1k_2),
\eea
where we used the fact that $\omega$ is coclosed and
\begin{multline}
\frac{\om(g^{-1}k_1,g)\om(g^{-1}k_2,g)}{\om(g,g^{-1}k_1)\om(g,g^{-1}k_2)}\om(g^{-1}k_1g,g^{-1}k_2g)\\
=\frac{d\om(g,g^{-1}k_1,g)d\om(g,g^{-1}k_2,g)d\om(g,g^{-1}k_1g,g^{-1}k_2g)d\om(k_1,k_2,g)}{d\om(g,g^{-1}k_1k_2,g)d\om(k_1,g,g^{-1}k_2g)}
\\
\hspace*{0.5in}
\cdot \frac{\om(g^{-1}k_1k_2,g)\om(k_1,k_2)}{\om(g,g^{-1}k_1k_2)}.
\end{multline}

This suggests that we define the action of $G/K$ on $\hat{K}_{\iota^\ast\om}$ by $q\cdot[\phi]=[L_q\phi]$.  Indeed, we have that (abbreviating $s=s(1)\in K$)
\bea
(L_1\phi)(k) &=& \frac{\om(s^{-1}k,s)}{\om(s,s^{-1}k)}\phi(s^{-1}ks)
\: = \:
\frac{\om(s^{-1}k,s)}{\om(s,s^{-1}k)}\frac{\phi(s^{-1})\phi(ks)}{\om(s^{-1},ks)},
\non\\
&=& \frac{\om(s^{-1}k,s)}{\om(s,s^{-1}k)\om(s^{-1},ks)}\frac{\om(s,s^{-1})\phi(s)^{-1}\phi(k)\phi(s)}{\om(k,s)},
\non\\
&=& \frac{d\om(s,s^{-1}k,s)}{d\om(s,s^{-1},ks)}\phi(s)^{-1}\phi(k)\phi(s)
\: = \: 
\phi(s)^{-1}\phi(k)\phi(s),
\eea
where we used the cocycle condition on $\om$ (meaning the $d\om$'s evaluate to $1$) and the projective representation properties (\ref{eq:GeneralProjRepMult}) and (\ref{eq:GeneralProjRepInverse}).  So $L_1\phi$ is isomorphic to $\phi$ and hence $1\cdot[\phi]=[\phi]$.  Similarly, we can show that $L_{q_1}(L_{q_2}\phi)$ is isomorphic to $L_{q_1q_2}\phi$, that if $\phi_1$ and $\phi_2$ are isomorphic, then so are
$L_q\phi_1$ and $L_q\phi_2$, and finally that if $\phi$ is irreducible then so is $L_q\phi$.  This shows that we do have a good $G/K$ action on $\hat{K}_{\iota^\ast\om}$.

Note that even when $\iota^\ast\om$ is trivial and $K$ is a central subgroup of $G$ 
(so that the usual $G/K$ actions on $K$ and $\hat{K}$ are trivial), 
this action we have constructed is not necessarily trivial!  
In other words, the choice of prefactors required in (\ref{eq:ProjectiveLqDef}) 
to make the projectivity work out  in general have the side effect that they 
can lead to a nontrivial action even in the non-projective case, 
and this will turn out to be crucial in correctly accommodating some examples.

One example that we will explore in more detail below is to take $G=\Z_2\times\Z_2=\{1,a,b,c\}$, $K=\{1,a\}\cong\Z_2$, so $G/K=\{K,bK\}\cong\Z_2$.  We'll choose our section to be $s(K)=1$, $s(bK)=b$.  If we turn on the nontrivial discrete torsion (\ref{eq:Z2Z2NontrivialCocycle}) in $G$, then $\iota^\ast\om$ is trivial (since $\om(a,a)=1$).  In this case $\hat{K}=\{[\rho_+],[\rho_-]\}$, where $\rho_\pm$ are one-dimensional irreducible representations of $K$ defined by their action on the generator, $\rho_\pm(a)=\pm 1$.  The usual action of $G/K$ on $\hat{K}$, as constructed in the previous section, would be trivial, but instead the action we have defined above satisfies
\be
(L_{bK}\phi)(a)=\frac{\om(b^{-1}a,b)}{\om(b,b^{-1}a)}\phi(b^{-1}ab)=\frac{\om(c,b)}{\om(b,c)}\phi(a)=-\phi(a),
\ee
and hence $bK\cdot[\rho_\pm]=[\rho_\mp]$.  Thus, in this example, $L_{bK}$
exchanges different representations, and so
$\hat{K}$ consists of only a single $G/K$ orbit, with representative $[\rho_+]$ and trivial stabilizer subgroup $H_+\cong 1$.

More generally, define
\begin{equation}
\beta(\omega): \: G/K \times K \: \longrightarrow \: U(1)
\end{equation}
by
\begin{equation}
\label{eq:FirstBetaDef}
\beta(\omega)(q,k) \: = \: 
\frac{
\omega(k s(q), s(q)^{-1} )
}{
\omega(s(q)^{-1}, k s(q) )
}.
\end{equation}
(We will discuss this function in greater generality and detail
in appendix~\ref{sect:beta},
where we will see that for $\iota^* \omega$ trivial,
it represents an element of $H^1(G/K, H^1(K,U(1))$.)
(When $K$ is central, $\beta(\omega)$ is invariant under coboundaries
and so gives a well-defined $U(1)$ phase.)  In all cases,
we will see that $\beta$ also
defines a suitable homomorphism in appendix~\ref{sect:beta}.
We will also see in appendix~\ref{sect:beta} that the phase factor
in $L_q \phi$ is $\beta(\omega)^{-1}$, so that
\begin{equation}
(L_q \phi)(k) \: = \: \beta(\omega)(q,k)^{-1} 
\phi( s(q)^{-1} k s(q) ).
\end{equation}
Later we will use $\beta$ to give a more efficient approach to
decomposition with discrete torsion.

Now that we have defined our $G/K$ action, we can decompose $\hat{K}_{\iota^\ast\om}$ into $G/K$ orbits, and choose representatives $[\rho_a]$, where $\rho_a$ is an irreducible projective representation of $K$ chosen to stand in for its isomorphism class.  Let $H_a\subseteq G$ be the stabilizer subgroup of $[\rho_a]$.  Our conjecture is that the $G$-orbifold with discrete torsion $\om$ decomposes into a disjoint union of theories, one for each orbit $[\rho_a]$, with orbifold group $H_a$ and discrete torsion $\hat{\om}_a$ which we will construct below.  In other words, schematically we have
\begin{equation}   %\label{eq:case1:predict}
{\rm QFT}\left( [X/G]_{\omega} \right) \: = \:
{\rm QFT}\left( \left[ \frac{ X \times \hat{K}_{\iota^* \omega} }{ G/K }
\right]_{\hat{\omega}} \right),
\end{equation}
or with slightly more detail,
\be
[X/G]_{\om}=\coprod_a[X/H_a]_{\hat{\om}_a}.
\ee
It remains to construct the cocycles $\hat{\om}_a\in Z^2(H_a,\U(1))$.

We will follow a similar approach to what we did in the previous section.  Given the projective irreps $\rho_a$ of $K$, we will construct projective irreps $\widetilde{\rho}_a$ for $G_a=\pi^{-1}(H_a)$.  The steps will all be the same, but we will have to be more careful with $\om$-related phase factors.  The statement that $H_a$ is the stabilizer of $[\rho_a]$ means that there exist intertwiners $f_a(h)\in\GL(V_a)$ for each $h\in H_a$ such that the diagrams
\begin{equation} %\label{eq:intertwine:omega}
\xymatrixcolsep{5pc}
\xymatrix{
V_a \ar[r]^{\rho_a(k)} \ar[d]_{f_a(h)} & V_a \ar[d]^{f_a(h)} \\
V_a \ar[r]_{(L_h \rho_a)(k)} & V_a
}
\end{equation}
commute for each $k\in K$.  In other words, the intertwiners satisfy
\be
\rho_a(k)f_a(h)^{-1}=\frac{\om(s(h)^{-1}k,s(h))}{\om(s(h),s(h)^{-1}k)}f_a(h)^{-1}\rho_a(s(h)^{-1}ks(h)).
\ee
As before, each element $g$ of $G_a$ can uniquely be written as $g=s(h)k$ for some $h\in H_a$ and $k\in K$, and we can define a map $\widetilde{\rho}_a:G_a\rr\GL(V_a)$ by
\be
\label{eq:TildeRhoWithPhaseDef}
\widetilde{\rho}_a(s(h)k)=\om(s(h),k)^{-1}f_a(h)^{-1}\rho_a(k).
\ee
The phase $\om(s(h),k)^{-1}$ is chosen to make the formulas below cleaner.  A different choice would lead to a projective representation whose cocycle differed by a coboundary.

To check that this is indeed a projective representation and to find the associated cocycle, we first compute
\bea
\widetilde{\rho}_a(g_1g_2) &=& \om(s_{12},s_{12}^{-1}s_1k_1s_2k_2)^{-1}f_{12}^{-1}\rho_a(s_{12}^{-1}s_1k_1s_2k_2)\non\\
&=& \om(s_{12},s_{12}^{-1}s_1k_1s_2k_2)^{-1}\om(s_{12}^{-1}s_1s_2,s_2^{-1}k_1s_2k_2)^{-1}\om(s_2^{-1}k_1s_2,k_2)^{-1}\non\\
&& \qquad\times f_{12}^{-1}\rho_a(s_{12}^{-1}s_1s_2)\rho_a(s_2^{-1}k_1s_2)\rho_a(k_2),
\eea
where we used (\ref{eq:GeneralProjRepMult}) twice to factorize $\rho_a(s_{12}^{-1}s_1k_1s_2k_2)$.  Then we define an operator
\be
\label{eq:GeneralCDef}
C_a(g_1,g_2)=\widetilde{\rho}_a(g_1)\widetilde{\rho}_a(g_2)\widetilde{\rho}_a(g_1g_2)^{-1}.
\ee

One can show that this operator is a scalar multiple of the identity, 
\begin{equation}
C_a(g_1,g_2) \: = \: \widetilde{\om}_a(g_1,g_2)\mathbf{1},
\end{equation}
specifically,
\be
\label{eq:Ingredient3}
C_a(g_1,g_2)=\frac{\om(s_1k_1,s_2k_2)\om(s_1s_2s_{12}^{-1},s_{12})}{\om(s_1,s_2)\om(s_{12}s_2^{-1}s_1^{-1},s_1s_2s_{12}^{-1})}f_1^{-1}f_2^{-1}f_{12}\,\rho_a(s_{12}s_2^{-1}s_1^{-1}),
\ee
hence $\widetilde{\rho}_a$ is a projective representation of $G_a$ with cocycle $\widetilde{\om}_a$.  Since the details of the calculation are not particularly enlightening, they are relegated to appendix~\ref{app:SomeCocycleCalculations}.

It is convenient to write the cocycle which appears as
\be
\label{eq:Ingredient2}
\widetilde{\om}_a(g_1,g_2)=\frac{\om(g_1,g_2)}{\hat{\om}_a(h_1,h_2)},
\ee
where $\hat{\om}_a$ can be defined by combining (\ref{eq:Ingredient2}) and (\ref{eq:Ingredient3}) and using (\ref{eq:GeneralProjRepInverse}) to get
\be
\label{eq:GeneratHatOmegaDef}
\hat{\om}_a(h_1,h_2)\mathbf{1}=\frac{\om(s_1,s_2)}{\om(s_1s_2s_{12}^{-1},s_{12})}\rho_a(s_1s_2s_{12}^{-1})f_{12}^{-1}f_2f_1,
\ee
generalizing expression~(\ref{eq:decomp:cocycle}) for ordinary
decomposition.

From the definition (\ref{eq:GeneralCDef}) of $C_a(g_1,g_2)$, it now follows that $\widetilde{\rho}_a$ is a projective representation on $G_a$ with respect to the cocycle $\widetilde{\om}_a$,
\be
\widetilde{\rho}_a(g_1) \, \widetilde{\rho}_a(g_2)
\: = \: 
\widetilde{\om}_a(g_1,g_2) \, \widetilde{\rho}_a(g_1g_2).
\ee
The manifest associativity of the matrices $\widetilde{\rho}_a(g)$ implies the cocycle condition for $\widetilde{\om}_a$, and combined with the fact that $\om$ is also coclosed we learn that $\hat{\om}_a$ is a cocycle and hence defines a class in $H^2(H_a,\U(1))$.  This is the discrete torsion which will appear in the correspoonding factor of the decomposition.

Let's note some aspects of this result.  The factor $\om(s_1,s_2)$ which appears in $\hat{\om}_a$ is essentially the pullback of $\om$ along the section $s$, i.e.\ $\om(s_1,s_2)=(s^\ast\om)(h_1,h_2)$.  Suppose that $\om=\pi^\ast\bar{\om}$ is the pullback of some cocycle $\bar{\om}\in H^2(G/K,\U(1))$.  In that case we have $s^\ast\om=\bar{\om}$ and the other $\om$ factor in (\ref{eq:GeneratHatOmegaDef}) equals one, so we see that $\hat{\om}_a$ is simply $\bar{\om}$ multiplied by the result (\ref{eq:decomp:cocycle}) that we obtained in the ordinary decomposition case.  It's interesting to observe also that in this case $\om=\pi^\ast\bar{\om}$ drops out entirely from $\widetilde{\om}_a$; the projective representation $\widetilde{\rho}_a$ is insensitive to the presence of $\bar{\om}$.

\subsection{Summary}

As this has been a somewhat long-winded discussion,
in this section we summarize the highlights.
We can essentially break this into three cases,
determined by the image of the discrete torsion $\omega \in
H^2(G,U(1))$ under various maps defined by the short exact
sequence~(\ref{eq:ext}).

\begin{enumerate}
\item If $\iota^* \omega \neq 0$,
then
\begin{equation}   \label{eq:case1:predict}
{\rm QFT}\left( [X/G]_{\omega} \right) \: = \:
{\rm QFT}\left( \left[ \frac{ X \times \hat{K}_{\iota^* \omega} }{ G/K }
\right]_{\hat{\omega}} \right),
\end{equation}
where $\hat{K}_{\iota^* \omega}$ denotes the set of irreducible
projective representations of $K$, defined with respect to
$\iota^* \omega \in H^2(K,U(1))$, and $\hat{\omega}$ refers to
discrete torsion we defined in the previous subsection.

Without the twisting, when the cocycle $\iota^* \omega$ vanishes,
the number of irreducible ordinary representations
of a finite group is counted by the number of
conjugacy classes.  As reviewed in appendix~\ref{subapp:ProjReps},
the number of irreducible
projective representations of a finite group $K$ twisted by
some $\omega \in H^2(K,U(1))$ is equal to the number of conjugacy
classes in $K$ which consist of $g \in K$ such that
$\omega(g,h) = \omega(h,g)$ for all $h \in K$ that commute with $g$:
\begin{equation}   \label{eq:countprojirreps}
| \hat{K}_{\omega} | \: = \: | \{ [g] \, | \, \omega(g,h) = \omega(h,g)
\mbox{ for all } h \mbox{ s.t. } hg = gh \} |.
\end{equation}

We can slightly simplify the prediction~(\ref{eq:case1:predict})
as follows.
Let $\{ \rho_a \}$ be a collection of representatives of the
orbits of $G/K$ on $\hat{K}_{\iota^* \omega}$.
For each $\rho_a$, let $H_a \subset G/K$ be the stabilizer of
$[\rho_a]$ in $\hat{K}_{\iota^* \omega}$.  
Then,
\begin{equation}
\left[ \frac{ X \times \hat{K}_{\iota^* \omega} }{ G/K }
\right]_{\hat{\omega}}
\: = \:
\coprod_a \left[ X / H_a \right]_{\hat{\omega}_a},
\end{equation}
where the details of the discrete torsion $\hat{\omega}_a$ was
described in the previous subsection.

\item Suppose that $\omega$ is annihilated by $\iota^*$.  As we have
seen, the $G/K$ action on the irreducible representations can
still involve phases, which can exchange
representations.  Those phases are 
determined by $\beta(\omega) \in H^1(G/K, H^1(K,U(1))$,
as described in appendix~\ref{sect:beta}, where following
\cite{hochschild}, and as we describe in detail in
appendix~\ref{sect:beta}, $\beta$ itself
is the map in 
the following\footnote{
This is one piece of the seven-term exact sequence
(see e.g. \cite{hochschild})
\begin{eqnarray}
\lefteqn{
0 \: \longrightarrow \: H^1(G/K, U(1)) \: \stackrel{\pi^*}{\longrightarrow} \:
H^1(G, U(1)) \: \stackrel{\iota^*}{\longrightarrow} \:
H^1(K, U(1)) \: \stackrel{d_2}{\longrightarrow} \:
H^2(G/K, U(1))
} \nonumber \\
& &
 \: \stackrel{\pi^*}{\longrightarrow} \:
{\rm Ker}\left( \iota^*  \right)
\: \stackrel{\beta}{\longrightarrow} \:
H^1( G/K, H^1(K,U(1)))
\: \stackrel{d_2}{\longrightarrow} \:
H^3(G/K, U(1))
\end{eqnarray}
that is a consequence of the Lyndon-Hochschild-Serre spectral sequence
\cite{hochserre},
slightly generalizing the inflation-restriction exact sequence
(see e.g. \cite{hochserre},
\cite[example 6.8.3]{weibel}, \cite[section I.6]{neukirch},
\cite[section 3.3]{gille-szamuely})
\begin{equation}
0 \: \longrightarrow \: H^1(G/K, U(1)) \: \stackrel{\pi^*}{\longrightarrow} \:
H^1(G, U(1)) \: \stackrel{\iota^*}{\longrightarrow} \:
H^1(K, U(1)) \: \stackrel{d_2}{\longrightarrow} \:
H^2(G/K, U(1)) \: \stackrel{\pi^*}{\longrightarrow} \: H^2(G,U(1))
\end{equation}
(where we have specialized to trivial action on the coefficients).
} exact sequence:
\begin{equation} \label{eq:3termessential}
H^2(G/K, U(1)) \: \stackrel{\pi^*}{\longrightarrow} \:
{\rm Ker}\left(\iota^*:  H^2(G,U(1)) \longrightarrow H^2(K,U(1)) \right)
\: \stackrel{\beta}{\longrightarrow} \: H^1(G/K, H^1(K,U(1))).
\end{equation}

If $\beta(\omega)$ is nontrivial in cohomology, and if $K$ is
in the center of $G$ (so that in particular $\hat{K} = H^1(K,U(1))$), then our prescription can
be efficiently summarized as
\begin{equation}
{\rm QFT}\left( [X/G]_{\omega} \right) \: = \:
{\rm QFT}\left( \left[ \frac{X \times  {\rm Coker}(\beta(\omega)) } {
{\rm Ker}(\beta(\omega)) } \right]_{\hat{\omega}} \right),
\end{equation} 
where we interpret $\beta(\omega)$ as a homomorphism
$G/K \rightarrow \hat{K}\cong H^1(K,U(1))$, the irreducible representations of
 $K$, and where we gave
the discrete torsion $\hat{\omega}$ in the previous subsection.
(If $K$ is not central then we can still appeal to the methods of the
previous subsection to understand decomposition; however,
we will still classify examples according to the (non)triviality of
$\beta(\omega)$ in cohomology when $\iota^* \omega$ is trivial.)

We can simplify this expression slightly as follows.
Let $\{ \rho_a \}$ be a collection of representatives of the
orbits of Ker $\beta(\omega)$ on the set of irreducible (honest) representations
of the cokernel.  For each $\rho_a$, 
let $H_a \subset {\rm Ker} \, \beta(\omega)$ be the stabilizer of $\rho_a$,
then
\begin{equation}
\left[ \frac{X \times  {\rm Coker}(\beta(\omega)) } {
{\rm Ker}(\beta(\omega)) } \right]_{\hat{\omega}} 
\: = \:
\coprod_a [X/H_a]_{\hat{\omega}_a}.
\end{equation}

We shall argue in section~\ref{sect:exs:b} 
that in fact $\beta$ defines an analogue
of a quantum symmetry.
\item Finally, suppose $\iota^* \omega = 0$ and $\beta(\omega) = 0$.  Then,
there exists $\overline{\omega} \in
H^2(G/K,U(1))$ such that $\omega = \pi^* \overline{\omega}$.
In this case,
\begin{equation}
{\rm QFT}\left( [X/G]_{\omega} \right) \: = \:
{\rm QFT}\left( \left[ \frac{ X \times \hat{K}}{ G/K } 
\right]_{\hat{\omega}} \right).
\end{equation}

We will describe the discrete torsion $\hat{\omega}$ later in this
section.  Briefly, if $K$ is abelian, then 
\begin{equation}
\hat{\omega} \: = \: \overline{\omega} \: + \: \hat{\omega}_0,
\end{equation}
where $\hat{\omega}_0$ is the discrete torsion predicted by
ordinary decomposition as in section~\ref{sect:rev},
and $\omega = \pi^* \overline{\omega}$.

For example, if $K$ is in the center of $G$, so that $G/K$ acts
trivially on $\hat{K}$, then
\begin{equation}
\left[ \frac{ X \times \hat{K}}{ G/K } 
\right]_{\overline{\omega} + \hat{\omega}_0}
\: = \: \coprod_{\rho \in \hat{K}} [X / (G/K) ]_{\overline{\omega} + \hat{\omega}_0(
\rho) },
\end{equation}
where $\hat{\omega}_0(\rho)$ is the image under $\rho$ of the extension class
of $G$ in $H^2(G/K,K)$:
\begin{equation}
\rho: \: H^2(G/K,K) \: \longrightarrow \: H^2(G/K,U(1)).
\end{equation}
\end{enumerate}

In the special case $\omega = 0$, this reduces to the
ordinary decomposition story reviewed in section~\ref{sect:rev}.

The third case above ($\omega = \pi^* \overline{\omega}$)
can be viewed more formally as a $K$-gerbe over 
$[X/G]_{\overline{\omega}}$, and so this case
can be thought of
as another example of decomposition, in the spirit of
\cite{Hellerman:2006zs}.

One of the motivations for the original decomposition conjecture
was the open string sector.  A group that acts trivially on
bulk states, may act nontrivially on boundary states, and so the
boundary states decompose into representations of the group.
For example, K theory on gerbes decomposes, as was reviewed
in \cite{Hellerman:2006zs}.  Here, we have a slightly more complicated
situation, in that discrete torsion in open strings `twists' ordinary
representations into projective representations 
\cite{Douglas:1998xa,Sharpe:2000ki}.  An open-string interpretation
of that result over gerbes in the second two cases is somewhat beyond
the scope of this article, but its interpretation is immediate in the
first case, and is the underlying physics reason for the appearance
of $\hat{K}_{\iota^* \omega}$, the set of irreducible projective 
representations of the trivially-acting subgroup.

A quick consistency check closely akin to that for ordinary
decomposition also applies to these theories.
Given a decomposition of the form
\begin{equation}
{\rm QFT}\left( [X/G]_{\omega} \right) \: = \: 
{\rm QFT}\left( \coprod_i [X/G_i]_{\omega_i} \right),
\end{equation}
then it should be true that the total number of 
$\omega$-twisted irreducible representations
of $G$ should match the sum of the number of $\omega_i$-twisted
irreducible representations of each $G_i$.  (This is a consequence
of the special case that $X$ is a point.) 
This observation can sometimes be handy
when double-checking expressions for decomposition.

Many of the resulting theories have multiple disconnected components.
This reflects some subtle ``one-form'' symmetries of the theory,
as can typically be seen in two ways:
\begin{itemize}
\item An orbifold has a one-form symmetry denoted $BH$ when its genus-one
twisted sectors can be exchanged by tensoring\footnote{
`Tensoring' is perhaps not precisely the right term, as these are
principal bundles not vector bundles, but nevertheless there is a product
operation defined in the obvious way.
} with $H$ bundles, for $H$ abelian.  In the presence of discrete torsion,
one must be careful, as discrete torsion weights different twisted sectors
with different phases, and so can break such symmetries, but there can be
residual symmetries remaining.  
\item A sigma model on a disjoint union of $k$ identical target spaces
has a $B H$ symmetry for $|H| = k$, as explained in
\cite{Sharpe:2019ddn}.
\end{itemize}
One consistency test of the proposal above is that, in examples,
both sides (the original orbifold and its final simplified description)
have the same one-form symmetry.
We shall discuss this phenomenon in examples as it arises.

In the remainder of this paper, we will work through
a number of examples, to illustrate and test the various features
of the prediction above.
Among other things,
we will see that in cases where a choice of normal subgroups $K$ 
is ambiguous,
the resulting prediction for the field theories for all choices
of $K$ will be the
same, so that the physics is well-defined.  
For example, the ${\mathbb Z}_2 \times {\mathbb Z}_2$ orbifold
of a point, with discrete torsion, can be considered either as an
example with $K = {\mathbb Z}_2 \times {\mathbb Z}_2$ and 
$\iota^* \omega \neq 0$, as we discuss in section~\ref{sect:ex:pt},
or as an example with $K = {\mathbb Z}_2$, $\iota^* \omega = 0$, and
$\beta(\omega) \neq 0$, as we discuss in section~\ref{sect:z2z2kz2}.
From both perspectives, we will get the same physics, that the theory
is equivalent to a sigma model on a single point.  The latter perspective
ties into quantum symmetries in abelian orbifolds, as we discuss
in section~\ref{sect:z2z2kz2}.

In previous work such as \cite{Hellerman:2006zs}, decomposition in
orbifolds was tested extensively, by e.g. computing not just genus-one
partition functions, but also partition functions at all genera,
as well as projection operators, massless spectra, comparing open string
states and K theory, and more.  However, as the basic point of decomposition
now seems well-established, in this paper for brevity
we will test our claims merely by
computing genus-one partition functions.

\section{Examples in which $\iota^* \omega \neq 0$}
\label{sect:exs:iw}

In this section we will consider examples of orbifolds
$[X/G]$ with discrete torsion $\omega$, where
$\iota^* \omega \neq 0$.

\subsection{Orbifold of a point with discrete torsion}
\label{sect:ex:pt}

Consider an orbifold of a point by a finite group $G$ with
discrete torsion.  Since all of $G$ acts trivially, we have
$K=G$, hence $\iota^* \omega = \omega \neq 0$ (by assumption).

From section~\ref{sect:conj}, this theory is predicted to be the same
as that of $N$ points, where $N$ is the number of irreducible projective
representations of $G$ with respect to $\omega$.
Using \cite[equ'n (6.40)]{Dijkgraaf:1989pz} 
\begin{equation}
N \: = \: \frac{1}{|G|} \sum_{gh = hg} 
\frac{ \omega(g,h) }{\omega(h,g)},
\end{equation}
we see immediately that the genus-one partition function of the
$G$ orbifold with discrete torsion is
always the same as that of $N$
points.

We collect here a number of special cases, to more explicitly
check this claim.
\begin{itemize}
\item $G = {\mathbb Z}_2 \times {\mathbb Z}_2$.  As discussed
in appendix~\ref{app:z2z2}, there is only one irreducible projective
representation of $G$, and given the cocycle given there, it is
also straightforward to compute that the genus-one partition function is $1$.
\item $G = ({\mathbb Z}_2)^3$.  Assume that the discrete torsion is in
one ${\mathbb Z}_2 \times {\mathbb Z}_2$ factor.  Then, the total number
of irreducible projective representations is two -- the tensor product
of one irreducible projective representation of 
${\mathbb Z}_2 \times {\mathbb Z}_2$ and two irreducible honest
representations of ${\mathbb Z}_2$ -- and it is straightforward to
compute that the genus-one partition function is $2$.

The reader should also note in this case that because
of the ${\mathbb Z}_2$ factor,
this theory has a $B {\mathbb Z}_2$ symmetry, which is consistent
with the fact that it decomposes into two equal factors 
\cite{Sharpe:2019ddn}.
\item $G = {\mathbb Z}_2 \times {\mathbb Z}_4$.  As discussed
in appendix~\ref{app:z2z4}, there are two irreducible projective
representations of ${\mathbb Z}_2 \times {\mathbb Z}_4$ with
nontrivial discrete torsion, and utilizing the phases in
table~\ref{table:z2z4-phases}, it is straightforward to compute
that the genus-one partition function is
\begin{equation}
Z \: = \: \frac{1}{|{\mathbb Z}_2 \times {\mathbb Z}_4|}
\sum_{gh = hg} \epsilon(g,h) \: = \:
\frac{1}{8} (16) \: = \: 2 \: = \: Z\left( \mbox{two points} \right),
\end{equation}
where $\e(g,h)=\om(g,h)/\om(h,g)$, 
in agreement with the prediction.

The reader should also note that the 
discrete torsion phases in table~\ref{table:z2z4-phases} are
periodic under multiplication of group elements by
the subgroup $\langle b^2 \rangle \cong {\mathbb Z}_2$.
As a result, this theory has a $B {\mathbb Z}_2$ (one-form) symmetry,
interchanging twisted sectors weighted by the same phase, which is
consistent with the fact that this theory describes a disjoint union
of two objects (points) \cite{Sharpe:2019ddn}.
\item $G = D_4$.  As discussed in appendix~\ref{app:d4},
there are two irreducible projective representations of $D_4$,
and given the phases in table~\ref{table:d4-phases}, 
it is also straightforward to compute
the genus-one partition function.  Given that there are $28$ ordered
commuting pairs for which $\epsilon(g,h) = +1$,
and 12 ordered commuting pairs for which $\epsilon(g,h) = -1$,
one finds
\begin{equation}
Z([{\rm point}/D_4]) \: = \: \frac{1}{| D_4| } \left( 28 - 12 \right) \: = \: 2
\: = \: Z( \mbox{two points} ),
\end{equation}
agreeing with the prediction for this case.

As the result has two equivalent disconnected components,
this theory has a $B {\mathbb Z}_2$ symmetry, just like the previous
case.  Unlike the previous case, this $B {\mathbb Z}_2$ does not seem
to have a group-theoretic origin in $D_4$ or its discrete torsion itself.
Instead, because this is an orbifold of a point, the contribution of
each twisted sector is determined solely by the discrete torsion phase
$\epsilon(g,h)$ -- so there is room for other permutations of twisted
sectors, unrelated to group theory, which appears to be what is happening
in this case, and one expects, in most cases of orbifolds of points.  
More to the point, it also seems to be true of the other examples in this
list, so we will not explicit discuss these symmetries further in
this section.
\item $G = {\mathbb Z}_4 \rtimes {\mathbb Z}_4$.  
As discussed in appendix~\ref{app:z4sz4}, there are four irreducible
projective representations of ${\mathbb Z}_4 \rtimes {\mathbb Z}_4$.
Utilizing the phases in table~\ref{table:z4z4-phases}, it is 
straightforward to compute that the genus-one partition function is
\begin{equation}
Z \: = \: \frac{1}{| {\mathbb Z}_4 \rtimes {\mathbb Z}_4 | }
\sum_{gh=hg} \epsilon(g,h) \: = \: 4 \: = \: Z\left( \mbox{four points} \right),
\end{equation}
in agreement with the prediction.
\item $G = S_4$.  As discussed in appendix~\ref{app:s4},
there are three irreducible projective representations of $S_4$.
Adding up the contributions, weighted by signs as given
in table~\ref{table:s4-nums}, one finds
\begin{equation}
Z\left( [ {\rm point}/S_4] \right) \: = \:
 \frac{1}{|S_4|} (72) Z({\rm point})
\: = \: (3) Z({\rm point}),
\end{equation}
agreeing with the prediction.
\end{itemize}

\subsection{$[X/{\mathbb Z}_4 \rtimes {\mathbb Z}_4]$ with discrete torsion
and trivially-acting ${\mathbb Z}_2 \times {\mathbb Z}_4$ subgroup}
\label{sect:iw:z4sz4-z2z4}

In this section we consider an orbifold by the semidirect product of
two copies of ${\mathbb Z}_4$, ${\mathbb Z}_4 \rtimes {\mathbb Z}_4$.
It can be shown (see appendix~\ref{app:z4sz4}) that $H^2(
{\mathbb Z}_4 \rtimes {\mathbb Z}_4,U(1)) = {\mathbb Z}_2$, so there
is one nontrivial value $\omega$ of discrete torsion, which we turn on in this
orbifold.  Furthermore, we take the action of the subgroup
$K = \langle x^2, y \rangle \cong {\mathbb Z}_2 \times {\mathbb Z}_4$
(in the notation of appendix~\ref{app:z4sz4}) to be trivial.

Now, $H^2(K,U(1)) = {\mathbb Z}_2$ in this case (see
appendix~\ref{app:z2z4}), so in principle, $\iota^* \omega$
could be nonzero.  To check, we note that, from table~\ref{table:z4z4-phases},
for $\iota^* \omega$ the genus-one phase $\epsilon(x^2,y) = -1$,
so $\iota^* \omega$ will be nontrivial, and as $H^2(K,U(1)) = {\mathbb Z}_2$,
we see that there is only one nontrivial choice.

For completeness, let us also compute the full cocycle.
Applying table~\ref{table:z4z4-cocycle}, the full cocycle for
$\iota^* \omega$ is given in table~\ref{table:pullback-z4sz4-cocycle}.

\begin{table}[h]
\begin{center}
\begin{tabular}{c|cccccccc}
& $1$ & $x^2$ & $y$ & $x^2y$ & $y^2$ & $x^2 y^2$ & $y^3$ & $x^2y^3$ \\ \hline
$1$ & $1$ & $1$ & $1$ & $1$ & $1$ & $1$ & $1$ & $1$ \\
$x^2$ & $1$ & $1$ & $+i$ & $-i$ & $1$ & $1$ & $+i$ & $-i$ \\
$y$ & $1$ & $-i$ & $1$ & $+i$ & $1$ & $-i$ & $1$ & $+i$ \\
$x^2y$ & $1$ & $+i$ & $-i$ & $1$ & $1$ & $+i$ & $-i$ & $1$ \\
$y^2$ & $1$ & $1$ & $1$ & $1$ & $1$ & $1$ & $1$ & $1$ \\
$x^2y^2$ & $1$ & $1$ & $+i$ & $-i$ & $1$ & $1$ & $+i$ & $-i$ \\
$y^3$ & $1$ & $-i$ & $1$ & $+i$ & $1$ & $-i$ & $1$ & $+i$ \\
$x^2y^3$ & $1$ & $+i$ & $-i$ & $1$ & $1$ & $+i$ & $-i$ & $1$
\end{tabular}
\caption{Cocycle for $\iota^* \omega$ for $\omega$ the nontrivial
element of $H^2({\mathbb Z}_4 \rtimes {\mathbb Z}_4,U(1))$,
and $\iota: {\mathbb Z}_2 \times {\mathbb Z}_4 \hookrightarrow
{\mathbb Z}_4 \rtimes {\mathbb Z}_4$.
\label{table:pullback-z4sz4-cocycle}
}
\end{center}
\end{table}

The cocycle for $\iota^* \omega$ matches the cocycle for the
nontrivial element of $H^2({\mathbb Z}_2 \times {\mathbb Z}_4,U(1))$
given in table~\ref{table:z2z4-cocycle}.  (In principle, they only needed
to match up to a coboundary; it is a reflection of our conventions that they
happen to match on the nose.)

Thus, we see that $\iota^* \omega \neq 0$.  Furthermore, $G/K = 
{\mathbb Z}_2$.  Our conjecture of section~\ref{sect:conj} then predicts
that
\begin{equation}
{\rm QFT}\left( [X/{\mathbb Z}_4 \rtimes {\mathbb Z}_4]_{\omega}
\right) \: = \: 
{\rm QFT}\left( \left[ \frac{ X \times \hat{K}_{\iota^* \omega} }{
G/K} \right] \right)
\: = \:
{\rm QFT}\left( \left[ \frac{ X \times \hat{K}_{\iota^* \omega} }{
{\mathbb Z}_2} \right] \right).
\end{equation}
As discussed in appendix~\ref{app:z2z4}, there are only two
irreducible projective representations of $K = {\mathbb Z}_2 \times
{\mathbb Z}_4$, corresponding to the conjugacy classes
$\{1\}$, $\{b^2 = y^2\}$.  Furthermore, the prefactors (\ref{eq:FirstBetaDef}) which appears in the definition of $(L_q\phi)(k)$ is unity for all $q$ and $k$.  This combined with the fact that 
conjugation by $G$ leaves
both conjugacy classes invariant, implies\footnote{\label{footnote:GModKActionSubtlety} This is a consequence of the fact that the decomposition of a projective representation into irreducible projective representations is determined by the projective characters which are given by tracing over the matrices of the representation.  One can show that the projective characters vanish except on $\om$-trivial conjugacy classes and hence two irreducible projective representations are isomorphic if and only if their characters agree on all $\om$-trivial conjugacy classes.  In particular, if $\beta(\om)(q,k)=1$ for all $q$ and $k$ and the $G/K$ action preserves all of the $\iota^\ast\om$-trivial conjugacy classes of $K$, then $L_q\phi$ and $\phi$ will be isomorphic.}  that $G/K$ acts trivially on
$\hat{K}_{\iota^* \omega}$.  Therefore, we make the prediction that
\begin{equation}
{\rm QFT}\left( [X/{\mathbb Z}_4 \rtimes {\mathbb Z}_4]_{\omega}
\right) \: = \: 
{\rm QFT}\left( \left[ \frac{ X \times \hat{K}_{\iota^* \omega} }{
G/K} \right] \right)
\: = \:
{\rm QFT}\left( [X/{\mathbb Z}_2] \, \coprod \, [X/{\mathbb Z}_2] \right),
\end{equation}
two copies of the orbifold $[X/{\mathbb Z}_2]$.

It is straightforward to check this in genus-one partition functions.
If we let $\overline{x}$ denote the generator of $G/K = {\mathbb Z}_2$,
then using the ${\mathbb Z}_4 \rtimes {\mathbb Z}_4$ discrete torsion
phases in table~\ref{table:z4z4-phases}, we find
\begin{eqnarray}
Z\left( [X/{\mathbb Z}_4 \rtimes {\mathbb Z}_4]_{\omega} \right)
& = & 
\frac{1}{| {\mathbb Z}_4 \rtimes {\mathbb Z}_4 |} 
\sum_{gh = hg} \epsilon(g,h) \, {\scriptstyle g} \square_h,
\\
& = & \frac{16}{16} \left[
{\scriptstyle 1} \square_1 \: + \:
{\scriptstyle 1} \square_{\overline{x}} \: + \:
{\scriptstyle \overline{x}} \square_1 \: + \:
{\scriptstyle \overline{x}} \square_{\overline{x}} \right],
\\
& = & 2 Z\left( [X/{\mathbb Z}_2] \right) \: = \:
Z\left(  [X/{\mathbb Z}_2] \, \coprod \,  [X/{\mathbb Z}_2] \right),
\end{eqnarray}
matching the prediction.

As this theory has two equal disconnected components, the two
copies of $[X/{\mathbb Z}_2]$, it has a $B {\mathbb Z}_2$ symmetry,
which is reflected in the fact that $\iota^* \omega(g,h)$ is invariant
under the subgroup $\langle y^2 \rangle \cong {\mathbb Z}_2$,
as is visible in table~\ref{table:pullback-z4sz4-cocycle}.

\subsection{$[X/S_4]$ with discrete torsion and trivially-acting
${\mathbb Z}_2 \times {\mathbb Z}_2$ subgroup}

Consider $[X/S_4]$ with nontrivial discrete torsion,
and with ${\mathbb Z}_2 \times {\mathbb Z}_2 \subset S_4$ acting
trivially on $X$.

The normal ${\mathbb Z}_2 \times {\mathbb Z}_2$ subgroup of $S_4$ has elements
\begin{equation}
1, \: \: \: (12)(34), \: \: \: (13)(24), \: \: \: (14)(23),
\end{equation}
and
the coset $S_4/{\mathbb Z}_2 \times {\mathbb Z}_2 = S_3$,
with elements
\begin{eqnarray}
1 & = & \{ 1, (12)(34), (13)(24), (14)(23) \},
\\
a & = & \{ (123), (134), (243), (142) \},
\\
b & = & \{ (132), (143), (234), (124) \},
\\
c & = & \{ (12), (34), (1423), (1324) \},
\\
d & = & \{ (13), (24), (1234), (1432) \},
\\
e & = & \{ (14), (23), (1243), (1342) \}.
\end{eqnarray}
As elements of $S_3$, $b = a^{-1}$ and
\begin{equation}
a^3 \: = \: b^3 \: = \: 1, \: \: \:
c^2 \: = \: d^2 \: = \: e^2 \: = \: 1.
\end{equation}
The only distinct elements that commute with one another are $a$ and $b$.

Analyzing this formally, define $K = {\mathbb Z}_2 \times {\mathbb Z}_2$, 
$G = S_4$.  The restriction of the nontrivial element of $H^2(S_4,U(1))$
to $K \subset S_4$, $\iota^* \omega$,
is the nontrivial element of $H^2({\mathbb Z}_2 \times
{\mathbb Z}_2,U(1))$.

From section~\ref{sect:conj},
since there is only a single projective irreducible representation of
$K$ with nontrivial $H^2(K,U(1))$ in this case, we predict that
\begin{equation}
{\rm QFT}\left( [X/S_4]_{\rm d.t.} \right) \: = \:
{\rm QFT}\left( [X/S_3] \right).
\end{equation}
Since $H^2(S_3,U(1))=0$, there is no possibility of discrete torsion
in the $[X/S_3]$ orbifold.

We can confirm this at the level of genus-one partition functions.
Using table~\ref{table:s4-phases}, it is straightforward to compute that
\begin{eqnarray}
Z\left( [X / S_4]_{\rm d.t.} \right) & = &
\frac{1}{| S_4 |} \sum_{g h = h g} \epsilon(g,h) {\scriptstyle g} \square_h,
\\ 
& = & \frac{1}{| S_3 | } \sum_{\overline{g} \overline{h} = 
\overline{h} \overline{g} } 
{\scriptstyle \overline{g} } \square_{\overline{h} },
\\
& = & Z\left( [X / S_3 ] \right),
\end{eqnarray}
where $\overline{g}, \overline{h} \in S_3$.
Thus, in this case, the $S_4$ orbifold with discrete torsion and
trivially-acting ${\mathbb Z}_2 \times {\mathbb Z}_2$ has the
same partition function as the $S_4 / {\mathbb Z}_2 \times {\mathbb Z}_2
\cong S_3$ orbifold.

\subsection{$[X/S_4]$ with discrete torsion and trivially-acting
$A_4$ subgroup}

Consider $[X/S_4]$ where $K = A_4 \subset S_4$ acts
trivially and the orbifold has discrete torsion.
The coset $S_4/A_4 = {\mathbb Z}_2$.

The elements of $A_4$ are the even permutations, which are transpositions
of the form $1$, $(ab)(cd)$.  The odd permutations are of the form
$(ab)$, $(abcd)$.

Let us first work out the prediction of section~\ref{sect:conj}, 
then compare to
physics.  First,
it can be shown that
\begin{equation}
H^2(A_4, U(1)) \: = \: {\mathbb Z}_2,
\end{equation}
(with trivial action on the coefficients,)
and the restriction of the nontrivial element of $H^2(S_4,U(1))$,
$\iota^* \omega$, is
the nontrivial element of $H^2(A_4,U(1))$, as can be
seen by restricting the genus-one phases in
table~\ref{table:s4-phases}.

Since $\iota^* \omega \neq 0$, in general terms, section~\ref{sect:conj}
predicts
\begin{equation}
{\rm QFT}\left( [X/S_4]_{\omega} \right) \: = \:
{\rm QFT}\left( \left[ \frac{X \times \hat{K}_{\iota^* \omega} }{ G/K }
\right]_{\hat{\omega}} \right).
\end{equation}
In this case, $G/K = {\mathbb Z}_2$, and as $H^2({\mathbb Z}_2,U(1)) = 0$,
there will not be any discrete torsion contributions $\hat{\omega}$,
but we do need to compute the number of irreducible projective 
representations of $K = A_4$ and the action of $G/K$ on those representations.

Now, let us compute the irreducible projective representations of $A_4$.
The group $A_4$ has four conjugacy classes, which we list below:
\begin{equation}
\begin{array}{c}
\{ 1 \}, \\
\{ (12)(34), (13)(24), (14)(23) \}, \\
\{ (123), (421), (243), (341) \}, \\
\{ (132), (412), (234), (314) \},
\end{array}
\end{equation}
and from table~\ref{table:s4-phases}, the conjugacy classes consisting of
$g \in A_4$ such that $\omega(g,h) = \omega(h,g)$ for all $h \in A_4$
commuting with $g$ are the first one and the last two,
\begin{equation}
\begin{array}{c}
\{ 1 \}, \\
\{ (123), (421), (243), (341) \}, \\
\{ (132), (412), (234), (314) \}.
\end{array}
\end{equation}
For example,
\begin{equation}
\omega( (12)(34), (13)(24) ) \: = \: 
- \omega( (13)(24), (12)(34) ).
\end{equation}
Thus, $A_4$ has three irreducible projective representations with respect
to the nontrivial element of $H^2(A_4,U(1))$.

Conjugating by odd elements of $S_4$ exchanges the two nontrivial
conjugacy classes, for example:
\begin{equation}
(12)(123)(12) \: = \: (132).
\end{equation}
Since the prefactors $\beta(\om)(q,k)$ are trivial, logic similar to footnote~\ref{footnote:GModKActionSubtlety} allows us to conclude that the $S_4/A_4 = {\mathbb Z}_2$ action will exchange two of those
irreducible projective representations.

As mentioned previously, section~\ref{sect:conj} predicts
\begin{equation}
{\rm QFT}\left( [X/S_4]_{\rm d.t.} \right)
 \: = \: {\rm QFT}\left( \left[ \frac{ X \times \hat{A}_{4, \iota^* \omega} }{ {\mathbb Z}_2 }
\right] \right),
\end{equation}
for $\hat{A}_{4,\iota^* \omega}$ 
the set of irreducible projective representations with
respect to the restriction of $\omega \in H^2(S_4,U(1))$.
As a set, $\hat{A}_{4,\iota^* \omega}$ 
has three elements, but two are interchanged by
the ${\mathbb Z}_2$, so that
\begin{equation}
{\rm QFT}\left( [X/S_4]_{\rm d.t.} \right)
 \: = \: {\rm QFT}\left( \left[ \frac{ X \times \hat{A}_{4, \iota^* \omega} }{ {\mathbb Z}_2 }
\right] \right)
 \: = \: {\rm QFT}\left( X \, \coprod \, [X/{\mathbb Z}_2] \right).
\end{equation}

Computing the genus-one partition function, we find
\begin{eqnarray}
Z\left( [X / S_4]_{\rm d.t.} \right) & = &
\frac{1}{|S_4|} \sum_{g h = h g} \epsilon(g,h) 
{\scriptstyle g} \square_h,
\\
& = & \frac{1}{|S_4|} \left[
(36) {\scriptstyle 1} \square_1 \: + \:
(12) {\scriptstyle 1} \square_{\xi} \: + \:
(12) {\scriptstyle \xi} \square_1 \: + \:
(12) {\scriptstyle \xi} \square_{\xi} \right],
\\
& = & Z(X) \: + \: Z([X/{\mathbb Z}_2]).
\end{eqnarray}
Thus, at least at the level of partition functions, we see
\begin{equation}
[X/S_4]_{\rm d.t.} \: = \: X \coprod [X/{\mathbb Z}_2],
\end{equation}
matching the prediction of section~\ref{sect:conj}.

By way of comparison, for the orbifold $[X/S_4]$ with trivially-acting $A_4$
and no discrete torsion, ordinary decomposition implies
\begin{equation}
{\rm QFT}\left( [X/S_4] \right) \: = \:
{\rm QFT}\left(  X \, \coprod \, [X/{\mathbb Z}_2] \, \coprod \,
[X/{\mathbb Z}_2] \right),
\end{equation}
i.e. it has one additional copy of $[X/{\mathbb Z}_2]$ relative to the
case with discrete torsion.

\subsection{$[X / D_4 \times ({\mathbb Z}_2)^2]$ with discrete torsion
and trivially-acting $({\mathbb Z}_2)^3$ subgroup}

Consider $[X/G]_{\omega}$, where $G = D_4 \times ( {\mathbb Z}_2 )^2$,
and $\omega = p_2^* \omega_0$,
where
\begin{equation}
p_2: \: D_4 \times ( {\mathbb Z}_2 )^2 \: \longrightarrow \:
({\mathbb Z}_2)^2
\end{equation}
is the projection map onto the second factor, and $\omega_0$ is the
nontrivial element of $H^2( {\mathbb Z}_2 \times {\mathbb Z}_2, U(1))$.
Let us assume that the trivially acting subgroup is $K = ({\mathbb Z}_2)^3 \subset G$,
where one ${\mathbb Z}_2$ factor is the center of $D_4$ and the remaining $({\mathbb Z}_2)^2$
factors match those in $G$.

Let $p_2': K \rightarrow ({\mathbb Z}_2)^2$ be the projection onto the
second two ${\mathbb Z}_2$ factors, so that the diagram
\begin{equation}
\xymatrix{
G \ar[r]^-{p_2} & {\mathbb Z}_2 \times {\mathbb Z}_2 
\\
K \ar[u]^{\iota} \ar[ru]_{p_2'}   
}
\end{equation}
commutes.
Then we have
\begin{equation}
\iota^* \omega \: = \: \iota^* p_2^* \omega_0 \: = \:
p_2'^* \omega_0 \: \neq \: 0.
\end{equation}

Following section~\ref{sect:conj}, next consider the
set $\hat{K}_{\iota^* \omega}$.  This is essentially a product
of two factors, ${\mathbb Z}_2$ and $({\mathbb Z}_2)^2$, 
with the discrete torsion
entirely in the $({\mathbb Z}_2)^2$ factor.  There is only one
irreducible projective representation of $({\mathbb Z}_2)^2$,
as can be seen by counting conjugacy classes, but 
the remaining ${\mathbb Z}_2$ factor has two honest representations.
Furthermore, since $K$ is in the center of $G$, $G/K$ acts trivially
on $\hat{K}_{\iota^* \omega}$.

Putting this together, we have the prediction
\begin{equation}
{\rm QFT}\left( [X/G]_{\omega} \right) \: = \:
{\rm QFT}\left( [X/{\mathbb Z}_2 \times {\mathbb Z}_2]_{\hat{\omega}_1}
 \, \coprod \,
[X/{\mathbb Z}_2 \times {\mathbb Z}_2]_{\hat{\omega}_2} \right),
\end{equation}
where $\hat{\omega}_{1,2}$ are elements of discrete torsion we determine
next.
We are given discrete torsion $\omega$ as the pullback of
the generator of $H^2( {\mathbb Z}_2 \times {\mathbb Z}_2, U(1))$.  
From equation~(\ref{eq:GeneratHatOmegaDef}) for $\hat{\omega}$, 
the ratio of $\omega$'s multiplying the
image of the extension class is determined by the values of $\omega$
on a section $s: D_4/{\mathbb Z}_2 \rightarrow D_4 \times ({\mathbb Z}_2)^2$.
We can choose the section so that $p_2(s(q))=1\in\Z_2\times\Z_2$ for all $q\in D_4/\Z_2$.  Then the $\omega$'s in (\ref{eq:GeneratHatOmegaDef}) 
are trivial, and the discrete torsion $\hat{\omega}$ is determined solely by
the image of the extension class.
Since $D_4$ is not a semidirect product
${\mathbb Z}_2 \rtimes ({\mathbb Z}_2)^2$, the extension class
$H^2(({\mathbb Z}_2)^2, {\mathbb Z}_2)$ is nontrivial, and applying the
two honest irreducible representations of ${\mathbb Z}_2$, we get that
one of $\hat{\omega}_{1,2}$ is trivial, and the other is nontrivial.
Thus, we predict
\begin{equation}
{\rm QFT}\left( [X/G]_{\omega} \right) \: = \:
{\rm QFT}\left( [X/{\mathbb Z}_2 \times {\mathbb Z}_2]
 \, \coprod \,
[X/{\mathbb Z}_2 \times {\mathbb Z}_2]_{\rm d.t.} \right),
\end{equation}
where exactly one of the $[X/{\mathbb Z}_2 \times
{\mathbb Z}_2]$ summands has discrete torsion.

Next, let us compare to physics.  We can view this example as a pair of
successive orbifolds, first $[X/D_4]$ without discrete torsion and with
trivially-acting ${\mathbb Z}_2$, and then a trivially-acting
$({\mathbb Z}_2)^2$ orbifold
with discrete torsion.  Applying the ordinary version of
decomposition \cite{Hellerman:2006zs}, we know that
\begin{equation}
{\rm QFT}\left( [X/D_4] \right) \: = \:
{\rm QFT}\left( [X / {\mathbb Z}_2 \times {\mathbb Z}_2] \, \coprod \,
[X/{\mathbb Z}_2 \times {\mathbb Z}_2]_{\rm d.t.} \right),
\end{equation}
a disjoint union of two $({\mathbb Z}_2)^2$ orbifolds where one copy
has discrete torsion and the other does not.
As we shall see in
section~\ref{sect:z2z2kz2}, a $({\mathbb Z}_2)^2$ orbifold with discrete torsion
in which either ${\mathbb Z}_2$ factor acts trivially is just a realization
of orbifolding by a quantum symmetry, and returns the original theory.
Putting this together, we find that
\begin{equation}
{\rm QFT}\left( [X/G] \right)_{\omega} \: = \:
{\rm QFT}\left( [X / {\mathbb Z}_2 \times {\mathbb Z}_2] \, \coprod \,
[X/{\mathbb Z}_2 \times {\mathbb Z}_2]_{\rm d.t.} \right).
\end{equation}

\section{Examples in which $\iota^* \omega = 0$ and
$\beta(\omega) \neq 0$}
\label{sect:exs:b}

We will see in this section that this case corresponds to one
generalization\footnote{
See also \cite{Bhardwaj:2017xup} 
for a different generalization of quantum symmetries to
nonabelian orbifolds.
} of quantum symmetries of orbifolds.
Ordinarily, in quantum symmetries in abelian orbifolds,
we have a $G/K$ orbifold, $G/K$ abelian, which when orbifolded by
a further $K$ (abelian), with an action only on twist fields and not the
underlying space, one recovers the original theory.  The action of
$K$ on $G/K$ can be encoded in a set of phases
\begin{equation}
K \times G/K \: \longrightarrow \: U(1),
\end{equation}
which are equivalent to a map
\begin{equation}
G/K \: \longrightarrow \: {\rm Hom}(K,U(1)).
\end{equation}
Taking into account the action of $G/K$ on $K$ and its induced action on $H^1(K,U(1))$, this map is a crossed homomorphism\footnote{Suppose a group $G$ acts on an abelian group $H$.  Then a crossed homomorphism $\phi:G\rr H$ is a map satisfying $\phi(g_1g_2)=g_1\cdot\phi(g_2)+\phi(g_1)$.}.
This is equivalent to an element of $H^1\left( G/K, H^1(K,U(1)) \right)$.

The map $\beta$ sends
\begin{equation}
\beta: \: H^2(G,U(1)) \: \longrightarrow \: 
H^1\left( G/K, H^1(K,U(1)) \right),
\end{equation}
and so it can be interpreted (at least in abelian cases) as
giving us the quantum symmetry action corresponding to an element of
discrete torsion in the extension $G$.

In more general cases, these maps all factor through abelianizations,
\begin{equation}
H^1( G/K, H^1(K, U(1))) \: = \:
H^1( (G/K)_{\rm ab}, H^1( K_{\rm ab}, U(1))),
\end{equation}
consistent with descriptions in the old literature of
quantum symmetries of a $G$ orbifold in terms\footnote{
Unlike abelian cases, orbifolding by the abelianization does not return
the original theory.  A fix for this was recently proposed in
\cite{Bhardwaj:2017xup}, in which the abelianization was extended to a 
unitary fusion category, orbifolding by which would then return the
original theory.  This is somewhat beyond the scope of this article, however.
} of the
abelianization of $G$, $G_{\rm ab} = G/[G,G]$.

We shall elaborate on this perspective in the examples.

\subsection{$[X/{\mathbb Z}_2 \times {\mathbb Z}_2]$ with
discrete torsion and trivially-acting ${\mathbb Z}_2$}
\label{sect:z2z2kz2}

Consider $[X / {\mathbb Z}_2 \times {\mathbb Z}_2]$
with nontrivial discrete torsion, and with one ${\mathbb Z}_2$
acting trivially on $X$.  In this case, $K = {\mathbb Z}_2$,
$G = {\mathbb Z}_2$, and since $H^2(K,U(1)) = 0$, we have
that $\iota^* \omega = 0$ trivially.  Furthermore, since
$H^2(G/K,U(1))=0$ also, we see that $\beta(\omega) \neq 0$.

In principle, to apply the analysis of 
section~\ref{sect:conj},
we need to compute $\beta$.  In this example,
since the phases are nontrivial, one can get to the result without
detailed computation, but to illustrate the method, we work through the
details here.

From appendix~\ref{sect:beta}, recall that for $q \in G/K$, $k \in K$,
\begin{equation}
\beta(\omega)(q,k) \: = \: \frac{
\omega( k s(q), s(q)^{-1} )
}{
\omega( s(q)^{-1}, k s(q) )
}.
\end{equation}
Here, write $K = {\mathbb Z}_2 = \langle a \rangle$,
$G/K = {\mathbb Z}_2 = \langle b \rangle$.  Without loss of generality,
we take the section $s: {\mathbb Z}_2  \rightarrow 
{\mathbb Z}_2 \times {\mathbb Z}_2$ to be given by,
$s(q) = q$.  Following 
the explicit cocycle given in appendix~\ref{app:z2z2},
it is straightforward to compute that $\beta(\omega)(q,k) = 1$ for 
either $q=1$ or $k=1$, and its only different value is for
$k = a$ and $q=b$, for which
\begin{equation}
\beta(\omega)(b,a) \: = \: 
\frac{ \omega(ab, b) }{ \omega(b, ab) } \: = \: -1.
\end{equation}

Using
\begin{equation}
H^1(G/K, H^1(K,U(1)) ) \: = \: H^1( {\mathbb Z}_2, {\mathbb Z}_2 )
\: = \: {\mathbb Z}_2,
\end{equation}
we interpret $\beta(\omega)$ as a map
$H^2(G,U(1)) \rightarrow H^1(G/K,H^1(K,U(1)))$.
From the analysis of $\beta$ above, we see
that $\beta(\omega)(q=1, -)$ is the trivial
map that sends all elements of $K$ to the identity,
and $\beta(\omega)(q=b, -)$ is the identity, sending any
element of $K = {\mathbb Z}_2$ to itself.  Thus, we see that 
$\beta(\omega)$ is an isomorphism
$H^2(G,U(1)) \rightarrow H^1(G/K,H^1(K,U(1)))$, hence
\begin{equation}
{\rm Ker}(\beta) \: = \: 0, \: \: \:
{\rm Coker}(\beta) \: = \: 0.
\end{equation}

Given the form for $\beta$, we predict from our
analysis in section~\ref{sect:conj} that
\begin{equation}
{\rm QFT}\left( [ X/ {\mathbb Z}_2 \times {\mathbb Z}_2 ]_{\omega}
\right) \: = \: X.
\end{equation}

It is straightforward to see this directly in physics.
In fact, this example is the same as orbifolding $[X/{\mathbb Z}_2]$
by the quantum symmetry ${\mathbb Z}_2$ \cite{Vafa:1989ih}:
\begin{equation}
[X / \hat{\mathbb Z}_2 \times {\mathbb Z}_2]_{\rm d.t.}
\: = \:
\left[ [X/{\mathbb Z}_2] / \hat{\mathbb Z}_2 \right].
\end{equation}
Following \cite[section 8.5]{Ginsparg:1988ui},
if we let primes denote the twisted sectors of the $\hat{\mathbb Z}_2$
orbifold, then
\begin{eqnarray}
{\scriptstyle +} \square_+' & = &
\frac{1}{2} \left(
{\scriptstyle +} \square_+ \: + \:
{\scriptstyle -} \square_+ \: + \:
{\scriptstyle +} \square_- \: + \:
{\scriptstyle -} \square_- \right),
\\
{\scriptstyle -} \square_+' & = &
\frac{1}{2} \left( 
{\scriptstyle +} \square_+ \: + \:
{\scriptstyle -} \square_+ \: - \:
{\scriptstyle +} \square_- \: - \:
{\scriptstyle -} \square_- \right),
\\
{\scriptstyle +} \square_-' & = &
\frac{1}{2} \left(
{\scriptstyle +} \square_+ \: - \:
{\scriptstyle -} \square_+ \: + \:
{\scriptstyle +} \square_- \: - \:
{\scriptstyle -} \square_- \right),
\\
{\scriptstyle -} \square_-' & = &
\frac{1}{2} \left(
{\scriptstyle +} \square_+ \: - \:
{\scriptstyle -} \square_+ \: - \:
{\scriptstyle +} \square_- \: + \:
{\scriptstyle -} \square_- \right).
\end{eqnarray}
Putting this together, the combination of the ${\mathbb Z}_2$
and $\hat{\mathbb Z}_2$ orbifolds can be expressed as a
single $\hat{\mathbb Z}_2 \times {\mathbb Z}_2$ orbifold with phases
$\epsilon(a,g; b,h)$ where $a,b \in \hat{\mathbb Z}_2$, $g,h \in {\mathbb Z}_2$,
given by
\begin{eqnarray}
\epsilon(a,+; b, +) & = & +1,
\\
\epsilon(a, -; b, -) & = & ab,
\\
\epsilon(a,+; b, -) & = & a,
\\
\epsilon(a,-; b,+) & = & b,
\end{eqnarray}
where we have taken $a, b \in \{\pm 1\}$.
This means that $\epsilon = -1$ for the following combinations:
\begin{equation}
( (+,-), (-,+) ), \: \: \:
( (+,-), (-,-) ), \: \: \:
( (-,+), (-,-) ),
\end{equation}
and the other three obtained by reversing the order.
For all other elements of $\hat{\mathbb Z}_2 \times {\mathbb Z}_2$,
$\epsilon = +1$.
These are precisely the phases assigned by discrete torsion.

Thus, we see that the orbifold $[ [X/{\mathbb Z}_2] / \hat{\mathbb Z}_2 ]$
is the same as $[X / (\hat{\mathbb Z}_2 \times {\mathbb Z}_2) ]$,
with phases given by discrete torsion.  In particular,
\begin{equation}
[ [X/{\mathbb Z}_2] / \hat{\mathbb Z}_2 ] \: = \:
[X / (\hat{\mathbb Z}_2 \times {\mathbb Z}_2) ]_{\rm d.t.} \: = \:
X,
\end{equation}
agreeing with our prediction of section~\ref{sect:conj}.

As a final consistency check, let us redo the computation
of section~\ref{sect:conj} without appealing to the description of
this case in terms of kernels and cokernels of $\beta(\omega)$.
Briefly, since $K$ is central in $G$, the action of $L_q$ for $q \in G/K$
is to map $\phi(k)$ to 
\begin{equation}
(L_q \phi)(k) \: = \: \beta(\omega)(q,k)^{-1} \phi(k).
\end{equation}
As computed above, and using the same section as there,
$\beta(\omega)(b,a) = -1$, so we find
\begin{equation}
L_b \phi(1) \: = \: + \phi(1),
\: \: \:
L_b \phi(a) \: = \: - \phi(a).
\end{equation}
In other words, the effect of $L_b$ is to exchange the two
irreducible representations of $K = {\mathbb Z}_2$: 
the trivial representation becomes nontrivial, and vice-versa.
As a result, there is only one orbit of $G/K$ on $\hat{K}$,
consisting of both the elements of $\hat{K}$, and so one recovers
the result above.  

In passing, the reader should note the critical role played by the
phases encoded in $\beta(\omega)$:  if those phases were all $+1$,
then $L_q \phi$ would be isomorphic to $\phi$ for all $q \in G/K$,
and we would have predicted instead that the resulting theory
be a disjoint union of two copies of $[X/ {\mathbb Z}_2]$, instead of
one copy of $X$, which is not what we see in physics.

\subsection{Example with nonabelian $K$}

In this subsection we will consider an example that is closely
related to the previous one, essentially adding a nonabelian factor to
$G$.

Consider $[X/G]$ where $G = \tilde{G} \times {\mathbb Z}_2 \times
{\mathbb Z}_2$, and $K = \tilde{G} \times {\mathbb Z}_2$.
Let $\alpha$ denote the nontrivial element of
$H^2({\mathbb Z}_2 \times {\mathbb Z}_2, U(1))$,
and let $p_2: G \rightarrow {\mathbb Z}_2 \times {\mathbb Z}_2$
be the projection, then take $\omega = p_2^* \alpha$.

Since $K$ is (potentially) nonabelian,
we cannot simply resort to a computation of the kernel and cokernel
of $\beta(\omega)$, but must work slightly harder.
Let $a$ generate the ${\mathbb Z}_2$ factor in $K$,
and $b$ generate the remaining ${\mathbb Z}_2$ in $G$,
and pick a section $s: G/K \rightarrow G$ such that
$s(1) = 1$, $s(bK) = b$.
In general,
\begin{equation}
(L_q \phi)(k) \: = \: 
\frac{ \omega( s(q)^{-1}, k s(q) ) }{
\omega( k s(q), s(q)^{-1} ) } \phi( s(q)^{-1} k s(q) ),
\end{equation}
and the nontrivial case, for which the phase factor (the ratio
of $\omega$'s) is nontrivial is $q = bK$:
\begin{eqnarray}
(L_q \phi)(ag) & = &
\frac{ \omega(b, agb) }{ \omega(agb, b) } \phi( ag )
\: = \: \frac{ \alpha(b, ab) }{ \alpha(ab,b) } \phi(ag )
\: = \: - \phi(ag),
\\
(L_q \phi)(g) & = &  
\frac{ \omega(b,gb) }{ \omega(gb,b) } \phi(g) \: = \:
\frac{ \alpha(b,b) }{ \alpha(b,b) } \phi(g) \: = \: + \phi(g),
\end{eqnarray}
for $g \in \tilde{G}$.  We see that the effect of $L_q$ is to
exchange two copies of the irreducible representations of $\hat{G}$,
indexed by the irreducible representations of ${\mathbb Z}_2$.
Taking that into account, we see that the prediction for physics
is that 
\begin{equation}
{\rm QFT}\left( [X/G]_{\omega} \right) \: = \:
{\rm QFT}\left( \coprod_{ \hat{G} } X \right),
\end{equation}
the same decomposition as for $[X/\tilde{G}]$ with a trivial
action of $\tilde{G}$ and no discrete torsion -- an example of
ordinary decomposition.  This is easily verified in partition
functions.

Before going on, let us quickly walk through the details of the
analogous computation in terms of kernels and cokernels of
$\beta(\omega)$, which is not applicable since $K$ is
(potentially) nonabelian, to illustrate the problem.
Let the section $s: G/K \rightarrow G$ be as above.
Then it is straightforward to compute that the only nontrivial
elements $\beta(\omega)(q,k)$ are
\begin{equation}
\beta(\omega)(bK,ag) \: = \: - 1
\end{equation}
for $a$ the generator of the ${\mathbb Z}_2$ factor in $K$,
and $g \in \tilde{G}$.  As a result,
$\beta(\omega)(1,-)$ is the identity element of 
$H^1(K,U(1))$, and $\beta(\omega)(bK,-)$ is the product of the
trivial element of Hom$(\tilde{G},U(1))$ and the nontrivial
element of $H^1( {\mathbb Z}_2,U(1))$.
As a result, $\beta(\omega)$ is giving a map $G/K \rightarrow
H^1(K,U(1))$ with zero kernel and cokernel Hom$(\tilde{G},U(1))$,
giving as many copies of $X$ as the number of one-dimensional
irreducible representations of $\tilde{G}$,
but the correct answer gives as many copies as
the number of all irreducible representations of $\tilde{G}$,
which need not be one-dimensional in general.

\subsection{$[X/{\mathbb Z}_2 \times {\mathbb Z}_4]$ with
discrete torsion and trivially-acting
${\mathbb Z}_4$ or ${\mathbb Z}_2 \times {\mathbb Z}_2$ subgroup}

Consider $[X/{\mathbb Z}_2 \times {\mathbb Z}_4]$ with
discrete torsion.  In this section we will consider three closely
related examples in which a subgroup $K$, either ${\mathbb Z}_4$
or ${\mathbb Z}_2 \times {\mathbb Z}_2$, acts trivially:
\begin{enumerate}
\item $K = \langle b \rangle \cong {\mathbb Z}_4$ (in the presentation
in appendix~\ref{app:z2z4}),
\item $K = \langle ab \rangle \cong {\mathbb Z}_4$,
\item $K = \langle a, b^2 \rangle \cong {\mathbb Z}_2 \times {\mathbb Z}_2$.
\end{enumerate}

In each of these three cases, $\iota^* \omega = 0$.
In the first two cases, this is for the trivial reason that
$H^2(K,U(1)) = 0$.  In the third case, $H^2(K,U(1)) \neq 0$; however,
when one computes $\iota^* \omega$ in the third case, from
table~\ref{table:z2z4-cocycle}, one finds it is given by
\begin{center}
\begin{tabular}{c|rrrr}
& $1$ & $a$ & $b^2$ & $ab^2$ \\ \hline
$1$ & $1$ & $1$ & $1$ & $1$ \\ 
$a$ & $1$ & $1$ & $1$ & $1$ \\
$b^2$ & $1$ & $1$ & $1$ & $1$ \\
$ab^2$ & $1$ & $1$ & $1$ & $1$ 
\end{tabular}
\end{center}
and so is trivial.

In each of these three cases, $G/K \cong {\mathbb Z}_2$,
where $G = {\mathbb Z}_2 \times {\mathbb Z}_4$.  As a result,
$H^2(G/K,U(1)) = 0$, and so the image of $\pi^*$ vanishes.
This means the map $\beta$ is injective, 
and so $\beta(\omega)$ is a nontrivial element of
\begin{equation}
H^1(G/K, H^1(K,U(1))) \: = \: {\rm Hom}\left( {\mathbb Z}_2, \hat{K} \right).
\end{equation}
In each case, 
\begin{equation}
{\rm Ker} \, \beta(\omega) \: = \: 0, \: \: \:
{\rm Coker} \, \beta(\omega) \: = \: {\mathbb Z}_2.
\end{equation}

Let us check this description of $\beta$ explicitly.
As the analysis for all three cases is nearly identical,
we give the details of $\beta$ explicitly for only the first case,
where $K = \langle b \rangle \cong {\mathbb Z}_4$.
We take the section $s: G/K \rightarrow G$ to be
$s(K) = 1$, $s(aK) = a$.  We then compute using
\begin{equation}
\beta(\omega)(q,k) \: = \: \frac{
\omega( k s(q), s(q)^{-1} )
}{
\omega( s(q)^{-1}, k s(q) )
}.
\end{equation}
For $q = 1$, one has immediately that $\beta = +1$,
and if $q=a$, $k=b$ or $b^2$, one also computes $\beta = +1$.
The only two nontrivial cases are as follows:
\begin{eqnarray}
\beta(\omega)(q=a,k=b) & = &
\frac{ \omega(ba,a) }{ \omega(a, ba) } \: = \: -1,
\\
\beta(\omega)(q=a,k=b^3) & = &
\frac{ \omega(b^3 a, a) }{ \omega(a,b^3a) } \: = \: -1,
\end{eqnarray}
using the cocycle representative in appendix~\ref{app:z2z4}.
As a result, $\beta(\omega)(q=1,-)$ is the element of $H^1(K,U(1))$ that
maps all elements of $K$ to the identity, while $\beta(\omega)(q=a,-)$ is
the element of $H^1(K,U(1))$ that maps ${\mathbb Z}_4$ onto ${\mathbb Z}_2
\subset U(1)$.
As a homomorphism from $G/K = {\mathbb Z}_2$ to $H^1(K,U(1)) = {\mathbb Z}_4$,
we see that $\beta(\omega)$ maps injectively into ${\mathbb Z}_4$,
hence the kernel and cokernel are as given above.

Putting this together, the conjecture of section~\ref{sect:conj} then
predicts
\begin{equation}
{\rm QFT}\left( [X/G]_{\omega} \right) \: = \: 
{\rm QFT}\left( \left[ \frac{X \times \widehat{ {\rm Coker} \, \beta(\omega) }
}{ {\rm Ker} \, \beta(\omega) } \right] \right) \: = \:
{\rm QFT}\left( X \times \hat{ {\mathbb Z} }_2 \right)
\: = \: {\rm QFT}\left( X \, \coprod \, X \right).
\end{equation}

Next, we shall check this prediction against partition functions.
Briefly, in each of these three cases,
\begin{eqnarray}
Z\left( [ X / {\mathbb Z}_2 \times {\mathbb Z}_4 ]_{\omega} \right)
& = &
\frac{1}{| {\mathbb Z}_2 \times {\mathbb Z}_4 |} \sum_{gh = hg}
\epsilon(g,h) \, {\scriptstyle g} \square_h,
\\
& = & \frac{16}{8} {\scriptstyle 1} \square_1,
\\
& = & 2 Z(X) \: = \: Z\left( X \, \coprod \, X \right),
\end{eqnarray}
agreeing with the prediction.

\subsection{$[X/{\mathbb Z}_2 \times {\mathbb Z}_4]$ with
discrete torsion and trivially-acting
${\mathbb Z}_2$ subgroup}

Consider again $[X/{\mathbb Z}_2 \times {\mathbb Z}_4]$ with
discrete torsion.  In this section we will consider two closely related
examples in which a subgroup $K \cong {\mathbb Z}_2$ acts trivially:
\begin{enumerate}
\item $K = \langle a \rangle$,
\item $K = \langle ab^2 \rangle$,
\end{enumerate}
(in the presentation of ${\mathbb Z}_2 \times {\mathbb Z}_4$ described
in appendix~\ref{app:z2z4}).

In both cases, $\iota^* \omega = 0$, for the trivial reason that
$H^2(K,U(1)) = 0$.
Furthermore, in both cases, $G/K \cong {\mathbb Z}_4$.
As a result, $H^2(G/K,U(1)) = 0$, hence the image of $\pi^*$ vanishes,
so the map $\beta$ is injective.  Thus, $\beta(\omega)$ is a nontrivial
element of
\begin{equation}
H^1(G/K, H^1(K,U(1))) \: = \: {\rm Hom}( {\mathbb Z}_4, {\mathbb Z}_2 ).
\end{equation}
In each case,
\begin{equation}
{\rm Ker} \, \beta(\omega) \: = \: {\mathbb Z}_2, \: \: \:
{\rm Coker} \, \beta(\omega) \: = \: 0.
\end{equation}

Let us pause for a moment to check this description of
$\beta(\omega)$ explicitly for the first case, for which
$K = \langle a \rangle$.  (The second case will be identical.)
We take the section $s: G/K \rightarrow G$ to be 
$s(b^n K) = b$ for $n < 4$, and then compute
\begin{equation}
\beta(\omega)(q,k) \: = \: \frac{
\omega( k s(q), s(q)^{-1} )
}{
\omega( s(q)^{-1}, k s(q) )
}.
\end{equation}
It is straightforward to see that $\beta(\omega)(1,k) = 1$ for all $k \in K$,
and $\beta(\omega)(q,1) = 1$ for all $q \in G/K$.
The nontrivial cases are as follows:
\begin{eqnarray}
\beta(\omega)(bK,a) & = & 
\frac{ \omega(a b, b^3) }{ \omega(b^3, ab) } \: = \: -1,
\\
\beta(\omega)(b^2K, a) & = &
\frac{ \omega(a b^2, b^2) }{ \omega(b^2, ab^2) } \: = \: +1,
\\
\beta(\omega)(b^3K, a) & = & 
\frac{ \omega(ab^3, b) }{ \omega(b, ab^3) } \: = \: -1,
\end{eqnarray}
using the explicit expression for the cocycle in appendix~\ref{app:z2z4}.
Thus, we see that $\beta(\omega)(1,-)$ and $\beta(b^2K,-)$ both are
the trivial element of $\hat{K}$, that maps both elements of 
$K = {\mathbb Z}_2$ to the identity,
and $\beta(\omega)(bK,-)$, $\beta(\omega)(b^3K,-)$ both are the
nontrivial element of $\hat{K} = \hat{\mathbb Z}_2$.
As a homomorphism from $G/K = {\mathbb Z}_4$ to $\hat{K} \cong 
{\mathbb Z}_2$, we see that $\beta(\omega)$ maps onto ${\mathbb Z}_2$,
and so has the kernel and cokernel given above.

Putting this together, the conjecture of section~\ref{sect:conj} then
predicts
\begin{equation}
{\rm QFT}\left( [X/G]_{\omega} \right) \: = \: 
{\rm QFT}\left( \left[ \frac{X \times \widehat{ {\rm Coker} \, \beta(\omega) }
}{ {\rm Ker} \, \beta(\omega) } \right] \right) \: = \:
{\rm QFT}\left( [X/{\mathbb Z}_2] \right).
\end{equation}
In the first case, the kernel of $\beta(\omega)$ is generated by
$b^2$, while in the second case, the kernel of $\beta(\omega)$ is
generated by
$\overline{a}^2$, where $\overline{a} = \{b, ab^3\}$ is the coset
generating $G/K \cong {\mathbb Z}_4$.

Next, comparing to partition functions, it is straightforward to
compute in both cases that
\begin{eqnarray}
Z\left( [ X / {\mathbb Z}_2 \times {\mathbb Z}_4 ]_{\omega} \right)
& = &
\frac{1}{| {\mathbb Z}_2 \times {\mathbb Z}_4 |} \sum_{gh = hg}
\epsilon(g,h) \, {\scriptstyle g} \square_h \: = \: 
Z\left( [X/{\mathbb Z}_2] \right),
\end{eqnarray}
with the effective ${\mathbb Z}_2$ orbifold action predicted above,
verifying the conjecture.

\subsection{$[X/D_4]$ with discrete torsion and
trivially-acting ${\mathbb Z}_2$ subgroup}

Consider $[X/D_4]$ with
center ${\mathbb Z}_2 \subset D_4$ acting trivially, and turn on discrete
torsion in the $D_4$ orbifold.

As before, we begin by computing the prediction of section~\ref{sect:conj}
for this case.  Here $K = {\mathbb Z}_2$ and $G = D_4$, 
so $G/K = {\mathbb Z}_2 \times {\mathbb Z}_2$,
and as
$H^2(K,U(1)) = 0$, we see immediately that $\iota^* \omega = 0$.
We also compute
\begin{eqnarray}
H^1( G/K, H^1(K, U(1)) ) & = & H^1( G/K, {\mathbb Z}_2) \: = \:
H^1( {\mathbb Z}_2 \times {\mathbb Z}_2, {\mathbb Z}_2),
\\
& = & {\mathbb Z}_2 \times {\mathbb Z}_2,
\\
H^2(G/K, U(1)) & = & {\mathbb Z}_2,
\end{eqnarray}
The exact sequence~(\ref{eq:3termessential}) becomes
\begin{equation}
H^2(G/K,U(1)) \: \stackrel{\pi^*}{\longrightarrow} \: 
{\mathbb Z}_2 \: \stackrel{\beta}{\longrightarrow} \: H^1(G/K,H^1(K,U(1))),
\end{equation}
From table~\ref{table:d4-phases}, since for example
$\epsilon(a,1) \neq \epsilon(a,z)$, we see that
$\omega$ is not the pullback of an element of discrete torsion
in $G/K = {\mathbb Z}_2 \times {\mathbb Z}_2$.  Thus, $\omega$ must map
to a nontrivial element of $H^1(G/K,H^1(K,U(1))$, a nontrivial homomorphism
$G/K \rightarrow {\mathbb Z}_2$.
From table~\ref{table:d4-cocycle}, 
we see that $\omega$ is trivial on the subgroup
generated by $b$, but not that generated by $a$, from which we infer that
the kernel of the map $G/K \rightarrow {\mathbb Z}_2$ is the ${\mathbb Z}_2$
generated by $\overline{b}$ (the image of $b$ in $G/K$), and there is no
cokernel.  Thus, we predict
\begin{equation}
{\rm QFT}\left( [X/D_4]_{\omega} \right) \: = \: 
{\rm QFT}\left( [X/{\mathbb Z}_2] \right),
\end{equation}
where the ${\mathbb Z}_2$ is generated by $\overline{b} \in G/K$.

Before going on to check the physics, let us take a moment to explicitly
check the description of $\beta(\omega)$ above.  
In the present case, $G/K = {\mathbb Z}_2 \times {\mathbb Z}_2 =
\langle \overline{a}, \overline{b} \rangle$, and we pick a section
$s: G/K \rightarrow G$ given by 
\begin{equation}
s(1) \: = \: 1,
 \: \: \:
s(\overline{a}) \: = \: a,
\: \: \:
s(\overline{b}) \: = \: b,
\: \: \:
s( \overline{a} \overline{b} ) \: = \: ab,
\end{equation}
in the conventions of appendix~\ref{app:d4}.
As before,
\begin{equation}
\beta(\omega)(q,k) \: = \: \frac{
\omega( k s(q), s(q)^{-1} )
}{
\omega( s(q)^{-1}, k s(q) )
},
\end{equation}
and 
it is straightforward to see that $\beta(\omega)(1,k) = 1$ for all $k \in K$,
and $\beta(\omega)(q,1) = 1$ for all $q \in G/K$.
The nontrivial cases are as follows:
\begin{eqnarray}
\beta(\omega)(\overline{a},z) & = &
\frac{ \omega(z a, a) }{ \omega(a,az) } \: = \: -1,
\\
\beta(\omega)(\overline{b},z) & = &
\frac{ \omega(z b, bz) }{ \omega(bz, zb) } \: = \: +1,
\\
\beta(\omega)(\overline{a} \overline{b}, z) & = &
\frac{ \omega(z ab, ab) }{ \omega(ab, zab) } \: = \: -1,
\end{eqnarray}
using table~\ref{table:d4-phases}.
From this we read off that $\beta(\omega)(1,-)$ and $\beta(\omega)(\overline{b},
-)$ are the trivial elements in $\hat{K}$, that map all elements of
$K$ to the identity,
and $\beta(\omega)(\overline{a},-)$ and $\beta(\omega)(\overline{a}
\overline{b},-)$ are the nontrivial elements of $\hat{K}$.
Thus, we see that $\beta(\omega)$ maps $G/K = {\mathbb Z}_4$
surjectively onto $\hat{K} = \hat{\mathbb Z}_2$, as predicted above,
and so the kernel and cokernel are as above.

Now, let us compare to physics.
Using the phases in table~\ref{table:d4-phases}, 
it is straightforward to compute that
\begin{eqnarray}
Z\left( [X/D_4]_{\rm d.t.} \right)
& = &
\frac{1}{|D_4|} \sum_{g h = h g} \epsilon(g,h) {\scriptstyle g} \square_h,
\\
& = &
\frac{1}{2} \left[ {\scriptstyle 1} \square_1 \: + \:
{\scriptstyle 1} \square_{\overline{b}} \: + \:
{\scriptstyle \overline{b}} \square_1 \: + \:
{\scriptstyle \overline{b}} \square_{\overline{b}} \right],
\\
& = & 
Z \left( [X/{\mathbb Z}_2] \right),
\end{eqnarray}
where the ${\mathbb Z}_2$ appearing is a subgroup of the effectively-acting
${\mathbb Z}_2 \times {\mathbb Z}_2$ which is generated by
$\overline{b}$.  This matches our prediction.

\subsection{$[X/{\mathbb Z}_4 \rtimes {\mathbb Z}_4]$ with discrete torsion
and trivially-acting subgroups}

In this example we study three closely related examples,
orbifolds by ${\mathbb Z}_4 \rtimes {\mathbb Z}_4$
(the semidirect product of two copies of ${\mathbb Z}_4$, discussed
in appendix~\ref{app:z4sz4}) with discrete torsion, and the
following trivially-acting subgroups:
\begin{enumerate}
\item $K = \langle x^2 \rangle \cong {\mathbb Z}_2$,
\item $K = \langle x\rangle \cong {\mathbb Z}_4$,
\item $K = \langle y^2, x \rangle \cong {\mathbb Z}_2 \times {\mathbb Z}_4$,
(a different ${\mathbb Z}_2 \times {\mathbb Z}_4$ 
subgroup than was considered in
section~\ref{sect:iw:z4sz4-z2z4})
\end{enumerate}
in the presentation of appendix~\ref{app:z4sz4}.

First, in each case, $\iota^* \omega = 0$.  In the first two cases,
this is a trivial consequence of the fact that $H^2(K,U(1)) = 0$.
In the third case, $H^2({\mathbb Z}_2 \times {\mathbb Z}_4,U(1)) \neq 0$,
so in principle $\iota^* \omega$ could be nonzero.
To see that in fact $\iota^* \omega = 0$ in this case as well,
one can compute the pullback of the genus-one phases, to find that
they are all equal to one.  (In fact, one can also compute
$\iota^* \omega$ directly from the representation of $\omega$ in
table~\ref{table:z4z4-cocycle}.  They only need to all be one up to a cocycle,
but in fact, one finds that $\iota^* \omega(g,h) = 1$ on the nose for
all $g, h \in K$.)

Furthermore, in each case one can also show that $\omega \neq 
\pi^* \overline{\omega}$ for an $\overline{\omega} \in
H^2(G/K,U(1))$:
\begin{enumerate}
\item For $K = \langle x^2 \rangle$, $G/K = {\mathbb Z}_2 \times
{\mathbb Z}_4$, which does admit discrete torsion.  However,
from table~\ref{table:z4z4-phases}, the genus-one phases are not
invariant under $x^2$:  for example, $\epsilon(1,y) \neq \epsilon(x^2,y)$,
and so $\omega$ cannot be a pullback from $H^2({\mathbb Z}_2 \times
{\mathbb Z}_4,U(1))$.
\item For $K = \langle x \rangle$, $G/K = {\mathbb Z}_4$,
for which $H^2(G/K,U(1)) = 0$, so $\omega \neq \pi^* \overline{\omega}$
for any $\overline{\omega}$ since $\omega \neq 0$.
\item For $K = {\mathbb Z}_2 \times {\mathbb Z}_4$,
$G/K = {\mathbb Z}_2$, for which $H^2(G/K,U(1)) = 0$, so again we see
that $\omega \neq \pi^* \overline{\omega}$ for an
$\overline{\omega} \in H^2(G/K,U(1))$.
\end{enumerate}

In each of these three cases, $\beta(\omega) \neq 0$,
and determines the prediction.
\begin{enumerate}
\item $K = \langle x^2 \rangle$ is contained in the center of $G$, so $\beta(\om)$ will be a nontrivial homomorphism from $G/K\cong\Z_2\times\Z_4$ to $H^1(K,U(1))\cong\Z_2$.
So we see that Coker $\beta(\omega) = 0$ and the kernel is either
${\mathbb Z}_4$ or ${\mathbb Z}_2 \times {\mathbb Z}_2$.  By looking at the cocycles in table~\ref{table:z4z4-cocycle}, one can explicitly verify that $\beta(\om)(yK,x^2)=-1$, so $yK\notin\ker\beta(\om)$, and hence the kernel must be $\langle xK,y^2K\rangle\cong\Z_2\times\Z_2$.

We can see this more explicitly as follows.  Let $s: G/K \rightarrow G$
be the section 
\begin{equation}
s(y^nK)=y^n,\quad s(xy^nK)=xy^n,\quad\mbox{ for } 0\le n<4,
\end{equation}
in the conventions of appendix~\ref{app:z4sz4}.
As before,
\begin{equation}
\beta(\omega)(q,k) \: = \: \frac{
\omega( k s(q), s(q)^{-1} )
}{
\omega( s(q)^{-1}, k s(q) )
},
\end{equation}
and it is straightforward to compute that the nontrivial elements are
\begin{eqnarray}
\beta(\omega)(xK, x^2) & = &
\frac{ \omega(x^2 x, x^3) }{ \omega(x^3, x^3) } \: = \: +1,
\\
\beta(\omega)(yK,x^2) & = &
\frac{ \omega(x^2 y, y^3) }{ \omega(y^3, x^2 y) } \: = \: -1,
\\
\beta(\omega)(xyK, x^2) & = &
\frac{ \omega( x^2 xy, xy^3) }{ \omega(x y^3, x^3 y) } \: = \: -1,
\\
\beta(\omega)(y^2 K, x^2) & = &
\frac{ \omega(x^2 y^2, y^2) }{ \omega(y^2, x^2 y^2) } \: = \: +1,
\\
\beta(\omega)(xy^2K, x^2) & = &
\frac{ \omega(x^2 xy^2, x^3 y^2) }{ \omega(x^3 y^2, x^3 y^2)} \: = \: +1,
\\
\beta(\omega)(y^3K, x^2) & = &
\frac{ \omega(x^2 y^3, y) }{ \omega(y, x^2 y^3) } \: = \: -1,
\\
\beta(\omega)(xy^3K, x^2) & = &
\frac{ \omega(x^2 xy^3, xy) }{ \omega(xy, x^3 y^3) } \: = \: -1,
\end{eqnarray}
using table~\ref{table:z4z4-cocycle}.  From this it is straightforward to
see that
\begin{equation}
\beta(\omega)(1,-), \: \: \:
\beta(\omega)(xK,-), \: \: \:
\beta(\omega)(y^2K,-), \: \: \:
\beta(\omega)(xy^2K,-)
\end{equation}
all are the trivial element of $\hat{K} = \hat{\mathbb Z}_2$,
while
\begin{equation}
\beta(\omega)(yK,-), \: \: \:
\beta(\omega)(xyK,-), \: \: \:
\beta(\omega)(y^3K,-), \: \: \:
\beta(\omega)(xy^3K,-)
\end{equation}
all are the nontrivial element of $\hat{K}$, so $\beta(\omega):
G/K \rightarrow \hat{K}$ is a surjective map.  Since
multiplying by $xK$ leaves the map invariant, the kernel is
$\langle xK, y^2 K \rangle = {\mathbb Z}_2 \times {\mathbb Z}_2$,
as anticipated above.  The cokernel, trivially, vanishes.

We'll have no discrete torsion $\hat{\om}$ in this case, since the cokernel vanishes.  Equivalently, the two representations in $\hat{K}$ fall into a single $G/K$ orbit, so we can take our representative to be the trivial homomorphism that sends $K$ to $1\in U(1)$, and the stabilizer is $H_0=\ker\beta(\om)=\Z_2\times\Z_2$.  If there was nontrivial discrete torsion, we would be able to detect it by computing
\be
\hat{\e}_0(xK,y^2K)=\frac{\hat{\om}_0(xK,y^2K)}{\hat{\om}_0(y^2K,xK)}=\frac{\om(x,y^2)}{\om(y^2,x)}=1,
\ee
where we have made the last step using table~\ref{table:z4z4-cocycle}.  So $\hat{\om}_0$ cannot be the nontrivial class in $H^2(\Z_2\times\Z_2,U(1))$.

In summary, for this case we predict
\begin{equation}
{\rm QFT}\left( [X/ {\mathbb Z}_4 \rtimes {\mathbb Z}_4]_{\omega} \right)
\: = \: 
{\rm QFT}\left( [X/{\mathbb Z}_2 \times {\mathbb Z}_2] \right),
\end{equation}
a ${\mathbb Z}_2 \times {\mathbb Z}_2$ orbifold without discrete torsion.

\item 
The subgroup $K=\langle x\rangle$ is not contained in the center of $G$, so we can't really use the cokernel of the map $\beta(\om)$.  Instead we will proceed more directly.  First of all, we have $G/K=\langle yK\rangle\cong\Z_4$.  We can choose a section $s(y^nK)=y^n$.  Note that this is actually a homomorphism, so the extension class is trivial, which is sensible since $G$ is precisely the semidirect product of $K$ with $G/K$.  Now $\hat{K}=\{[\rho_m]|0\le m<4\}$, where each $\rho_m$ is a homomorphism from $K$ to $U(1)$ defined by its action on the generator of $K$, $\rho_m(x)=i^m$.  We compute the action of $G/K$ on $\hat{K}$ using (\ref{eq:ProjectiveLqDef}).
\be
(L_{yK}\rho_m)(x)=\frac{\om(y^{-1}x,y)}{\om(y,y^{-1}x)}\rho_m(y^{-1}xy)=\frac{\om(x^3y^3,y)}{\om(y,x^3y^3)}\rho_m(x^3)=i^{-m-1}.
\ee
So we have
\be
yK\cdot[\rho_0]=[\rho_3],\ yK\cdot[\rho_1]=[\rho_2],\ yK\cdot[\rho_2]=[\rho_1],\ yK\cdot[\rho_3]=[\rho_0].
\ee
In other words, $\hat{K}$ breaks into two orbits under $G/K$, each with stabilizer $\langle y^2K\rangle\cong\Z_2$.  Since $H^2(\Z_2,U(1))=0$, we don't need to compute $\hat{\om}_a$.

We are left with a prediction of
\begin{equation}
{\rm QFT}\left( [X/ {\mathbb Z}_4 \rtimes {\mathbb Z}_4]_{\omega} \right)
\: = \: 
{\rm QFT}\left( [X/{\mathbb Z}_2] \, \coprod \, [X/{\mathbb Z}_2] \right),
\end{equation}
two copies of the $[X/{\mathbb Z}_2]$ orbifold.

\item For $K = \langle y^2, x \rangle$, again we are not in the center.  Here we have $G/K=\langle yK\rangle\cong\Z_2$, and we can choose the section $s(K)=1$, $s(yK)=y$ (note that in this case the extension class is nontrivial since $e(yK,yK)=y^2$).  The irreducible representations of $K$ are given by $\hat{K}=\{[\rho_{m,n}]|0\le m<4,0\le n<2\}$ where the homomorphisms $\rho_{m,n}$ are defined by
\be
\rho_{m,n}(x)=i^m,\qquad\rho_{m,n}(y^2)=(-1)^n.
\ee
To compute the action of $G/K$ on $\hat{K}$ we compute
\begin{eqnarray}
(L_{yK}\rho_{m,n})(x) & = &
\frac{\om(y^{-1}x,y)}{\om(y,y^{-1}x)}\rho_{m,n}(y^{-1}xy)=i^{-m-1},   
\\
(L_{yK}\rho_{m,n})(y^2) & = &
\frac{\om(y^{-1}y^2,y)}{\om(y,y^{-1}y^2)}\rho(y^{-1}y^2y)=(-1)^n.
\end{eqnarray}
From this we learn that $\hat{K}$ breaks into four distinct orbits $\{[\rho_{m,n}],[\rho_{3-m,n}]\}$, where both $m$ and $n$ can be $0$ or $1$.  Each orbit has a trivial stabilizer $H_{m,n}=1$.  Since the stabilizer groups are trivial, there is of course no possible discrete torsion $\hat{\om}$, and so our prediction is 
\begin{equation}
{\rm QFT}\left( [X/ {\mathbb Z}_4 \rtimes {\mathbb Z}_4]_{\omega} \right)
\: = \: 
{\rm QFT}\left( \coprod_4 X \right),
\end{equation}
a disjoint union of four copies of $X$.
\end{enumerate}

In each case, these predictions can be verified by genus-one partition
function computations.  Briefly,
given
\begin{equation}
Z\left( [ X / {\mathbb Z}_4 \rtimes {\mathbb Z}_4 ]_{\omega} \right)
\: = \: 
\frac{1}{| {\mathbb Z}_4 \rtimes {\mathbb Z}_4 |}
\sum_{gh = hg} \epsilon(g,h) \, {\scriptstyle g} \square_h,
\end{equation}
and using the discrete torsion phases for ${\mathbb Z}_4 \rtimes
{\mathbb Z}_4$ given in table~\ref{table:z4z4-phases}, it is straightforward
to show that
\begin{enumerate}
\item For $K = \langle x^2 \rangle$,
\begin{equation}
Z\left( [ X / {\mathbb Z}_4 \rtimes {\mathbb Z}_4 ]_{\omega} \right)
\: = \: Z\left( [X/{\mathbb Z}_2 \times {\mathbb Z}_2 ] \right),
\end{equation}
a ${\mathbb Z}_2 \times {\mathbb Z}_2$ orbifold without
discrete torsion,
\item For $K = \langle x \rangle$,
\begin{equation}
Z\left( [ X / {\mathbb Z}_4 \rtimes {\mathbb Z}_4 ]_{\omega} \right)
\: = \:
2 Z\left( [X/{\mathbb Z}_2] \right) \: = \:
Z\left( [X/{\mathbb Z}_2] \, \coprod \, [X/{\mathbb Z}_2] \right),
\end{equation}
\item For $K = \langle y^2, x \rangle$,
\begin{equation}
Z\left( [ X / {\mathbb Z}_4 \rtimes {\mathbb Z}_4 ]_{\omega} \right)
\: = \:
4 Z(X) \: = \: Z\left( \coprod_4 X \right),
\end{equation}
\end{enumerate}
in each case matching the prediction.

\section{Examples in which $\iota^* \omega = 0$ and $\omega =
\pi^* \overline{\omega}$}
\label{sect:exs:pi}

In this section we will look at examples of $G$ orbifolds in which
the discrete torsion is a pullback from $G/K$, where the subgroup
$K$ acts trivially.  Note that in this case we have $\beta(\om)=0$, i.e.\ the prefactors in (\ref{eq:ProjectiveLqDef}) are trivial.

\subsection{$[X/{\mathbb Z}_2 \times {\mathbb Z}_4]$ with discrete
torsion and trivially-acting ${\mathbb Z}_2$ subgroup}

In this example we consider $[X/{\mathbb Z}_2 \times {\mathbb Z}_4]$
where the subgroup $\langle b^2 \rangle \cong {\mathbb Z}_2$
(in the conventions of appendix~\ref{app:z2z4}) acts trivially.

As given, $K = {\mathbb Z}_2 = \langle b^2 \rangle$ and
$G = {\mathbb Z}_2 \times {\mathbb Z}_4$,,
so as $H^2(K,U(1)) = 0$, we have that $\iota^* \omega = 0$ trivially.
Furthermore $G/K = {\mathbb Z}_2 \times {\mathbb Z}_2$,
and as both $H^2(G/K,U(1)) = {\mathbb Z}_2$ and
$H^2(G,U(1)) = {\mathbb Z}_2$, there is a chance that the
discrete torsion in ${\mathbb Z}_2 \times {\mathbb Z}_4$ is a pullback
from ${\mathbb Z}_2 \times {\mathbb Z}_2$.

Write $G/K = {\mathbb Z}_2 \times {\mathbb Z}_2 = \langle \overline{a},
\overline{b} \rangle$, so that 
\begin{equation}
\pi(a) \: = \: \pi(ab^2) \: = \: \overline{a}, \: \: \:
\pi(b) \: = \: \pi(b^3) \: = \: \overline{b}, \: \: \:
\pi(ab) \: = \: \pi(ab^3) \: = \: \overline{a} \overline{b},
\: \: \:
\pi(b^2) \: = \: 1.
\end{equation}
Using the cocycle for the nontrivial element of
$H^2(G,U(1))$ given in table~\ref{table:z2z4-cocycle},
we see that,
\begin{equation}
\omega(1,g) \: = \: \omega(b^2,g), \: \: \:
\omega(a,g) \: = \: \omega(ab^2,g), \: \: \:
\omega(b,g) \: = \: \omega(b^3,g), \: \: \: 
\omega(ab,g) \: = \: \omega(ab^3,g)
\end{equation}
(and symmetrically for $\omega(y,x)$ instead of $\omega(x,y)$)
for any element $g \in G = {\mathbb Z}_2 \times {\mathbb Z}_4$,
which indicates that $\omega = \pi^* \overline{\omega}$ for
$\overline{\omega} \in H^2(G/K, U(1))$, which is given explicitly
in the table below:
\begin{center}
\begin{tabular}{c|rrrr}
& $1$ & $\overline{a}$ & $\overline{b}$ & $\overline{a} \overline{b}$ \\ \hline
$1$ & $1$ & $1$ & $1$ & $1$ \\
$\overline{a}$ & $1$ & $1$ & $i$ & $-i$ \\
$\overline{b}$ & $1$ & $-i$ & $1$ & $i$ \\
$\overline{a} \overline{b}$ & $1$ & $i$ & $-i$ & $1$
\end{tabular}
\end{center}
This is a representative of the nontrivial element of
$H^2({\mathbb Z}_2 \times {\mathbb Z}_2,U(1))$ (see for
example appendix~\ref{app:z2z2}), so we see that
$\omega = \pi^* \overline{\omega}$ for the nontrivial element
$\overline{\omega} \in H^2(G/K,U(1))$.

Since $G/K$ acts trivially on $\hat{K}$, the conjecture of
section~\ref{sect:conj} is going to
predict two components, two copies of $[X/{\mathbb Z}_2 \times {\mathbb Z}_2]$.
We need to compute the discrete torsion in each component.
One contribution to that discrete torsion will be from $\overline{\omega}$.
There is another potential contribution, from the image of the
extension class of $G$, an element of $H^2(G/K,K)$, under each
$\rho \in \hat{K}$.  
Since $G$ is abelian, the argument below (\ref{eq:HatRhoForAbelianK}) shows that this additional contribution vanishes.
As a result, the only contribution to discrete torsion on each
$[X/{\mathbb Z}_2 \times {\mathbb Z}_2]$ component is from
$\overline{\omega}$.

Putting this together, from the conjecture of section~\ref{sect:conj},
we predict that 
\begin{equation}
{\rm QFT}\left( [X/G]_{\omega} \right) \: = \: {\rm QFT}\left(
\left[ \frac{X \times \hat{K} }{G/K} \right]_{\overline{\omega}} \right)
\: = \: {\rm QFT}\left( 
[X / {\mathbb Z}_2 \times {\mathbb Z}_2]_{\rm d.t.} \, \coprod \,
[X / {\mathbb Z}_2 \times {\mathbb Z}_2]_{\rm d.t.} \right),
\end{equation}
or more simply, two copies of the ${\mathbb Z}_2 \times {\mathbb Z}_2$
orbifold with discrete torsion $\overline{\omega} \in
H^2({\mathbb Z}_2 \times {\mathbb Z}_2,U(1))$.

Next, we shall check this prediction at the level of 
partition functions.  Using table~\ref{table:z2z4-phases},
it is straightforward to compute that the genus-one partition function
of the ${\mathbb Z}_2 \times {\mathbb Z}_4$ orbifold with discrete torsion
$\omega$ is given by
\begin{eqnarray}
\lefteqn{
Z\left( [X/{\mathbb Z}_2 \times {\mathbb Z}_4]_{\omega} \right)
} \nonumber \\
& = &
\frac{1}{| {\mathbb Z}_2 \times {\mathbb Z}_4|}
\sum_{gh = hg} \epsilon(g,h) \, {\scriptstyle g} \square_h,
\\
& = &
\frac{1}{2} \left[ 
{\scriptstyle 1} \square_1 \: + \:
{\scriptstyle 1} \square_{\overline{a}} \: + \:
{\scriptstyle 1} \square_{\overline{b}} \: + \:
{\scriptstyle 1} \square_{\overline{a} \overline{b}} \: + \:
{\scriptstyle \overline{a}} \square_1 \: + \:
{\scriptstyle \overline{a}} \square_{\overline{a}} \: + \:
{\scriptstyle \overline{b}} \square_1 \: + \:
{\scriptstyle \overline{b}} \square_{\overline{b}} \: + \:
{\scriptstyle \overline{a} \overline{b}} \square_1 \: + \:
{\scriptstyle \overline{a} \overline{b}} \square_{\overline{a} \overline{b}}
\right. \nonumber \\
& & \hspace*{0.5in} \left.
\: - \:
{\scriptstyle \overline{a}} \square_{\overline{b}} \: - \: 
{\scriptstyle \overline{a}} \square_{\overline{a} \overline{b}} \: - \:
{\scriptstyle \overline{b}} \square_{\overline{a}} \: - \:
{\scriptstyle \overline{b}} \square_{\overline{a} \overline{b}} \: - \:
{\scriptstyle \overline{a} \overline{b}} \square_{\overline{a}} \: - \:
{\scriptstyle \overline{a} \overline{b}} \square_{\overline{b}}
\right],
\\
& = & 2 Z\left( [X / {\mathbb Z}_2 \times {\mathbb Z}_2]_{\rm d.t.} \right),
\\
& = & Z\left(  [X / {\mathbb Z}_2 \times {\mathbb Z}_2]_{\rm d.t.}
\, \coprod \,
 [X / {\mathbb Z}_2 \times {\mathbb Z}_2]_{\rm d.t.} \right),
\end{eqnarray}
precisely matching the prediction for this case.

By way of comparison, ordinary decomposition ($\omega = 0$) 
says something very similar in this case:
\begin{equation}
Z\left( [X/{\mathbb Z}_2 \times {\mathbb Z}_4] \right) \: = \:
Z\left( \coprod_2 [X/{\mathbb Z}_2 \times {\mathbb Z}_2] \right),
\end{equation}
with no discrete torsion on either side.
This is consistent with our computation that the image of
the extension class of ${\mathbb Z}_2 \times {\mathbb Z}_4$,
an element of $H^2({\mathbb Z}_2 \times {\mathbb Z}_2, {\mathbb Z}_2)$,
in $H^2({\mathbb Z}_2 \times {\mathbb Z}_2,U(1))$, necessarily vanishes
for all $\rho \in \hat{K}$.  Thus, in both this case and in the
example above, the two copies of $[X/{\mathbb Z}_2 \times {\mathbb Z}_2]$
have the same discrete torsion.

\subsection{$[X/{\mathbb Z}_4 \rtimes {\mathbb Z}_4]$ with discrete torsion
and trivially-acting ${\mathbb Z}_2$ subgroup}

In this section we consider a ${\mathbb Z}_4 \rtimes {\mathbb Z}_4$ orbifold
(the semidirect product of two copies of ${\mathbb Z}_4$)
with discrete torsion in which a subgroup $K = \langle y^2 \rangle \cong
{\mathbb Z}_2$ acts trivially, where ${\mathbb Z}_4 \rtimes {\mathbb Z}_4$
is generated by $x$ and $y$ in the notation of appendix~\ref{app:z4sz4}.
There is only one nontrivial value of discrete torsion since
$H^2({\mathbb Z}_4 \rtimes {\mathbb Z}_4,U(1)) = {\mathbb Z}_2$.
Also, it will be useful to observe that $K = \langle y^2 \rangle$
is a subgroup of the center of ${\mathbb Z}_4 \rtimes {\mathbb Z}_4$.

First, note that $G/K = D_4$.  Relating to the notation of
appendix~\ref{app:d4}, in which $D_4$ has generators $a$, $b$,
we identify $a = \{y, y^3\}$, $b = \{x, xy^2\}$.  It is straightforward
to check, for example, that $b^2 = \{ x^2, x^2 y^2 \}$ generates the
center, and that
\begin{equation}
a^2 \: = \: 1 \: = \: b^4, \: \: \:
ba \: = \: \{xy, xy^3\} \: = \: ab^3,
\end{equation}
as expected for $D_4$.

Next, since $H^2(K,U(1)) = 0$, we trivially have $\iota^* \omega = 0$.

Since $H^2(G/K,U(1)) = {\mathbb Z}_2$, the same as $H^2(G, U(1))$, it is possible that the element of discrete torsion
in the ${\mathbb Z}_4 \rtimes {\mathbb Z}_4$ orbifold is a pullback
from $D_4$.  Indeed, the phases in table~\ref{table:z4z4-phases}
are symmetric under $K$:  for example, 
\begin{equation}
\epsilon(g,h) \: = \: \epsilon(g y^2, h)
\end{equation}
and symmetrically, hence they pull back from $G/K = D_4$.  If we let
$\overline{\omega}$ be the nontrivial element of discrete torsion
in $D_4$, represented by a cocycle given in table~\ref{table:d4-cocycle},
then the pullback $\pi^* \overline{\omega}$ is given in 
table~\ref{table:pullback-d4-cocycle}.

\begin{table}[h] 
\begin{center}
\begin{tabular}{c|cccccccc}
& $1$ & $b=x$ & $b^2 = x^2$ & $b^3=x^3$ & $a=y$ & $ab = x^3y$ &
$ab^2 = x^2y$ & $ab^3 = xy$ \\ \hline
$1$ & $1$ & $1$ & $1$ & $1$ & $1$ & $1$ & $1$ & $1$ \\
$b=x$ & $1$ & $1$ & $1$ & $1$ & $\xi$ & $-\xi$ & $-\xi$ & $-\xi$ \\
$b^2=x^2$ & $1$ & $1$ & $1$ & $1$ & $+i$ & $+i$ & $-i$ & $-i$ \\
$b^3=x^3$ & $1$ & $1$ & $1$ & $1$ & $-\xi^*$ & $-\xi^*$ & $-\xi^*$ &
$+\xi^*$ \\
$a=y$ & $1$ & $-\xi$ & $-i$ & $\xi^*$ & $1$ & $-\xi^*$ & $+i$ & $\xi$ \\
$ab=x^3y$ & $1$ & $-\xi$ & $-i$ & $-\xi^*$ & $-\xi$ & $1$ & $-\xi^*$ & $+i$ \\
$ab^2=x^2y$ & $1$ & $-\xi$ & $+i$ & $-\xi^*$ & $-i$ & $-\xi$ & $1$ & $-\xi^*$\\
$ab^3=xy$ & $1$ & $+\xi$ & $+i$ & $-\xi^*$ & $\xi^*$ & $-i$ & $-\xi$ & $1$
\end{tabular}
\caption{Cocycle $\pi^* \overline{\omega}$ for $\overline{\omega}$ the
nontrivial element of $D_4$ discrete torsion and $\pi: 
{\mathbb Z}_4 \rtimes {\mathbb Z}_4 \rightarrow D_4$.  We define
$\xi = \exp(+6 \pi i/8)$, matching the $\xi$ of table~\ref{table:z4z4-cocycle}.
\label{table:pullback-d4-cocycle}
}
\end{center}
\end{table}

Taking into account the $y^2$ periodicities, 
it is straightforward to verify that
table~\ref{table:pullback-d4-cocycle}, describing $\pi^* \overline{\omega}$,
matches the cocycle for the nontrivial element of
$H^2({\mathbb Z}_4 \rtimes {\mathbb Z}_4,U(1))$ given in
table~\ref{table:z4z4-cocycle}.  (Of course, they need only match up to
coboundaries, but our conventions are such that they match on the nose.)
Therefore, we see explicitly that the
${\mathbb Z}_4 \rtimes {\mathbb Z}_4$ discrete torsion
$\omega = \pi^* \overline{\omega}$.

Applying section~\ref{sect:conj}, we can now see that the
$[X/{\mathbb Z}_4 \rtimes {\mathbb Z}_4]$ orbifold with
discrete torsion should be equivalent to a disjoint union of
two $[X/D_4]$ orbifolds, using the fact that
in this case,
$D_4 = G/K$ acts trivially\footnote{
This is because the two conjugacy classes corresponding to elements of
$\hat{K}$, namely $\{1\}$ and $\{y^2\}$, are invariant under conjugation by
elements of $G$.
} on $\hat{K}$.

It remains to compute the discrete
torsion on each $[X/D_4]$ summand.  Part of that discrete torsion will
be $\overline{\omega}$, but there could also be another 
contribution, the image of the extension class $H^2(D_4,{\mathbb Z}_2)$
under each of the irreducible representations of $K = {\mathbb Z}_2$.
Since ${\mathbb Z}_4 \rtimes {\mathbb Z}_4$ is not 
${\mathbb Z}_2 \rtimes D_4$ (in fact,
${\mathbb Z}_2 \times D_4$, since the ${\mathbb Z}_2$ is
central), the extension class in $H^2(D_4,{\mathbb Z}_2)$
is nontrivial. However, we claim that any map induced by a
representation into $H^2(D_4,U(1))$ vanishes.

Indeed, to compute the extension class, we must first pick a section $s:G/K\rightarrow G$ (so $\pi(s(q))=q$ for all $q\in G/K$), and then define the extension class by
\begin{equation}
e(q_1,q_2)=s(q_1)s(q_2)s(q_1q_2)^{-1}\in K,\qquad\mathrm{for\ all\ }q_1,q_2\in G/K.
\end{equation}
Different choices of section will give extension classes that differ by something exact.  For the current example, we could take the section $s(a^nb^m)=y^nx^m$, where $0\le n<2$ and $0\le m<4$.  Then we compute
\begin{equation}
e(b^k,b^m)=e(ab^k,b^m)=e(b^k,ab^m)=1,\quad e(ab^k,ab^m)=y^2.
\end{equation}
Now there are two possible irreducible representations of $K=\mathbb{Z}_2$, a trivial one $\rho_0$ which sends $y^2$ to $1$, and a nontrivial one $\rho_1$ which sends $y^2$ to $-1$.  Applying these to the extension class $e$ gives us a pair of 2-cocycles, $\hat{\omega}_a(q_1,q_2)=\rho_a(e(q_1,q_2))$.  Clearly $\hat{\omega}_0$ is the trivial 2-cocycle.  The other possibility $\hat{\omega}_1$ is not trivial, but it is exact, and in particular it is symmetric in its arguments, $\hat{\omega}_1(q_1,q_2)=\hat{\omega}_1(q_2,q_1)$, which implies that the corresponding discrete torsion contribution vanishes.

Putting these pieces together, we can now
make the prediction that
\begin{equation}
{\rm QFT}\left( [X/{\mathbb Z}_4 \rtimes {\mathbb Z}_4]_{\omega} \right)
\: = \:
{\rm QFT}\left( \left[ \frac{ X \times \hat{K} }{ D_4 } \right]_{
\overline{\omega}} \right) \: = \:
{\rm QFT}\left( [X/D_4]_{\overline{\omega}} \, \coprod \,
[X/D_4]_{\overline{\omega}} \right).
\end{equation}

The fact that the QFT decomposes into two identical
summands (universes), and so has a $B {\mathbb Z}_2$ symmetry,
reflects the group-theoretic fact that the discrete torsion
$\omega$ in ${\mathbb Z}_4 \rtimes {\mathbb Z}_4$ is invariant
under $\langle y^2 \rangle \cong {\mathbb Z}_2$: 
\begin{equation}
\omega(g,h) \: = \: \omega(g y^2, h) \: = \: \omega(g, h y^2)
\: = \: \omega(g y^2, h y^2),
\end{equation}
as can be seen in table~\ref{table:z4z4-cocycle}.
(Of course, such a relation need only hold up to cocycles,
but in our conventions it holds on the nose.)

Next, we shall compare this prediction to physics results.
Using table~\ref{table:z4z4-phases} of phases for discrete torsion
in ${\mathbb Z}_4 \rtimes {\mathbb Z}_4$ orbifolds,
it is straightforward to compute that
the genus-one orbifold partition function is
\begin{eqnarray}
Z\left( [X/{\mathbb Z}_4 \rtimes {\mathbb Z}_4]_{\omega} \right)
& = &
\frac{1}{| {\mathbb Z}_4 \rtimes {\mathbb Z}_4 |}
\sum_{gh = hg} \epsilon(g,h) \, {\scriptstyle g} \square_h,
\\
& = & 2 Z\left( [X/D_4]_{\overline{\omega}} \right)
\: = \: Z\left(  [X/D_4]_{\overline{\omega}} \, \coprod \,
 [X/D_4]_{\overline{\omega}} \right),
\end{eqnarray}
or two copies of the $D_4$ orbifold, each with discrete torsion,
confirming the prediction.

By way of comparison, ordinary decomposition ($\omega = 0$)
says something very similar in this case:
\begin{equation}
Z\left( [X/{\mathbb Z}_4 \rtimes {\mathbb Z}_4] \right) \: = \:
Z\left( \coprod_2 [X/D_4] \right),
\end{equation}
with no discrete torsion on either side.

\section{Mixed examples}
\label{sect:exs:mixed}

In this section we will describe examples that encompass further
aspects of the conjecture, depending upon the value of 
discrete torsion.  

Each of these examples will be a 
${\mathbb Z}_k \times {\mathbb Z}_k$ orbifold.  To that end, recall that
the phase assigned by discrete torsion $\omega \in {\mathbb Z}_k$ in a
${\mathbb Z}_k \times {\mathbb Z}_k$ orbifold is determined as follows.
Let $g \in {\mathbb Z}_k \times {\mathbb Z}_k$ be determined by
two integers $(a,b)$, $a, b \in \{0, \cdots, k-1\}$.
Then, associated to the twisted sector $(g,h) = (a,b; a', b')$
is the phase \cite[equ'n (2.2)]{Vafa:1994rv}
\begin{equation}  \label{eq:zkzkdt}
\epsilon(a,b; a', b') \: = \: \xi^{a b' - b a'},
\end{equation}
where $\xi$ is a $k$th root of unity corresponding to the value
of discrete torsion.

\subsection{${\mathbb Z}_4 \times {\mathbb Z}_4$ orbifold with
discrete torsion and trivially-acting ${\mathbb Z}_2 \times
{\mathbb Z}_2$}

Consider the case $G = {\mathbb Z}_4 \times {\mathbb Z}_4$,
with trivially-acting $K = {\mathbb Z}_2 \times {\mathbb Z}_2$,
so that $G/K = {\mathbb Z}_2 \times {\mathbb Z}_2$.  Suppose the
orbifold has discrete torsion,
$\omega \in H^2({\mathbb Z}_4 \times {\mathbb Z}_4, U(1)) = {\mathbb Z}_4$,
which we encode in a fourth root of unity labelled $\xi$.

Let us first compute our predictions from section~\ref{sect:conj} 
for the various
possible values of discrete torsion.  Let $\xi$ be a fourth root of unity
which encodes the value of discrete torsion.

Note that for any value of $\xi$, $\iota^* \omega = 0$.
To see this, it suffices to consider the genus-one phases~(\ref{eq:zkzkdt}).
If $(\overline{a},\overline{b}) \in {\mathbb Z}_2 \times {\mathbb Z}_2$,
with $\overline{a}, \overline{b} \in \{0, 1 \}$,
then the inclusion $\iota$ is given by
\begin{equation}
\iota( \overline{a}, \overline{b} ) \: = \: (2 \overline{a}, 2 \overline{b}),
\end{equation}
and so as a result,
\begin{equation}
\iota^* \epsilon( \overline{a}, \overline{b}; \overline{a}', \overline{b}')
\: = \: \epsilon(2 \overline{a}, 2 \overline{b}; 2 \overline{a}',
2 \overline{b}') \: = \:
\xi^{4 (\overline{a} \overline{b}' - \overline{b} \overline{a}') }.
\end{equation}
Since 
\begin{equation}
\overline{a} \overline{b}' - \overline{b} \overline{a}'
\: \in \: \{-1, 0, +1 \},
\end{equation}
and $\xi^4 = 1$, we see that all the phases $\iota^* \epsilon(
\overline{a}, \overline{b}; \overline{a}', \overline{b'}) = 1$,
so the cocycle $\iota^* \omega$ must be trivial.

Next, let us consider the four possible values of discrete torsion.
\begin{itemize}
\item First, consider the case of vanishing discrete torsion.
Then, decomposition \cite{Hellerman:2006zs} applies, and we have
\begin{equation}
{\rm QFT}\left( [X / {\mathbb Z}_4 \times {\mathbb Z}_4] \right)
\: = \: 
{\rm QFT}\left( \left[ \frac{ X \times \hat{K} }{ {\mathbb Z}_2 \times
{\mathbb Z}_2 } \right] \right)
\: = \: 
{\rm QFT}\left( \coprod_4 [ X / {\mathbb Z}_2 \times {\mathbb Z}_2 ]
\right),
\end{equation}
a disjoint union of four copies of $[X/{\mathbb Z}_2 \times
{\mathbb Z}_2]$, where here $\hat{K}$ denotes the set of
irreducible honest representations of $K$.
\item $\xi = -1$.  In this case, using the phases
in equation~(\ref{eq:zkzkdt}), it is straightforward to see that 
the discrete torsion in
${\mathbb Z}_4 \times {\mathbb Z}_4$ is a pullback
of discrete torsion in ${\mathbb Z}_2 \times {\mathbb Z}_2$.
(Since $\xi^2 = +1$, the phases~(\ref{eq:zkzkdt}) are invariant under
shifting any of $a$, $b$, $a'$, $b'$ by $2$, or more simply, invariant
under $K$, reflecting an underlying
$BK$ one-form symmetry.)
As a result, decomposition \cite{Hellerman:2006zs} applies again, and we have
\begin{equation}
{\rm QFT}\left( [X / {\mathbb Z}_4 \times {\mathbb Z}_4]_{\omega} \right)
\: = \: 
{\rm QFT}\left( \left[ \frac{ X \times \hat{K} }{ {\mathbb Z}_2 \times
{\mathbb Z}_2 } \right]_{\rm d.t.} \right)
\: = \: 
{\rm QFT}\left( \coprod_4 [ X / {\mathbb Z}_2 \times {\mathbb Z}_2 ]_{\rm d.t.}
\right),
\end{equation}
a disjoint union of four copies of $[X/{\mathbb Z}_2 \times
{\mathbb Z}_2]_{\rm d.t.}$, where here $\hat{K}$ denotes the set of
irreducible honest representations of $K$.
\item $\xi^2 = -1$.  In these cases, the phases~(\ref{eq:zkzkdt}) are not
invariant under $K$, since shifting any of $a$, $b$, $a'$, $b'$ changes
the genus-one phase~(\ref{eq:zkzkdt}) by a sign.
For this reason, the cocycle $\omega \neq \pi^* \overline{\omega}$ for
any $\overline{\omega} \in H^2({\mathbb Z}_2 \times {\mathbb Z}_2, U(1))$.
Instead, in these cases, $\beta(\omega) \neq 0$.  Now, $\beta(\omega)$ is
an element of
\begin{equation}
H^1(G/K, H^1(K, U(1))) \: = \: {\rm Hom}\left( {\mathbb Z}_2 \times
{\mathbb Z}_2, {\mathbb Z}_2 \times {\mathbb Z}_2 \right),
\end{equation}
so enumerating possibilities, if $\beta(\omega)$ is nontrivial, one quickly
deduces that 
\begin{equation}
{\rm Ker}\, \beta(\omega) \: = \: {\rm Coker}\, \beta(\omega) \: \in \:
\{ 1, {\mathbb Z}_2, {\mathbb Z}_2 \times {\mathbb Z}_2 \}.
\end{equation}
Given that the genus-one phases~(\ref{eq:zkzkdt}) are entirely nontrivial,
they appear to define isomorphisms, hence
\begin{equation}
{\rm Ker}\, \beta(\omega) \: = \: {\rm Coker}\, \beta(\omega) \: = \: 0,
\end{equation}
and so for these cases, we predict
\begin{equation}
{\rm QFT}\left( [X/{\mathbb Z}_4 \times {\mathbb Z}_4]_{\omega} \right)
\: = \:
{\rm QFT}\left( \left[ \frac{ X \times \widehat{ {\rm Coker}\, \beta(\omega) }
}{ {\rm Ker}\, \beta(\omega) } \right] \right)
\: = \:
{\rm QFT}\left( X \right).
\end{equation}
\end{itemize}

Next, we compute the genus-one partition functions in the different
cases, to confirm the predictions above.  
In the current example, since the ${\mathbb Z}_2 \times {\mathbb Z}_2 \subset
{\mathbb Z}_4 \times {\mathbb Z}_4$ acts trivially, we should compute
the numerical factors multiplying each effective
${\mathbb Z}_2 \times {\mathbb Z}_2$ orbifold twisted sector.
These are listed in table~\ref{table:z4z4:z2z2mult}, where we use the notation
$(a,b; a', b')$ for $a, b, a', b' \in \{0, 1\}$ to indicate
an effective ${\mathbb Z}_2 \times {\mathbb Z}_2$ twisted sector,
and where $\xi$ is a fourth root of unity, corresponding to the choice
of ${\mathbb Z}_4 \times {\mathbb Z}_4$ discrete torsion.

\begin{table}[h]
\begin{center}
\begin{tabular}{c|c||c|c}
$(0,0; 0,0)$ & $16$ & $(1,0; 1,0)$ & $8 \xi^3(1 + \xi^2)$ \\
$(1,0; 0,0)$ & $8(1 + \xi^2)$ & $(1,1; 0,1)$ & $8 \xi(1+\xi^2)$ \\
$(0,1; 0,0)$ & $8(1+\xi^2)$ & $(1,0; 1,1)$ & $8\xi(1+\xi^2)$ \\
$(0,0; 1,0)$ & $8(1+\xi^2)$ & $(0,1; 1,1)$ & $8\xi^3(1+\xi^2)$ \\
$(0,0; 0,1)$ & $8(1+\xi^2)$ & $(1,1; 1,1)$ & $8(1+\xi^2)$ \\
$(1,1; 0,0)$ & $8(1+\xi^2)$ & & \\
$(1,0; 1,0)$ & $8(1+\xi^2)$ & & \\
$(1,0; 0,1)$ & $8\xi(1+\xi^2)$ & & \\
$(0,1; 1,0)$ & $8\xi^3(1+\xi^2)$ & & \\
$(0,1; 0,1)$ & $8(1+\xi^2)$ & & \\
$(0,0; 1,1)$ & $8(1+\xi^2)$ & &
\end{tabular}
\caption{Multiplicities and phases of effective ${\mathbb Z}_2 \times
{\mathbb Z}_2$ twisted sectors in ${\mathbb Z}_4 \times {\mathbb Z}_4$
orbifold with discrete torsion.
\label{table:z4z4:z2z2mult} }
\end{center}
\end{table}

Our results for genus-one partition functions are as follows.
\begin{itemize}
\item Case $\xi=1$.  In this case, there is no discrete torsion, and
using the results in table~\ref{table:z4z4:z2z2mult},
it is straightforward to show that
\begin{eqnarray}
Z\left( [ X / {\mathbb Z}_4 \times {\mathbb Z}_4 ]_{\rm d.t.} \right)
& = &
\frac{1}{| {\mathbb Z}_4 \times {\mathbb Z}_4 | }
\sum_{a, b, a', b' \in \{0, \cdots, 3\} }
\epsilon(a, b; a', b') {\scriptstyle (a,b)} \square_{a',b'},
\\
& = & (4) Z\left( [X/{\mathbb Z}_2 \times {\mathbb Z}_2] \right),
\end{eqnarray}
consistent with the prediction of decomposition \cite{Hellerman:2006zs}
that this orbifold
be equivalent to a disjoint union of four copies of an effective
$[ X / {\mathbb Z}_2 \times {\mathbb Z}_2 ]$ orbifold.
\item Case $\xi = -1$.  In this case, 
using the results in table~\ref{table:z4z4:z2z2mult},
it is straightforward to show that
\begin{eqnarray}
Z\left( [ X / {\mathbb Z}_4 \times {\mathbb Z}_4 ]_{\rm d.t.} \right)
& = &
\frac{1}{| {\mathbb Z}_4 \times {\mathbb Z}_4 | }
\sum_{a, b, a', b' \in \{0, \cdots, 3\} }
\epsilon(a, b; a', b') {\scriptstyle (a,b)} \square_{a',b'},
\\
& = &
(4) Z\left( [ X / {\mathbb Z}_2 \times {\mathbb Z}_2 ]_{\rm d.t.} \right),
\end{eqnarray}
consistent with the statement that in this case,
the ${\mathbb Z}_4 \times {\mathbb Z}_4$ orbifold with discrete torsion
is equivalent to a disjoint union of four copies of an effective
${\mathbb Z}_2 \times {\mathbb Z}_2$ orbifold with discrete torsion.
\item Cases $\xi^2 = -1$.  In these cases,
using the results in table~\ref{table:z4z4:z2z2mult},
it is straightforward to show that
\begin{eqnarray}
Z\left( [ X / {\mathbb Z}_4 \times {\mathbb Z}_4 ]_{\rm d.t.} \right)
& = &
\frac{1}{| {\mathbb Z}_4 \times {\mathbb Z}_4 | }
\sum_{a, b, a', b' \in \{0, \cdots, 3\} }
\epsilon(a, b; a', b') {\scriptstyle (a,b)} \square_{a',b'},
\\
& = & \frac{1}{16} \left[ (16) {\scriptstyle (0,0)} \square_{(0,0)}
 \right] \: = \: Z(X),
\end{eqnarray}
consistent with the statement that in these two cases,
the ${\mathbb Z}_4 \times {\mathbb Z}_4$ orbifold with discrete torsion
is equivalent to no orbifold at all.
\end{itemize}
This confirms our predictions.

In the first two cases, in which $\xi^2 = +1$,
the ${\mathbb Z}_4 \times {\mathbb Z}_4$
orbifold has a natural $B({\mathbb Z}_2 \times {\mathbb Z}_2)$ symmetry,
as the discrete torsion phases $\epsilon(a,b;a',b')$ are invariant
under incrementing any of $a, b, a', b'$ by $2$.
This is a symmetry-based reason why there is a decomposition
\cite{Hellerman:2006zs} in these
cases.

Thus, depending upon the value of discrete torsion, we see that the
${\mathbb Z}_4 \times {\mathbb Z}_4$ orbifold is equivalent to one of
the following three possibilities:
\begin{equation}
\coprod_4 \left[ X / {\mathbb Z}_2 \times {\mathbb Z}_2 \right],
\: \: \:
\coprod_4 \left[ X / {\mathbb Z}_2 \times {\mathbb Z}_2 \right]_{\rm d.t.},
\: \: \:
X.
\end{equation}

\subsection{${\mathbb Z}_8 \times {\mathbb Z}_8$ orbifold with
discrete torsion and trivially-acting ${\mathbb Z}_2 \times {\mathbb Z}_2$}

In this section, we will consider a potentially more complex
case of a ${\mathbb Z}_k \times {\mathbb Z}_k$ orbifold.  Here,
there are more
possible values of discrete torsion in the effectively-acting orbifold
summands/universes than in the previous (set of) examples, which will enable
us to conduct more thorough tests of predictions for discrete torsion
in summands/universes.

Much as in the last section, here $\iota^* \omega = 0$ in all cases.
We can see this as follows.  The map $\iota$ embeds
${\mathbb Z}_2 \times {\mathbb Z}_2$ in
${\mathbb Z}_8 \times {\mathbb Z}_8$ as
\begin{equation}
\iota( \overline{a}, \overline{b} ) \: = \: (4 \overline{a}, 
4 \overline{b}),
\end{equation}
hence from equation~(\ref{eq:zkzkdt}), we see that the pullback of the
genus-one phase is
\begin{equation}
(\iota^* \epsilon)( \overline{a}, \overline{b};
\overline{a}', \overline{b}') \: = \:
\epsilon( 4 \overline{a}, 4 \overline{b};
4 \overline{a}', 4 \overline{b}') \: = \:
\xi^{16 (\overline{a} \overline{b}' - \overline{a}' \overline{b})}
\: = \: 1,
\end{equation}
as in this case $\xi^8 = 1$ in all cases.  Since the phases are trivial,
we see that $\iota^* \omega = 0$.

Next, let us (briefly) predict the decomposition for various cases,
following section~\ref{sect:conj}:
\begin{enumerate}
\item $\xi^4 = +1$.  In this case, as for example
\begin{equation}
\epsilon(a+4, b; a', b') \: = \:
\xi^{(a+4)b' - a' b} \: = \: \xi^{a b' - a' b} \: = \:
\epsilon(a, b; a', b'),
\end{equation}
we see that $\omega = \pi^* \overline{\omega}$,
where $\overline{\omega}$ is discrete torsion in the
effectively-acting ${\mathbb Z}_4 \times {\mathbb Z}_4$ orbifold.

In principle, the discrete torsion is then given by $\overline{\omega} +
\hat{\omega}_0$, where $\hat{\omega}_0$ is that predicted by the
original decomposition.  In this case, the group $G$ is abelian,
hence $\hat{\omega} = 0$ on each component, as observed in
section~\ref{sect:rev}.  Mathematically, the same result can be obtained
by observing that since in this case the extension class is symmetric,
as can be made manifest by picking matching sections, one has 
\begin{equation}
\hat{\omega}_0(\overline{a}, \overline{b})
\: = \:
\hat{\omega}_0(\overline{b}, \overline{a}),
\end{equation}
and so the corresponding genus-one phases are trivial.

Therefore, we predict that for $\xi^4 = +1$,
\begin{equation}
{\rm QFT}\left( [ X / {\mathbb Z}_8 \times {\mathbb Z}_8]_{\omega} \right)
\: = \:
{\rm QFT} \left( \coprod_4 [X / {\mathbb Z}_4 \times {\mathbb Z}_4]_{
\overline{\omega}} \right).
\end{equation}

\item $\xi^4 = -1$.  For similar reasons as the case above,
here $\omega \neq \pi^* \overline{\omega}$, so instead $\beta(\omega) \neq 0$.
Judging from the phases, one expects that $\beta$ describes
a map of the generators of each ${\mathbb Z}_4$ to the generator of
each ${\mathbb Z}_2$, so that Coker $\beta(\omega) = 1$, and
Ker $\beta(\omega) = {\mathbb Z}_2 \times {\mathbb Z}_2$.
Thus, we predict
\begin{equation}
{\rm QFT}\left(  [ X / {\mathbb Z}_8 \times {\mathbb Z}_8]_{\omega} \right)
\: = \:
{\rm QFT} \left( [X / {\mathbb Z}_2 \times {\mathbb Z}_2]_{\hat{\omega}} 
\right),
\end{equation}
a single copy of $[X/{\mathbb Z}_2 \times {\mathbb Z}_2]$.  

To complete the prediction, we need to compute the discrete torsion,
following equation~(\ref{eq:GeneratHatOmegaDef}).  What really matters are the phases $\hat{\e}(q_1,q_2)=\hat{\om}(q_1,q_2)/\hat{\om}(q_2,q_1)$, where $q_1,q_2\in\ker\beta(\om)$.  But since $G$ is abelian, it's easy to see that the only part of the expression~(\ref{eq:GeneratHatOmegaDef}) that doesn't drop out of $\hat{\e}$ is the contribution of the $\om(s_1,s_2)$ factor.  In other words, we have
\be
\hat{\e}(q_1,q_2)=\frac{\om(s_1,s_2)}{\om(s_2,s_1)}.
\ee
Now Ker $\beta(\om)$ here consists of $(\overline{a},\overline{b})=(2c,2d)$, where $c,d\in\{0,1\}$, and from (\ref{eq:zkzkdt}) we then have $\hat{\e}(q_1,q_2)=(-1)^{c_1d_2-c_2d_1}$.  In other words, $\hat{\om}$ is in the nontrivial class of $H^2(\Z_2\times\Z_2,U(1))$.
\end{enumerate}

Next, we compute partition functions, to compare to the results above.
To that end, it is helpful to first write the genus-one partition function
of the ${\mathbb Z}_8 \times {\mathbb Z}_8$ orbifold in the form
\begin{eqnarray}
Z & = &
\frac{ 1 }{| {\mathbb Z}_8 \times {\mathbb Z}_8 |}
\sum_{a, b, a', b' = 0}^7
\xi^{a b' - a' b} \, 
{\scriptstyle (a,b)} \square_{(a',b')},
\\
& = &
\frac{1}{64} \sum_{\overline{a}, \overline{b}, \overline{a}',
\overline{b}' = 0}^3 \left( 
\xi^{ \overline{a} \overline{b}' - \overline{a}' \overline{b} } \right)
\left( 1 \: + \: ( \xi^4 )^{ \overline{b}' } \: + \:
( \xi^4 )^{\overline{a}} \: + \:
( \xi^4 )^{ \overline{b} } \: + \:
( \xi^4 )^{ \overline{a}' } \: + \:
( \xi^4 )^{\overline{a} + \overline{b}'} \: + \:
( \xi^4 )^{ \overline{a} + \overline{a}'}
\right. \nonumber \\
& & \hspace*{1.75in} \left. 
 \: + \:
( \xi^4 )^{ \overline{a} + \overline{b} } \: + \:
( \xi^4 )^{ \overline{b}' + \overline{a}'} \: + \:
( \xi^4 )^{ \overline{b}' + \overline{b} } \: + \:
( \xi^4 )^{ \overline{a}' + \overline{b}} \: + \:
( \xi^4 )^{ \overline{a} + \overline{b}' + \overline{a}'}
\right. \nonumber \\
& & \hspace*{1.75in} \left.
 \: + \:
( \xi^4 )^{ \overline{a} + \overline{b}' + \overline{b} } \: + \:
( \xi^4 )^{ \overline{a} + \overline{a}' + \overline{b} } \: + \:
( \xi^4 )^{ \overline{b}' + \overline{a}' + \overline{b} } \: + \:
( \xi^4 )^{ \overline{a} + \overline{b}' + \overline{a}' + \overline{b} }
\right)
\nonumber \\
& & \hspace*{3.5in} 
\left(
{\scriptstyle (\overline{a}, \overline{b})} \square_{
(\overline{a}', \overline{b}') } 
\right).
\end{eqnarray}
In table~\ref{table:z8z8:z4z4mult} 
we collect multiplicities of a few sectors of the effective
${\mathbb Z}_4 \times {\mathbb Z}_4$ orbifold.  (As the total number of
twisted sectors in the effective ${\mathbb Z}_4 \times {\mathbb Z}_4$
orbifold is $16^2 = 256$, we only list some representative examples,
instead of trying to list every case.)

\begin{table}[h]
\begin{center}
\begin{tabular}{c|c||c|c}
$(0,0;0,0)$ & $16$ & $(2,0;0,0)$ & $16$ \\
$(1,0;0,0)$ & $8(1 + \xi^4)$ & $(3,0;0,0)$ & $8(1+\xi^4)$ \\
$(1,1;0,0)$ & $8(1+\xi^4)$ & $(2,1;0,0)$ & $8(1+\xi^4)$ \\
$(3,1;0,0)$ & $8(1+\xi^4)$ & $(2,2;0,0)$ & $16$ \\
$(1,0;0,1)$ & $8 \xi (1+\xi^4)$ & $(2,0;0,2)$ & $16 \xi^4$ \\
$(2,0;0,1)$ & $8 \xi^2 (1 + \xi^4)$ & $(2,0;0,3)$ & $8 \xi^6 (1 + \xi^4)$ \\
$(3,0;0,1)$ & $8 \xi^3(1 + \xi^4)$ & $(3,0;0,2)$ & $8 \xi^6(1+\xi^4)$ \\
$(3,0;0,3)$ & $8 \xi (1+ \xi^4)$ & & \\
$(0,1;1,0)$ & $8 \xi^{-1}(1 + \xi^4)$ & $(0,2;2,0)$ & $16 \xi^4$ \\
$(0,2;1,0)$ & $8 \xi^{-6} (1 + \xi^4)$ & &
\end{tabular}
\caption{Multiplicities and phases of effective
${\mathbb Z}_4 \times {\mathbb Z}_4$ twisted sectors in
${\mathbb Z}_8 \times {\mathbb Z}_8$ orbifold with discrete torsion
for some representative sectors.
\label{table:z8z8:z4z4mult}
}
\end{center}
\end{table}

Assembling these results, we can now compute genus-one partition functions.
\begin{enumerate}
\item $\xi^4 = +1$.  In this case,
\begin{eqnarray}
Z\left( [X/{\mathbb Z}_8 \times {\mathbb Z}_8]_{\omega} \right) & = &
\frac{1}{| {\mathbb Z}_8 \times {\mathbb Z}_8 |}
\sum_{a,b,a',b' = 0}^7 \xi^{a b' - a' b}\left(
{\scriptstyle (a,b)} \square_{(a',b')} \right),
\\
& = &
\frac{4^2}{64} \sum_{\overline{a}, \overline{b}, \overline{a}',
\overline{b}' = 0}^3
\xi^{ \overline{a} \overline{b}' - \overline{a}' \overline{b}}
\left( 
{\scriptstyle (\overline{a}, \overline{b}) } 
\square_{ (\overline{a}', \overline{b}') } \right),
\\
& = &
4 Z\left( [X/ {\mathbb Z}_4 \times {\mathbb Z}_4]_{\overline{\omega}} \right)
\: = \:
Z\left( \coprod_4  [X/ {\mathbb Z}_4 \times {\mathbb Z}_4]_{\overline{\omega}} \right),
\end{eqnarray}
where the discrete torsion $\omega = \pi^* \overline{\omega}$,
confirming our earlier prediction.
\item $\xi^4 = -1$.  In this case, from e.g. table~\ref{table:z8z8:z4z4mult},
the only effective genus-one ${\mathbb Z}_4 \times {\mathbb Z}_4$
twisted sectors that survive are of the form
\begin{equation}
(0,0;0,0), \: \: \:
(2,0;0,0), \: \: \:
(2,2;0,0), \: \: \: \cdots
\end{equation}
of weight $16$, and
\begin{equation}
(2,0;0,2), \: \: \:
(0,2;2,0), \: \: \: \cdots
\end{equation}
of weight $16 \, \xi^4$.  Thus, the only genus-one sectors that survive are
those of a ${\mathbb Z}_2 \times {\mathbb Z}_2$ orbifold, weighted
with sums as appropriate for discrete torsion.  Taking into account
multiplicities, we find
\begin{eqnarray}
Z\left( [X/{\mathbb Z}_8 \times {\mathbb Z}_8]_{\omega} \right) & = &
\frac{1}{| {\mathbb Z}_8 \times {\mathbb Z}_8 |}
\sum_{a,b,a',b' = 0}^7 \xi^{a b' - a' b}\left(
{\scriptstyle (a,b)} \square_{(a',b')} \right),
\\
& = &
\frac{16}{64} (4) Z\left( [ X / {\mathbb Z}_2 \times {\mathbb Z}_2]_{\rm d.t.}
\right),
\\
& = & Z\left(  [ X / {\mathbb Z}_2 \times {\mathbb Z}_2]_{\rm d.t.}
\right),
\end{eqnarray}
confirming our prediction.
\end{enumerate}

\section{Conclusions}

In this paper we have generalized decomposition \cite{Hellerman:2006zs} in
orbifolds with trivially-acting subgroups to include orbifolds with
discrete torsion.  Although the discrete torsion breaks (much of) any
original one-form symmetry, there is nevertheless a 
decomposition-like story.  We have described a general prediction for
all cases, which we have checked in numerous examples.

\section*{Acknowledgements}

We would like to thank R.~Donagi, T.~Pantev, and Y.~Tachikawa for useful
conversations.  D.R. was partially supported by 
NSF grant PHY-1820867.
E.S. was partially supported by NSF grants
PHY-1720321
and PHY-2014086.

\appendix

\section{Group cohomology and projective representations}
\label{app:GroupCohomologyAndProjReps}

In this appendix we give a very brief review of group cohomology and projective representations for finite groups, primarily to establish notation and nomenclature.

\subsection{Group cohomology}
\label{subapp:GroupCohomology}

Suppose we have a finite group $G$ and a $G$-module $M$.  Then for $n\ge 0$ we define the space $C^n(G,M)$ of $M$-valued $n$-cochains on $G$ as maps from the direct product of $n$ copies of $G$ to $M$.  Of course since these cochains are $M$-valued, $C^n(G,M)$ also forms a $G$-module, whose zero is simply the map sending every element of $G^n$ to $0\in M$.  Note that when talking about the general case of arbitrary $G$-modules it is standard convention to use additive notation in $M$ and multiplicative notation for $G$.  This helps clarify the relative structures, so we do so here at the beginning of this appendix.  However, in the rest of the paper our module $M$ is always either $\U(1)$ (most of the time) or an abelian subgroup of $G$, and in either case multiplicative notation is more natural, so these formulae should be retranscribed using multiplicative notation for $M$.  For the case of $M=\U(1)$ with trivial $G$-action, we do so at the end of this section.

We can construct a coboundary map $d_n:C^n(G,M)\rr C^{n+1}(G,M)$ by taking, for $\om\in C^n(G,M)$,
\bea
(d_n\om)(g_1,\cdots,g_{n+1}) &=& g_1\cdot\om(g_2,\cdots,g_{n+1})+\lp -1\rp^{n+1}\om(g_1,\cdots,g_n)\non\\
&& \qquad +\sum_{i=1}^n\lp -1\rp^i\om(g_1,\cdots,g_{i-1},g_ig_{i+1},g_{i+2},\cdots,g_{n+1}).
\eea
When the context is clear we will often drop the $n$ subscript on $d_n$.

A cochain $\om$ satisfying $d\om=0$ is said to be coclosed, and is called a cocycle (the equation $d\om=0$ is often called the cocycle condition on $\om$).  The space of $M$-valued $n$-cocycles of $G$ is thus defined to be $Z^n(G,M)=\ker(d_n)$.  It can be verified that the coboundary maps satisfy $d_{n+1}d_n=0$, and hence the image $d_{n-1}(C^{n-1}(G,M))$ (known as the space of coboundaries) is a submodule of $Z^n(G,M)$, and so we can define cohomology groups by taking the quotient,
\be
H^n(G,M)=Z^n(G,M)/d_{n-1}(C^{n-1}(G,M)).
\ee

We say that an $n$-cochain $\om$ is normalized if $\om(g_1,\cdots,g_n)$ vanishes whenever any one of its arguments is the identity element of $G$.  It can be easily verified that if $\om$ is normalized, then so is $d\om$ and with a bit more work one can show that every cohomology class contains a normalized representative.  For instance, consider the case $n=2$.  Then the cocycle condition is
\be
(d\om)(g_1,g_2,g_3)=g_1\cdot\om(g_2,g_3)-\om(g_1g_2,g_3)+\om(g_1,g_2g_3)-\om(g_1,g_2).
\ee
By setting $g_1=g_2=1$, we learn that a cocycle satisfies $\om(1,g)=\om(1,1)$ for all $g$, and setting $g_2=g_3=1$ tells us that $\om(g,1)=g\cdot\om(1,1)$.  Now pick any map $\m:G\rr M$ satisfying $\m(1)=-\om(1,1)$ and define $\om'=\om+d\m$.  Then this new cocycle satisfies
\be
\om'(1,1)=\om(1,1)+(d\m)(1,1)=\om(1,1)+1\cdot\m(1)-\m(1)+\m(1)=0,
\ee
and from the previous argument we also have $\om'(1,g)=\om'(g,1)=0$, showing that the cohomology class of $\om$ contains a normalized representative $\om'$.  Throughout the paper we will assume that our cochains and cocycles are normalized.

The case of greatest interest in this paper involves taking $M$ to be $\U(1)$ with trivial $G$-action, and $n=2$.  Switching now to multiplicative notation for $\U(1)$, a 2-cochain $\om$ is normalized if $\om(1,g)=\om(g,1)=1$ for all $g\in G$, and the cocycle condition is
\be
1=(d\om)(g_1,g_2,g_3)=\frac{\om(g_2,g_3)\om(g_1,g_2g_3)}{\om(g_1g_2,g_3)\om(g_1,g_2)}.
\ee
Shifting a 2-cocycle $\om$ by a coboundary $d\mu$ gives a new 2-cocycle $\om'$,
\be
\om'(g_1,g_2)=\om(g_1,g_2)\m(g_1)\m(g_2)\m(g_1g_2)^{-1},
\ee
where $\m:G\rr\U(1)$ is any map.

The finite cyclic groups have no 2-cohomology, $H^2(\Z_N,\U(1))\cong 1$.  Direct products of cyclic groups satisfy $H^2(\Z_N\times\Z_M,\U(1))\cong\Z_{\gcd(N,M)}$.  Some other examples include $H^2(S_3,\U(1))\cong 1$, $H^2(S_4,\U(1))\cong\Z_2$, $H^2(D_4,\U(1))\cong\Z_2$, and $H^2(\Z_4\rtimes\Z_4,\U(1))\cong\Z_2$.

\subsection{Projective representations}
\label{subapp:ProjReps}

Again let $G$ be a group.  A projective representation of $G$ with respect to a $\U(1)$-valued 2-cocycle $\om$ is a vector space $V$ and a map $\phi:G\rr\GL(V)$ satisfying
\be
\label{eq:GeneralProjRepMult}
\phi(g_1)\phi(g_2)=\om(g_1,g_2)\phi(g_1g_2).
\ee
Of course the case when $\om$ is trivial corresponds to an ordinary representation of $G$.  Associativity of $\phi$ implies the cocycle condition on $\om$ since
\be
\lp\phi(g_1)\phi(g_2)\rp\phi(g_3)=\om(g_1,g_2)\phi(g_1g_2)\phi(g_3)=\om(g_1,g_2)\om(g_1g_2,g_3)\phi(g_1g_2g_3),
\ee
\be
\phi(g_1)\lp\phi(g_2)\phi(g_3)\rp=\om(g_2,g_3)\om(g_1,g_2g_3)\phi(g_1g_2g_3),
\ee
but since $\phi(g_i)$ are simply matrices in $\GL(V)$ whose multiplication is associative, these two expressions must be equal, and multiplying by $\phi(g_1g_2g_3)^{-1}$ gives us the cocycle condition.

Note that if $\om$ is normalized, then any projective representation $\phi$ necessarily sends the identity element to the identity matrix, $\phi(1)=\mathbf{1}$.  Also note that as a consequence of (\ref{eq:GeneralProjRepMult}), the relation between inversion in $G$ and inversion of $\GL(V)$ picks up a phase\footnote{For any class in $H^2(G,\U(1))$ it is always possible to pick a representative $\om$ that is both normalized and also satisfies $\om(g,g^{-1})=1$ for all $g\in G$.  If we insisted on this property (essentially a sort of gauge choice), then this expression and some others in the paper would simplify, but since we are able to keep it more general we will only ask that our cocycles be normalized in the usual sense.}
\be
\label{eq:GeneralProjRepInverse}
\phi(g^{-1})=\om(g,g^{-1})\phi(g)^{-1}.
\ee

As long as $\om$ is coclosed, there will exist projective representations with respect to $\om$.  Indeed, we can define a regular projective representation by taking a vector space $V_r$ with a basis $\{v_g|g\in G\}$ and defining a map $\phi_r:G\rr\GL(V_r)$ by its action on the basis,
\be
\phi_r(g)(v_h)=\om(g,h)v_{gh}.
\ee
Then
\be
(\phi_r(g_1)\phi_r(g_2))(v_h)=\om(g_1,g_2h)\om(g_2,h)v_{g_1g_2h},
\ee
and
\be
\phi_r(g_1g_2)(v_h)=\om(g_1g_2,h)v_{g_1g_2h}.
\ee
Thus
\be
\phi_r(g_1)\phi_r(g_2)=\frac{\om(g_1,g_2h)\om(g_2,h)}{\om(g_1g_2,h)}\phi_r(g_1g_2)=d\om(g_1,g_2,h)\om(g_1,g_2)\phi_r(g_1g_2),
\ee
so if $\om$ is coclosed then we see that $\phi_r$ is a projective representation with respect to $\om$.

A subspace $U$ of $V$ is said to be invariant if $\phi(g)(U)\subseteq U$ for all $g\in G$.  If the only invariant subspaces of $V$ are $0$ and $V$ itself, then we say that $\phi$ is irreducible.  We say that two projective representations $\phi_1:G\rr\GL(V_1)$ and $\phi_2:G\rr\GL(V_2)$ with respect to the same $\om$ are isomorphic if there exists a vector space isomorphism $f:V_1\rr V_2$ so that $\phi_2(g)=f\circ\phi_1(g)\circ f^{-1}$ for all $g\in G$.  Of course irreducibility is preserved by isomorphism, so given a cocycle $\om\in Z^2(G,\U(1))$ we can talk about the set $\hat{G}_\om$ of isomorphism classes of irreducible projective representations of $G$ with respect to $\om$.

Just as there is a (not necessarily canonical) 
one-to-one correspondence between the isomorphism classes 
of ordinary irreducible representations, $\hat{G}$, 
and conjugacy classes of $G$ \cite[section 2.5]{serre}, 
there is also a one-to-one correspondence between $\hat{G}_\om$ and 
conjugacy classes $[g]$ of $G$ which additionally satisfy
\cite[prop. 2.6]{cheng},
\cite{costache,karpilovsky,schur1,schur2,schur3}, 
for any element $g$ in the conjugacy class, that $\om(g,g')=\om(g',g)$ for all $g'\in G$ such that $gg'=g'g$.  We will call these $\om$-trivial conjugacy classes, and their elements will be $\om$-trivial elements.

As an example, consider $G=\Z_2\times\Z_2=\{1,a,b,c\}$.  Then $H^2(G,\U(1))\cong\Z_2$, and a representative $\om$ for the nontrivial class can be defined by
\be
\om(1,g)=\om(g,1)=\om(g,g)=1,\qquad g\in G,\non
\ee
\be
\label{eq:Z2Z2NontrivialCocycle}
\om(a,b)=\om(b,c)=\om(c,a)=i,\qquad\om(a,c)=\om(b,a)=\om(c,b)=-i.
\ee
An example of a projective representation is
\be
\phi(1)=\lp\begin{matrix} 1 & 0 \\ 0 & 1 \end{matrix}\rp,\quad\phi(a)=\lp\begin{matrix} 0 & 1 \\ 1 & 0 \end{matrix}\rp,\quad\phi(b)=\lp\begin{matrix} 0 & -i \\ i & 0 \end{matrix}\rp,\quad\phi(c)=\lp\begin{matrix} 1 & 0 \\ 0 & -1 \end{matrix}\rp.
\ee
Up to isomorphism, this is the only irreducible projective representation with respect to $\om$.  This is consistent with the observation that the only $\om$-trivial element of $G$ is the identity element.   If one works out the regular representation for this $G$ and $\om$, the resulting four-dimensional projective representation can be shown to be isomorphic to the direct sum of two copies of the irreducible representation.

\section{Some calculations with cocycles}
\label{app:SomeCocycleCalculations}

In this appendix we provide the details of our calculation of $\widetilde{\om}_a$ and $\hat{\om}_a$ in section~\ref{sect:conj}. 

Taking the definitions (\ref{eq:TildeRhoWithPhaseDef}) and (\ref{eq:GeneralCDef}), we have
\bea
C_a(g_1,g_2) &=& \frac{\om(s_{12},s_{12}^{-1}s_1k_1s_2k_2)\om(s_{12}^{-1}s_1s_2,s_2^{-1}k_1s_2k_2)\om(s_2^{-1}k_1s_2,k_2)}{\om(s_1,k_1)\om(s_2,k_2)}f_1^{-1}\rho_a(k_1)f_2^{-1}\rho_a(k_2)\non\\
&& \qquad\times\ls f_{12}^{-1}\rho_a(s_{12}^{-1}s_1s_2)\rho_a(s_2^{-1}k_1s_2)\rho_a(k_2)\rs^{-1}\non\\
&=& \frac{\om(s_{12},s_{12}^{-1}s_1k_1s_2k_2)\om(s_{12}^{-1}s_1s_2,s_2^{-1}k_1s_2k_2)\om(s_2^{-1}k_1s_2,k_2)}{\om(s_1,k_1)\om(s_2,k_2)}\frac{\om(s_2^{-1}k_1,s_2)}{\om(s_2,s_2^{-1}k_1)}\non\\
&& \qquad\times f_1^{-1}f_2^{-1}\rho_a(s_2^{-1}k_1s_2)\rho_a(k_2)\rho_a(k_2)^{-1}\rho_a(s_2^{-1}k_1s_2)^{-1}\rho_a(s_{12}^{-1}s_1s_2)^{-1}f_{12}\non\\
&=& \frac{\om(s_{12},s_{12}^{-1}s_1k_1s_2k_2)\om(s_{12}^{-1}s_1s_2,s_2^{-1}k_1s_2k_2)\om(s_2^{-1}k_1s_2,k_2)\om(s_2^{-1}k_1,s_2)}{\om(s_1,k_1)\om(s_2,k_2)\om(s_2,s_2^{-1}k_1)}\non\\
&& \qquad\times f_1^{-1}f_2^{-1}\frac{\rho_a(s_2^{-1}s_1^{-1}s_{12})}{\om(s_2^{-1}s_1^{-1}s_{12},s_{12}^{-1}s_1s_2)}f_{12}\non\\
&=& \Om f_1^{-1}f_2^{-1}f_{12}\,\rho_a(s_{12}s_2^{-1}s_1^{-1}),
\eea
where
\be
\Om= \frac{\om(s_{12},s_{12}^{-1}s_1k_1s_2k_2)\om(s_{12}^{-1}s_1s_2,s_2^{-1}k_1s_2k_2)\om(s_2^{-1}k_1s_2,k_2)\om(s_2^{-1}k_1,s_2)\om(s_{12},s_2^{-1}s_1^{-1})}{\om(s_1,k_1)\om(s_2,k_2)\om(s_2,s_2^{-1}k_1)\om(s_{12}^{-1}s_1s_2,s_2^{-1}s_1^{-1}s_{12})\om(s_2^{-1}s_1^{-1},s_{12})}.
\ee
Fortunately, $\Om$ can be greatly simplified using cocycle conditions,
\begin{multline}
\Om=\frac{d\om(s_{12},s_{12}^{-1}s_1s_2,s_2^{-1}k_1s_2k_2)d\om(s_1s_2,s_2^{-1}k_1s_2,k_2)}{d\om(s_1,s_2,s_2^{-1}k_1)d\om(s_1k_1,s_2,k_2)}\\
\times\frac{d\om(s_1s_2,s_2^{-1}k_1,s_2)d\om(s_2^{-1}s_1^{-1},s_{12},s_{12}^{-1}s_1s_2)}{d\om(s_{12},s_2^{-1}s_1^{-1},s_1s_2)d\om(s_{12}s_2^{-1}s_1^{-1},s_1s_2s_{12}^{-1},s_{12})}\times\frac{\om(s_1k_1,s_2k_2)\om(s_1s_2s_{12}^{-1},s_{12})}{\om(s_1,s_2)\om(s_{12}s_2^{-1}s_1^{-1},s_1s_2s_{12}^{-1})},
\end{multline}
so we have
\be
C_a(g_1,g_2)=\frac{\om(s_1k_1,s_2k_2)\om(s_1s_2s_{12}^{-1},s_{12})}{\om(s_1,s_2)\om(s_{12}s_2^{-1}s_1^{-1},s_1s_2s_{12}^{-1})}f_1^{-1}f_2^{-1}f_{12}\,\rho_a(s_{12}s_2^{-1}s_1^{-1}).
\ee

To check that $C_a(g_1,g_2)$ is proportional to the identtiy, we first show that it commutes with all $\rho_a(k)$,
\bea
C_a(g_1,g_2)\rho_a(k) &=& \Om\om(s_{12}s_2^{-1}s_1^{-1},k)f_1^{-1}f_2^{-1}f_{12}\rho_a(s_{12}s_2^{-1}s_1^{-1}k)\non\\
&=& \Om\frac{\om(s_2^{-1}s_1^{-1}k,s_{12})\om(s_2,s_2^{-1}s_1^{-1}ks_{12}s_2^{-1})\om(s_1,s_1^{-1}ks_{12}s_2^{-1}s_1^{-1})}{\om(s_{12},s_2^{-1}s_1^{-1}k)\om(s_2^{-1}s_1^{-1}ks_{12}s_2^{-1},s_2)\om(s_1^{-1}ks_{12}s_2^{-1}s_1^{-1},s_1)}\non\\
&& \qquad\times\om(s_{12}s_2^{-1}s_1^{-1},k)\rho_a(ks_{12}s_2^{-1}s_1^{-1})f_1^{-1}f_2^{-1}f_{12}\non\\
&=& \Om\frac{\om(s_{12}s_2^{-1}s_1^{-1},k)\om(s_2^{-1}s_1^{-1}k,s_{12})\om(s_2,s_2^{-1}s_1^{-1}ks_{12}s_2^{-1})}{\om(k,s_{12}s_2^{-1}s_1^{-1})\om(s_{12},s_2^{-1}s_1^{-1}k)\om(s_2^{-1}s_1^{-1}ks_{12}s_2^{-1},s_2)}\non\\
&& \qquad\times\frac{\om(s_1,s_1^{-1}ks_{12}s_2^{-1}s_1^{-1})}{\om(s_1^{-1}ks_{12}s_2^{-1}s_1^{-1},s_1)}\rho_a(k)\rho_a(s_{12}s_2^{-1}s_1^{-1})f_1^{-1}f_2^{-1}f_{12}\non\\
&=& \Om'\rho_a(k)C_a(g_1,g_2),
\eea
where
\bea
\Om' &=& \frac{\om(s_{12}s_2^{-1}s_1^{-1},k)\om(s_2^{-1}s_1^{-1}k,s_{12})\om(s_2,s_2^{-1}s_1^{-1}ks_{12}s_2^{-1})\om(s_1,s_1^{-1}ks_{12}s_2^{-1}s_1^{-1})}{\om(k,s_{12}s_2^{-1}s_1^{-1})\om(s_{12},s_2^{-1}s_1^{-1}k)\om(s_2^{-1}s_1^{-1}ks_{12}s_2^{-1},s_2)\om(s_1^{-1}ks_{12}s_2^{-1}s_1^{-1},s_1)}\non\\
&& \qquad\times\frac{\om(s_1^{-1}s_{12}s_2^{-1}s_1^{-1},s_1)\om(s_2^{-1}s_1^{-1}s_{12}s_2^{-1},s_2)\om(s_{12},s_2^{-1}s_1^{-1})}{\om(s_1,s_1^{-1}s_{12}s_2^{-1}s_1^{-1})\om(s_2,s_2^{-1}s_1^{-1}s_{12}s_2^{-1})\om(s_2^{-1}s_1^{-1},s_{12})}\non\\
&=& d\om(s_{12}s_2^{-1}s_1^{-1},s_1s_2,s_2^{-1}s_1^{-1}k)d\om(s_2^{-1}s_1^{-1}k,s_{12}s_2^{-1},s_2)d\om(s_2,s_2^{-1}s_1^{-1}ks_{12}s_2^{-1}s_1^{-1},s_1)\non\\
&& \quad\times d\om(s_1,s_1^{-1}k,s_{12}s_2^{-1}s_1^{-1})d\om(s_1,s_2,s_2^{-1}s_1^{-1}k)d\om(s_2,s_2^{-1}s_1^{-1}k,s_{12}s_2^{-1}s_1^{-1})\non\\
&& \quad\times d\om(s_2^{-1}s_1^{-1}k,s_{12}s_2^{-1}s_1^{-1},s_1)d\om(s_{12}s_2^{-1}s_1^{-1},s_1,s_2)\non\\
&& \quad\times d\om(s_1^{-1}s_{12}s_2^{-1}s_1^{-1},s_{12},s_2^{-1}s_1^{-1}s_{12}s_2^{-1})d\om(s_{12},s_2^{-1}s_1^{-1}s_{12}s_2^{-1},s_1^{-1}s_{12}s_2^{-1}s_1^{-1})\non\\
&& \quad\times d\om(s_2^{-1}s_1^{-1}s_{12}s_2^{-1},s_1^{-1}s_{12}s_2^{-1}s_1^{-1},s_{12})\non\\
&=& 1.
\eea
Hence $C_a(g_1,g_2)$ commutes with all $\rho_a(k)$.  Then the irreducibility of $\rho_a$ and an application of Schur's lemma tell us that $C_a(g_1,g_2)$ is proportional to the identity matrix,
\be
\label{eq:Ingredient1}
C_a(g_1,g_2)=\widetilde{\om}_a(g_1,g_2)\mathbf{1}.
\ee

\section{Explicit realization of $\beta$}
\label{sect:beta}

The map 
\begin{equation}
\beta: \:
{\rm Ker}\, \iota^* \: \longrightarrow \: H^1(G/K, H^1(K, U(1))
\end{equation}
that we have utilized is described explicitly in
\cite[section 7]{hochschild} as the `reduction' map $r$,
but as the description there is in somewhat different language,
in this appendix we will unroll that definition to understand its
properties more explicitly.

First, instead of working directly with 2-cocycles $\omega$,
\cite{hochschild} manipulates extensions.  For our purposes,
following \cite[section IV.3]{brown},
$\omega \in H^2(G,U(1))$ corresponds to an extension $E$ of $G$ by $U(1)$,
\begin{equation}
1 \: \longrightarrow \: U(1) \: \longrightarrow \: E \: 
\stackrel{\sigma}{\longrightarrow} \: G \: \longrightarrow \: 1,
\end{equation}
where we can think of $E$ as $V \times G$, $V = U(1)$,
with product defined by
\begin{equation}
(v_1, g_1) \cdot (v_2, g_2) \: = \:
(v_1 + v_2 + \omega(g_1, g_2), g_1 g_2),
\end{equation}
which is associative so long as the 2-cocycle condition
\begin{equation}
\omega(g,h) \: + \: \omega(g h, k) \: = \:
\omega(h,k) \: + \: \omega(g,h k)
\end{equation}
is obeyed.  The projection map $\sigma: E \rightarrow G$ simply
maps $(v,g) \mapsto g$, and we assume $\omega$ is normalized so that
\begin{equation}
\omega(1,g) \: = \: \omega(g,1) \: = \: 0
\end{equation}
(in additive notation).

For this appendix, we specialize to the case that
$\iota^* \omega$ is trivial, where $\iota: K \hookrightarrow G$.
For this case, \cite{hochschild} defines $V \cdot K \subset E$,
which is naturally isomorphic to $\sigma^{-1}(K)$.

The reference \cite{hochschild} defines two further pertinent maps.
First, $\rho: G \rightarrow {\rm Aut}(V \cdot K)$,
which is given by
\begin{equation}
\rho(g)(v,k) \: = \: (g \cdot v, g k g^{-1}) \: = \:
(v, g k g^{-1})
\end{equation}
since the group action on the coefficients is trivial here.
The reference also defines
$\gamma_{V\cdot K, E}: E \rightarrow {\rm Aut}(V \cdot K)$,
by
\begin{equation}
\gamma_{V\cdot K, E}(e)(v,k) \: = \: 
e \cdot (v,k) \cdot e^{-1} \: = \: 
\left( v + \omega(g,k) - \omega(g^{-1}, g) + \omega(g k, g^{-1}), 
g k g^{-1} \right),
\end{equation}
for all $e = (\tilde{v},g) \in E$.  

Finally, the map $\beta$ is derived from the homomorphism
$f: E \rightarrow {\rm Aut}(V \cdot K)$ given by
\begin{equation}
f(e) \: = \: \gamma_{V\cdot K, E}(e) \circ \rho(\sigma(e^{-1})).
\end{equation}
Explicitly,
it is straightforward to compute that
\begin{equation}
f(e)(v,k) \: = \: \left( v + \omega(k g, g^{-1}) -
\omega(g^{-1}, kg) , k \right),
\end{equation}
and the reader will recognize
\begin{equation}
\omega(k g, g^{-1}) \: - \:
\omega(g^{-1}, kg),
\end{equation}
or in multiplicative notation,
\begin{equation}
\frac{ \omega(k g, g^{-1})
}{
\omega(g^{-1}, kg)
}
\end{equation}
as defining the map $\beta$ used in the text.

Before going on, let us illustrate some properties.
One of the essential properties of $f$ used in \cite{hochschild} is that
\begin{equation}
f(e_1 e_2) \: = \: f(e_1) \circ \rho( \sigma(e_1))^{-1} \circ
f(e_2) \circ \rho(\sigma(e_1)),
\end{equation}
which is a consequence of the identity
\begin{eqnarray}
\lefteqn{
\omega(k g_1 g_2, g_2^{-1} g_1^{-1}) \: - \:
\omega(g_2^{-1} g_1^{-1}, k g_1 g_2 )
}  \\
& = &
\omega( g_1^{-1} k g_1 g_2, g_2^{-1} ) \: + \:
\omega(k g_1, g_1^{-1}) \: - \:
\omega(g_1^{-1}, k g_1) \: - \:
\omega(g_2^{-1}, g_1^{-1} k g_1 g_2)
 \\
& &
\: + \:
(d \omega)(g_2^{-1}, g_1^{-1}, k g_1 g_2) \: + \:
(d \omega)(k g_1 g_2, g_2^{-1}, g_1^{-1}) \: + \:
(d \omega)(g_1^{-1}, k g_1 g_2, g_2^{-1} ).
\nonumber
\end{eqnarray}

In multiplicative notation, if
we define, for $g\in G$, $k\in K$,
\begin{equation}
\tilde{\beta}(\omega)(g,k) \: = \:
\frac{ \omega(k g, g^{-1})
}{
\omega(g^{-1}, kg)
},
\end{equation}
then this means that
\begin{equation}
\tilde{\beta}(\omega)(g_1 g_2,k) \: = \:
\tilde{\beta}(\omega)(g_1,k) 
\cdot
\tilde{\beta}(\omega)(g_2, g_1^{-1} k g_1).
\end{equation}

Next, we will define a related cocycle in
$H^1(G/K,H^1(K,\U(1)))$.
Specifically, for a section $s: G/K \rightarrow G$, and for
$q \in G/K$, $k \in K$, define
\begin{equation}
\label{eq:BetaOmegaDef}
\beta(\omega)(q,k) \: = \:
\frac{
\omega( k s(q), s(q)^{-1} )
}{
\omega( s(q)^{-1}, k s(q) )
}.
\end{equation}
We will check in detail that this is indeed a cocycle.

First, since the action on the $\U(1)$ coefficients is trivial, 
$H^1(K,\U(1))$ consists of maps $\phi:K\rr\U(1)$ such that 
\begin{equation}
1 \: = \: 
d\phi(k_1,k_2) \: = \:
\phi(k_1)\phi(k_2)\phi(k_1,k_2)^{-1}.
\end{equation}  
In other words, $H^1(K,\U(1))$ consists of group homomorphisms from $K$ to $\U(1)$.

The cochains $C^1(G/K,H^1(K,\U(1)))$ are maps from $G/K$ to $H^1(K,\U(1))$.  
Alternatively, we can think of these cochains as maps $\psi:G/K\times K\rr\U(1)$ such that for fixed $q\in G/K$, $\psi(q,-)$ acts as a homomorphism from $K$ to $\U(1)$.  Also, $G/K$ acts on $H^1(K,\U(1))$ via
\be
\lp q\cdot\phi\rp(k)
\: = \:
\phi(q^{-1}\cdot k)=\phi(s(q)^{-1} k s(q)),
\ee
which is induced by the standard action $q\cdot k=s(q) k s(q)^{-1}$ of $G/K$ 
on $K$, where $s:G/K\rr G$ is some choice of section.  Because of this nontrivial action on the coefficients, the cocycle condition for $\psi\in C^1(G/K,H^1(K,\U(1)))$ is (abbreviating $s_1=s(q_1)$)
\be
\label{eq:PsiCocycleCondition}
1 \: = \: (d\psi)(q_1,q_2,k)
\: = \:
\psi(q_2,s_1^{-1} k s_1) \,
\psi(q_1,k) \,
\psi(q_1q_2,k)^{-1}.
\ee
Moreover, if $\phi\in C^0(G/K,H^1(K,\U(1)))\cong H^1(K,\U(1))$ is a 0-cochain, then
\be
\label{eq:PsiCoboundaries}
(d\phi)(q,k) \: = \: (q\cdot\phi)(k) \, \phi(k)^{-1}
\: = \: 
\phi(s(q)^{-1} k s(q)) \, \phi(k)^{-1}.
\ee
Then $H^1(G/K,H^1(K,\U(1)))$ is of course the space of cocycles, i.e.\ cochains $\psi$ satisfying (\ref{eq:PsiCocycleCondition}), modulo coboundaries (\ref{eq:PsiCoboundaries}).

For the particular choice of 1-cochain (\ref{eq:BetaOmegaDef}), let's check some of these properties.  First of all, we check that for fixed $q$, $\beta(\om)(q,k)$ is a homomorphism in its second argument.  
Indeed, abbreviating $s(q)=g^{-1}$,
\begin{multline}
\frac{
\beta(\om)(q,k_1)
\beta(\om)(q,k_2)
}{
\beta(\om)(q,k_1k_2)
}
\: = \:
\frac{
\om(k_1g,g^{-1})  \om(k_2g,g^{-1})  \om(g^{-1},k_1k_2g)
}{
\om(g^{-1},k_1g)  \om(g^{-1},k_2g)  \om(k_1k_2g,g^{-1})
}
\\
= \:
\frac{
(d\om)(g^{-1},k_1g,g^{-1})  (d\om)(g^{-1},k_2g,g^{-1})
(d\om)(g^{-1}k_1g,g^{-1}k_2g,g^{-1})  (d\om)(g^{-1},k_1,k_2)
}{
(d\om)(g^{-1},k_1k_2g,g^{-1})  (d\om)(g^{-1}k_1g,g^{-1},k_2)
}\\
\times\frac{
\om(g^{-1}k_1g,g^{-1}k_2g)
}{
\om(k_1,k_2)
}.
\end{multline}
As long as $\iota^* \omega$ is trivial in cohomology, the right-hand side
is trivial in cohomology, and we have a homomorphism on cohomology
\begin{equation}
\beta(\omega)(q, k_1 k_2) \: = \: 
\beta(\omega)(q,k_1) \cdot \beta(\omega)(q, k_2).
\end{equation}
(In more detail, if $\om(k,k')=\m(k)\m(k')/\m(kk')$ for some
$\m:K\rr\U(1)$, then the factors of $\mu$ can be absorbed into
$\beta$, as $\beta'(\om)(q,k)=\beta(\om)(q,k)\m(s(q)^{-1}ks(q))\m(k)^{-1}$,
which defines the same cohomology class in $H^1(G/K,H^1(K,\U(1)))$.)

Next we check that it is actually a cocycle satisfying~(\ref{eq:PsiCocycleCondition}).
\begin{multline}
\frac{
\beta(\om)(q_2,s_1^{-1} k s_1 )
\beta(\om)(q_1,k)
}{
\beta(\om)(q_1q_2,k)
}
=
\frac{
\om(s_1^{-1} k s_1 s_2, s_2^{-1})
\om(k s_1, s_1^{-1})
\om(s_{12}^{-1}, k s_{12})
}{
\om(s_2^{-1}, s_1^{-1} k s_1 s_2)
\om(s_1^{-1}, k s_1 )
\om(k s_{12}, s_{12}^{-1})
} 
\\
=
\frac{
(d\om)(k s_1 s_2, s_2^{-1} s_1^{-1} s_{12},s_{12}^{-1})
(d\om)(s_2^{-1} s_1^{-1}, k s_1 s_2, s_2^{-1} s_1^{-1} s_{12})
(d\om)(s_2^{-1} s_1^{-1} s_{12}, s_{12}^{-1}, k s_{12} )
}{
(d\om)(k s_1 s_2, s_2^{-1}, s_1^{-1} )
(d\om)(s_2^{-1}, s_1^{-1}, k s_1 s_2 )
(d\om)(s_1^{-1}, k s_1 s_2, s_2^{-1} )
}
\\
\times
\frac{
\om(s_2^{-1} s_1^{-1} k s_1 s_2, s_2^{-1} s_1^{-1} s_{12} )
}{
\om(s_2^{-1} s_1^{-1} s_{12}, s_{12}^{-1} k s_{12} )
}.
\end{multline}
As before, if $\iota^\ast\om$ is trivial in cohomology,
then in cohomology,
\begin{equation}
\beta(\om)(q_1q_2,k) \: = \:
\beta(\om)(q_1,k) \cdot \beta(\om)(q_2,s_1^{-1} k s_1 ).
\end{equation}

If we shift $\om$ by a coboundary to $\om'(g_1,g_2)=\om(g_1,g_2)\la(g_1)\la(g_2)\la(g_1g_2)^{-1}$ for some $\la$, then $\beta(\om)$ shifts to $\beta(\om')$ given by
\begin{eqnarray}
\beta(\om')(q,k)
& = &
\beta(\om)(q,k)
\frac{
\la(k s(q) ) \la(s(q)^{-1})
}{
\la(k)
}
\frac{
\la(s(q)^{-1} k s(q))
}{
\la(s(q)^{-1})\la(ks(q))
},
\\
& = &
\beta(\om)(q,k)
\frac{
\la(s(q)^{-1} k s(q))
}{
\la(k)
}.
\end{eqnarray}
But this is precisely a coboundary shift, so $\beta(\om)$ and $\beta(\om')$ define the same class in the cohomology group $H^1(G/K,H^1(K,\U(1)))$.

Finally, let us compare $\beta(\omega)$ above to a phase factor
appearing in the definition of $L_q \phi$
in~(\ref{eq:ProjectiveLqDef}), namely
\begin{equation} \label{eq:LqDef-phase}
\frac{
\om(s(q)^{-1}k,s(q))
}{
\om(s(q),s(q)^{-1}k)
}.
\end{equation}
Using cocycle conditions, it is straightforward to check that
\begin{equation}
\frac{
\omega(s^{-1} k, s)
}{
\omega(s, s^{-1} k)
}
\: = \:
\frac{
(d \omega)(s, s^{-1} k, s) 
(d \omega)(s, s^{-1}, s)
}{
(d \omega)(ks, s^{-1}, s)
(d \omega)(s, s^{-1}, ks)
}
\,
\frac{
\omega(s^{-1}, ks)
}{
\omega(ks, s^{-1})
} \: = \: \beta(\omega)(q,k)^{-1},
\end{equation}
hence we see that the phase factor~(\ref{eq:LqDef-phase}) in $L_q \phi$
is identical to $\beta(\omega)^{-1}$.

\section{Pertinent group theory results}
\label{app:gpthy}

In this paper we perform rather detailed manipulations of and
computations utilizing representative cocycles of elements of
discrete torsion.  To that end,
in this section we will give explicit presentations and discrete torsion
cocycles for several finite groups which we use in this paper.

\subsection{${\mathbb Z}_2 \times {\mathbb Z}_2$}
\label{app:z2z2}

As is well-known, $H^2({\mathbb Z}_2 \times {\mathbb Z}_2, U(1)) = 
{\mathbb Z}_2$.

Denoting the two generators of ${\mathbb Z}_2 \times
{\mathbb Z}_2$ by $a$, $b$, then the
group 2-cocycle explicitly is
\begin{eqnarray}
\omega(a,b) \: = \: \omega(b,ab) \: = \: \omega(ab,a) & = & +i,
\\
\omega(b,a) \: = \: \omega(ab,b) \: = \: \omega(a,ab) & = & -i,
\end{eqnarray}
with $\omega(g,h) = +1$ for other $g, h$.
Given this cocycle, it is straightforward to see that the only conjugacy
class obeying the condition~(\ref{eq:countprojirreps}) is $\{ 1 \}$.

\subsection{${\mathbb Z}_2 \times {\mathbb Z}_4$}
\label{app:z2z4}

We present the group ${\mathbb Z}_2 \times {\mathbb Z}_4$ as
generated by $a$, $b$, where $a^2 = 1 = b^4$.
It can be shown that $H^2({\mathbb Z}_2 \times {\mathbb Z}_4, U(1)) = 
{\mathbb Z}_2$, and the cocycle and genus-one phases of the nontrivial
element are given in tables~\ref{table:z2z4-cocycle},
\ref{table:z2z4-phases}.

\begin{table}[h]
\begin{center}
\begin{tabular}{c|rrrrrrrr}
& $1$ & $b$ & $b^2$ & $b^3$ & $a$ & $ab$ & $ab^2$ & $ab^3$ \\ \hline
$1$ & $1$ & $1$ & $1$ & $1$ & $1$ & $1$ & $1$ & $1$ \\
$b$ & $1$ & $1$ & $1$ & $1$ & $-i$ & $i$ & $-i$ & $i$ \\
$b^2$ & $1$ & $1$ & $1$ & $1$ & $1$ & $1$ & $1$ & $1$ \\
$b^3$ & $1$ & $1$ & $1$ & $1$ & $-i$ & $i$ & $-i$ & $i$ \\
$a$ & $1$ & $i$ & $1$ & $i$ & $1$ & $-i$ & $1$ & $-i$ \\
$ab$ & $1$ & $-i$ & $1$ & $-i$ & $i$ & $1$ & $i$ & $1$ \\
$ab^2$ & $1$ & $i$ & $1$ & $i$ & $1$ & $-i$ & $1$ & $-i$ \\
$ab^3$ & $1$ & $-i$ & $1$ & $-i$ & $i$ & $1$ & $i$ & $1$
\end{tabular}
\caption{Table of values for $\omega(g,h)$ for a cocycle
representing the nontrivial element of 
$H^2({\mathbb Z}_2 \times {\mathbb Z}_4,U(1))$.
\label{table:z2z4-cocycle}
}
\end{center}
\end{table}

\begin{table}[h]
\begin{center}
\begin{tabular}{c|rrrrrrrr}
& $1$ & $b$ & $b^2$ & $b^3$ & $a$ & $ab$ & $ab^2$ & $ab^3$ \\ \hline
$1$ & $1$ & $1$ & $1$ & $1$ & $1$ & $1$ & $1$ & $1$ \\
$b$ & $1$ & $1$ & $1$ & $1$ & $-1$ & $-1$ & $-1$ & $-1$ \\
$b^2$ & $1$ & $1$ & $1$ & $1$ & $1$ & $1$ & $1$ & $1$ \\
$b^3$ & $1$ & $1$ & $1$ & $1$ & $-1$ & $-1$ & $-1$ & $-1$ \\
$a$ & $1$ & $-1$ & $1$ & $-1$ & $1$ & $-1$ & $1$ & $-1$ \\
$ab$ & $1$ & $-1$ & $1$ & $-1$ & $-1$ & $1$ & $-1$ & $1$ \\
$ab^2$ & $1$ & $-1$ & $1$ & $-1$ & $1$ & $-1$ & $1$ & $-1$ \\
$ab^3$ & $1$ & $-1$ & $1$ & $-1$ & $-1$ & $1$ & $-1$ & $1$
\end{tabular}
\caption{Table of values of $\epsilon(g,h) = \omega(g,h)/\omega(h,g)$
for a cocycle representing the nontrivial element of 
$H^2({\mathbb Z}_2 \times {\mathbb Z}_4, U(1))$.
\label{table:z2z4-phases}
}
\end{center}
\end{table}

Since ${\mathbb Z}_2 \times {\mathbb Z}_4$ is abelian, 
each element corresponds to
its own separate conjugacy class.  Of these, from
table~\ref{table:z2z4-phases}, only the conjugacy classes
$\{1 \}$, $\{ b^2 \}$ satisfy the 
condition~(\ref{eq:countprojirreps}), and so are associated with irreducible
projective representations.

\subsection{$D_4$}
\label{app:d4}

We present the eight-element dihedral group $D_4$ as generated
by $z, a, b$, where $z$ generates the center, $z^2 = 1$,
$a^2 = 1$, $b^2 = z$, so that the elements are described as
\begin{equation}
D_4 \: = \: \{ 1, z, a, b, az, bz, ab, ba = abz \}.
\end{equation}

It can be shown that $H^2(D_4,U(1)) = {\mathbb Z}_2$.
A group cochain representing the
nontrivial element is given as
\begin{equation}
\omega(g,h) \: = \: \exp(2 \pi i n(g,h)/8),
\end{equation}
where the $n(g,h)$ are given in table~\ref{table:d4-cocycle}.
Note that in the group multiplication, $b^2 = z$, $b^3 = bz$, and $ba = abz$.

\begin{table}[h]
\begin{center}
\begin{tabular}{c|rrrrrrrr}
 & $1$ & $b$ & $z$ & $bz$ & $a$ & $ba$ & $az$ & $ab$ \\
\hline
$1$ & $0$ & $0$ & $0$ & $0$ & $0$ & $0$ & $0$ & $0$ \\
$b$ & $0$ & $0$ & $0$ & $0$ & $3$ & $-1$ & $-1$ & $-1$ \\
$z$ & $0$ & $0$ & $0$ & $0$ & $2$ & $-2$ & $-2$ & $2$ \\
$bz$ & $0$ & $0$ & $0$ & $0$ & $1$ & $-3$ & $1$ & $1$ \\
$a$ & $0$ & $-1$ & $-2$ & $-3$ & $0$ & $3$ & $2$ & $1$ \\
$ba$ & $0$ & $3$ & $2$ & $1$ & $-3$ & $0$ & $-1$ & $-2$ \\
$az$ & $0$ & $-1$ & $2$ & $1$ & $-2$ & $1$ & $0$ & $-1$ \\
$ab$ & $0$ & $-1$ & $-2$ & $1$ & $-1$ & $2$ & $1$ & $0$
\end{tabular}
\caption{Table of values of $n(g,h)$, appearing in the cocycle representing
the nontrivial element of $H^2(D_4,U(1))$.
\label{table:d4-cocycle}
}
\end{center}
\end{table}

Using this, the phases weighting a given genus-one
twisted sector, associated to a commuting pair $(g,h)$, are given
as ratios
\begin{equation}
\epsilon(g,h) \: = \:
\frac{ \omega(g,h) }{ \omega(h,g) }.
\end{equation}
We list the phases in table~\ref{table:d4-phases}.

\begin{table}[h]
\begin{center}
\begin{tabular}{c|rrrrrrrr}
 & $1$ & $b$ & $z$ & $bz$ & $a$ & $ba$ & $az$ & $ab$ \\
\hline
$1$ & $1$ & $1$ & $1$ & $1$ & $1$ & $1$ & $1$ & $1$ \\
$b$ & $1$ & $1$ & $1$ & $1$ & $0$ & $0$ & $0$ & $0$ \\
$z$ & $1$ & $1$ & $1$ & $1$ & $-1$ & $-1$ & $-1$ & $-1$ \\
$bz$ & $1$ & $1$ & $1$ & $1$ & $0$ & $0$ & $0$ & $0$ \\
$a$ & $1$ & $0$ & $-1$ & $0$ & $1$ & $0$ & $-1$ & $0$ \\
$ba$ & $1$ & $0$ & $-1$ & $0$ & $0$ & $1$ & $0$ & $-1$ \\
$az$ & $1$ & $0$ & $-1$ & $0$ & $-1$ & $0$ & $1$ & $0$ \\
$ab$ & $1$ & $0$ & $-1$ & $0$ & $0$ & $-1$ & $0$ & $1$
\end{tabular}
\caption{Table of values of $T^2$ twisted sector phases in a $D_4$
orbifold with discrete torsion.  
Non-commuting pairs are indicated with $0$.
\label{table:d4-phases}
}
\end{center}
\end{table}

Note in passing that this is not the pullback from $D_4/{\mathbb Z}_2 = {\mathbb Z}_2
\times {\mathbb Z}_2$, as for example the discrete torsion there generates
phases solely in sectors that do not lift to $D_4$.  (In particular,
for the pullback of $H^2( {\mathbb Z}_2 \times {\mathbb Z}_2, U(1))$,
the discrete torsion phases are trivial, so for that element of $H^2(D_4,U(1))$,
the $D_4$ orbifold is the same as a $D_4$ orbifold with no discrete torsion
at all.

The conjugacy classes of $D_4$ are
\begin{equation}
\{ 1 \}, \: \: \:
\{ z \}, \: \: \:
\{a, az\}, \: \: \:
\{b, bz\}, \: \: \:
\{ab, ba\}.
\end{equation}
Of these, only the conjugacy classes $\{ 1 \}$, $\{ b, bz \}$ are associated
with projective representations.  We describe issues with the others below:
\begin{itemize}
\item $\omega(a,z) \neq \omega(z,a)$, so $\{ z \}$ is not a pertinent conjugacy
class.
\item $\omega(a,az) \neq \omega(az,a)$, so $\{a, az\}$ is not a pertinent
conjugacy class.
\item $\omega(ab,ba) \neq \omega(ba,ab)$, so $\{ab, ba\}$ is not a pertinent
conjugacy class.
\end{itemize}
Thus, we see that there are two irreducible projective representations
of $D_4$ with the nontrivial element of $H^2(D_4,U(1))$.
(This result is also given in \cite[example 3.12]{cheng}.)

\subsection{${\mathbb Z}_4 \rtimes {\mathbb Z}_4$}
\label{app:z4sz4}

We describe the group ${\mathbb Z}_4 \rtimes {\mathbb Z}_4$,
the semidirect product of two copies of ${\mathbb Z}_4$, as
generated by $x$, $y$ subject to the conditions $x^4 = y^4 = 1$
and $y = x y x$.

It can be shown that $H^2( {\mathbb Z}_4 \rtimes {\mathbb Z}_4, U(1)) = 
{\mathbb Z}_2$, with cocycles as given in table~\ref{table:z4z4-cocycle}.

\begin{sidewaystable}[h]
\begin{center}
\begin{tabular}{c|rrrrrrrrrrrrrrrr}
& $1$ & $x$ & $x^2$ & $x^3$ & $y$ & $xy$ & $x^2y$ & $x^3y$ &
$y^2$ & $xy^2$ & $x^2y^2$ & $x^3y^2$ & $y^3$ & 
$xy^3$ & $x^2y^3$ & $x^3y^3$ \\ \hline
$1$ & $1$ & $1$ & $1$ & $1$ & $1$ & $1$ & $1$ & $1$ & $1$ & $1$ & $1$ &
$1$ & $1$ & $1$ & $1$ & $1$ \\
$x$ & $1$ & $1$ & $1$ & $1$ & $\xi$ & $-\xi$ & $-\xi$ & $-\xi$ &
$1$ & $1$ & $1$ & $1$ & $\xi$ & $-\xi$ & $-\xi$ & $-\xi$ \\
$x^2$ & $1$ & $1$ & $1$ & $1$ & $i$ & $-i$ & $-i$ & $i$ & $1$ &
$1$ & $1$ & $1$ & $i$ & $-i$ & $-i$ & $i$ \\
$x^3$ & $1$ & $1$ & $1$ & $1$ & $-\xi^*$ & $\xi^*$ & $-\xi^*$ & $-\xi^*$ &
$1$ & $1$ & $1$ & $1$ & $-\xi^*$ & $\xi^*$ &  $-\xi^*$ & $-\xi^*$ \\
$y$ & $1$ & $-\xi$ & $-i$ & $\xi^*$ & $1$ & $\xi$ & $i$ & $-\xi^*$ & $1$ &
$-\xi$ & $-i$ & $\xi^*$ & $1$ & $\xi$ & $i$ & $-\xi^*$ \\
$xy$ & $1$ & $\xi$ & $i$ & $-\xi^*$ & $\xi^*$ & $1$ & $-\xi$ & $-i$ & 
$1$ & $\xi$ & $i$ & $-\xi^*$ & $\xi^*$ & $1$ & $-\xi$ & $-i$ \\
$x^2 y$ & $1$ & $-\xi$ & $i$ & $-\xi^*$ & $-i$ & $-\xi^*$ & $1$ &
$-\xi$ & $1$ & $-\xi$ & $i$ & $-\xi^*$ & $-i$ & $-\xi^*$ & $1$ &
$-\xi$ \\
$x^3 y$ & $1$ & $-\xi$ & $-i$ & $-\xi^*$ & $-\xi$ & $i$ & $-\xi^*$ & $1$ &
$1$ & $-\xi$ & $-i$ & $-\xi^*$ & $-\xi$ & $i$ & $-\xi^*$ & $1$ \\
$y^2$ & $1$ & $1$ & $1$ & $1$ & $1$ & $1$ & $1$ & $1$ & $1$ &
$1$ & $1$ & $1$ & $1$ & $1$ & $1$ & $1$ \\
$x y^2$ & $1$ & $1$ & $1$ & $1$ & $\xi$ & $-\xi$ & $-\xi$ & $-\xi$ & $1$ &
$1$ & $1$ & $1$ & $\xi$ & $-\xi$ & $-\xi$ & $-\xi$ \\
$x^2 y^2$ & $1$ & $1$ & $1$ & $1$ & $i$ & $-i$ & $-i$ & $i$ & $1$ &
$1$ & $1$ & $1$ & $i$ & $-i$ & $-i$ & $i$ \\
$x^3 y^2$ & $1$ & $1$ & $1$ & $1$ & $-\xi^*$ & $\xi^*$ & $-\xi^*$ & $-\xi^*$ &
$1$ & $1$ & $1$ & $1$ & $-\xi^*$ & $\xi^*$ & $-\xi^*$ & $-\xi^*$ \\
$y^3$ & $1$ & $-\xi$ & $-i$ & $\xi^*$ & $1$ & $\xi$ & $i$ & $-\xi^*$ & $1$ &
$-\xi$ & $-i$ & $\xi^*$ & $1$ & $\xi$ & $i$ & $-\xi^*$ \\
$x y^3$ & $1$ & $\xi$ & $i$ & $-\xi^*$ & $\xi^*$ & $1$ & $-\xi$ & $-i$ & $1$ &
$\xi$ & $i$ & $-\xi^*$ & $\xi^*$ & $1$ & $-\xi$ & $-i$ \\
$x^2 y^3$ & $1$ & $-\xi$ & $i$ & $-\xi^*$ & $-i$ & $-\xi^*$ & $1$ &
$-\xi$ & $1$ & $-\xi$ & $i$ & $-\xi^*$ & $-i$ & $-\xi^*$ & $1$ &
$-\xi$ \\
$x^3 y^3$ & $1$ & $-\xi$ & $-i$ & $-\xi^*$ & $-\xi$ & $i$ & $-\xi^*$ & $1$ &
$1$ & $-\xi$ & $-i$ & $-\xi^*$ & $-\xi$ & $i$ & $-\xi^*$ & $1$ 
\end{tabular}
\caption{Table of values of the nontrivial cocycle in $H^2(
{\mathbb Z}_4 \rtimes {\mathbb Z}_4,U(1))$.
Here, $\xi \equiv \exp(3 \pi i/4) = (-1 + i)/\sqrt{2}$.
\label{table:z4z4-cocycle}
}
\end{center}
\end{sidewaystable}

\begin{table}[h]
\begin{center}
\begin{tabular}{c|rrrrrrrrrrrrrrrr}
& $1$ & $x$ & $x^2$ & $x^3$ & $y$ & $xy$ & $x^2y$ & $x^3y$ &
$y^2$ & $xy^2$ & $x^2y^2$ & $x^3y^2$ & $y^3$ &
$xy^3$ & $x^2y^3$ & $x^3y^3$ \\ \hline
$1$ & $1$ & $1$ & $1$ & $1$ & $1$ & $1$ & $1$ & $1$ & $1$ & $1$ & $1$ &
$1$ & $1$ & $1$ & $1$ & $1$ \\
$x$ & $1$ & 
$1$ & $1$ & $1$ & $0$ & $0$ & $0$ & $0$ & $1$ & $1$ & $1$ & $1$ & $0$ &
$0$ & $0$ & $0$ \\
$x^2$ & $1$ & $1$ & $1$ & $1$ & $-1$ & $-1$ & $-1$ & $-1$ & $1$ & $1$ & $1$ 
& $1$ & $-1$ & $-1$ & $-1$ & $-1$ \\
$x^3$ & $1$ & $1$ & $1$ & $1$ & $0$ & $0$ & $0$ & $0$ & 
$1$ & $1$ & $1$ & $1$ & $0$ & $0$ & $0$ & $0$ \\
$y$ & $1$ & $0$ & $-1$ & $0$ & $1$ & $0$ & $-1$ & $0$ & $1$ & $0$ & $-1$
& $0$ & $1$ & $0$ & $-1$ & $0$ \\
$xy$ & $1$ & $0$ & $-1$ & $0$ & $0$ & $1$ & $0$ & $-1$ & $1$ & $0$ & $-1$ & $0$
& $0$ & $1$ & $0$ & $-1$ \\
$x^2 y$ & $1$ & $0$ & $-1$ & $0$ & $-1$ & $0$ & $1$ & $0$ & $1$ & $0$ & $-1$ &
$0$ & $-1$ & $0$ & $1$ & $0$ \\
$x^3 y$ & $1$ & $0$ & $-1$ & $0$ & $0$ & $-1$ & $0$ & $1$ & $1$ & $0$ & $-1$ &
$0$ & $0$ & $-1$ & $0$ & $1$ \\
$y^2$ & $1$ & $1$ & $1$ & $1$ & $1$ & $1$ & $1$ & $1$ & $1$ & $1$ & $1$ & $1$ &
$1$ & $1$ & $1$ & $1$ \\
$x y^2$ & $1$ & $1$ & $1$ & $1$ & $0$ & $0$ & $0$ & $0$ & 
$1$ & $1$ & $1$ & $1$ & $0$ & $0$ & $0$ & $0$ \\
$x^2 y^2$ & $1$ & $1$ & $1$ & $1$ & $-1$ & $-1$ & $-1$ & $-1$ &
$1$ & $1$ & $1$ & $1$ & $-1$ & $-1$ & $-1$ & $-1$ \\
$x^3 y^2$ & $1$ & $1$ & $1$ & $1$ & $0$ & $0$ & $0$ & $0$ &
$1$ & $1$ & $1$ & $1$ & $0$ & $0$ & $0$ & $0$ \\
$y^3$ & $1$ & $0$ & $-1$ & $0$ & $1$ & $0$ & $-1$ & $0$ &
$1$ & $0$ & $-1$ & $0$ & $1$ & $0$ & $-1$ & $0$ \\
$x y^3$ & $1$ & $0$ & $-1$ & $0$ & $0$ & $1$ & $0$ & $-1$ & $1$ & $0$ & $-1$ &
$0$ & $0$ & $1$ & $0$ & $-1$ \\
$x^2 y^3$ & $1$ & $0$ & $-1$ & $0$ & $-1$ & $0$ & $1$ & $0$ &
$1$ & $0$ & $-1$ & $0$ & $-1$ & $0$ & $1$ & $0$ \\
$x^3 y^3$ & $1$ & $0$ & $-1$ & $0$ & $0$ & $-1$ & $0$ & $1$ &
$1$ & $0$ & $-1$ & $0$ & $0$ & $-1$ & $0$ & $1$ 
\end{tabular}
\caption{Twisted sector phases for discrete torsion in
a ${\mathbb Z}_4 \rtimes {\mathbb Z}_4$ orbifold.  A zero entry
indicates a non-commuting pair of group elements.
\label{table:z4z4-phases}
}
\end{center}
\end{table}

The group ${\mathbb Z}_4 \rtimes {\mathbb Z}_4$ has ten conjugacy classes,
namely
\begin{equation}
\begin{array}{c}
\{ 1 \}, \: \: \:
\{ x^2 \}, \: \: \:
\{ x, x^3 \}, \: \: \:
\{ y, x^2 y \}, \: \: \:
\{ y^2 \}, %\: \: \:
\\
\{y^3, x^2 y^3 \}, \: \: \:
\{xy, x^3 y\}, \: \: \:
\{x y^2, x^3 y^2\}, \: \: \:
\{ x^2 y^2 \}, \: \: \:
\{xy^3, x^3 y^3 \}.
\end{array}
\end{equation}
Of these, the only ones that satisfy
condition~(\ref{eq:countprojirreps}) with respect to the cocycle in
table~\ref{table:z4z4-cocycle} are
\begin{equation}
\{ 1 \}, \: \: \:
\{ x, x^3 \}, \: \: \:
\{ y^2 \}, \: \: \:
\{x y^2, x^3 y^2 \}.
\end{equation}
As a result, there are four irreducible projective representations of
${\mathbb Z}_4 \rtimes {\mathbb Z}_4$.

\subsection{$S_4$}
\label{app:s4}

The group $S_4$ is the symmetric group on four objects.
Its group elements can be presented as transpositions,
of the form $1$, $(ab)$, $(abc)$, $(ab)(cd)$, and $(abcd)$,
where for example $(abc)$ indicates that $a$ maps to $b$,
$b$ maps to $c$, and $c$ maps to $a$, so that, for example,
$(abc) = (bca) = (cab)$.

It can be shown that $H^2(S_4,U(1)) = {\mathbb Z}_2$.  In this appendix
we collect an explicit representative of the nontrivial cocycle
and corresponding twisted sector phases.

\begin{sidewaystable}%[h!]
\begin{center}
\begin{tabular}{c|rrrrrrrrrrrrrrrrrrrrrrrr}
 & $1$ & $2$ & $3$ & $4$ & $5$ & $6$ & $7$ & $8$ & $9$ & $10$ & $11$
& $12$ & $13$ & $14$ & $15$ & $16$ & $17$ & $18$ & $19$ & $20$ & $21$ &
$22$ & $23$ & $24$ \\
\hline
$1$ & $1$ & $1$ & $1$ & $1$ & $1$ & $1$ & $1$ & $1$ & $1$ & $1$ & $1$ & $1$
& $1$ & $1$ & $1$ & $1$ & $1$ & $1$ & $1$ & $1$ & $1$ & $1$ & $1$ & $1$ \\
$2$ & $1$ & $1$ & $i$ & $-i$ & $i$ & $-i$ & $i$ & $-i$ & $-i$ & $i$ & $i$
&  $-i$ & $-i$ & $i$ & $-i$ & $i$ & $-1$ & $-1$ & $1$ & $1$ & $-i$
& $i$ & $-i$ & $i$ \\
$3$ & $1$ & $-i$ & $1$ & $i$ & $i$ & $-i$ & $-i$ & $i$ &
$-i$ & $-i$ & $i$ & $i$ & $-i$ & $i$ & $i$ & $-i$ &
$i$ & $i$ & $-i$ & $-i$ & $-1$ & $1$ & $-1$ & $1$
\\
$4$ & $1$ & $i$ & $-i$ & $1$ & $i$ & $i$ & $-i$ & $-i$ &
$-i$ & $i$ & $-i$ & $i$ & $i$ & $i$ & $-i$ & $-i$ &
$1$ & $1$ & $1$ & $1$ & $1$ & $1$ & $1$ & $1$ \\
$5$ & $1$ & $i$ & $i$ & $i$ & $i$ & $-i$ & $-i$ & $-i$ &
$1$ & $i$ & $i$ & $i$ & $-i$ & $i$ & $1$ & $1$ &
$i$ & $i$ & $1$ & $1$ & $i$ & $i$ & $1$ & $1$ \\
$6$ & $1$ & $i$ & $-i$ & $-i$ & $-i$ & $-i$ & $i$ & $-i$ &
$-i$ & $i$ & $-i$ & $1$ & $1$ & $1$ & $-i$ & $i$  &
$-i$ & $-i$ & $1$ & $1$ & $-i$ & $-i$ & $1$ & $1$ \\
$7$ & $1$ & $-i$ & $i$ & $-i$ & $-i$ & $-i$ & $-i$ & $i$ &
$-i$ & $1$ & $i$ & $-i$ & $-i$ & $-i$ & $-1$ & $1$ &
$-1$ & $-1$ & $-i$ & $-i$ & $-i$ & $i$ & $-1$ & $1$ \\
$8$ & $1$ & $-i$ & $-i$ & $i$ & $-i$ & $i$ & $-i$ & $-i$ &
$-i$ & $-i$ & $1$ & $i$ & $-1$ & $1$ & $-i$ & $-i$ &
$1$ & $1$ & $-i$ & $-i$ & $i$ & $-i$ & $-1$ & $1$ \\
$9$ & $1$ & $-i$ & $-i$ & $-i$ & $1$ & $-i$ & $-i$ & $-i$ &
$-i$ & $i$ & $i$ & $i$ & $-i$ & $1$ & $-i$ & $1$ &
$i$ & $1$ & $-i$ & $1$ & $-i$ & $1$ & $-i$ & $1$ \\
$10$ & $1$ & $-i$ & $i$ & $i$ & $i$ & $-i$ & $1$ & $i$ &
$i$ & $i$ & $i$ & $-i$ & $-i$ & $-1$ & $i$ & $1$ &
$i$ & $-1$ & $i$ & $1$ & $-1$ & $i$ & $-1$ & $i$ \\
$11$ & $1$ & $i$ & $-i$ & $i$ & $i$ & $i$ & $-i$ & $1$ &
$i$ & $-i$ & $i$ & $i$ & $-1$ & $i$ & $1$ & $-i$ &
$i$ & $1$ & $i$ & $-1$ & $i$ & $1$ & $i$ & $1$ \\
$12$ & $1$ & $i$ & $i$ & $-i$ & $i$ & $1$ & $i$ & $-i$ &
$i$ & $i$ & $-i$ & $i$ & $1$ & $i$ & $1$ & $i$ &
$-i$ & $1$ & $i$ & $1$ & $1$ & $i$ & $1$ & $i$ \\
$13$ & $1$ & $i$ & $i$ & $-i$ & $i$ & $1$ & $i$ & $-1$ &
$i$ & $i$ & $-1$ & $1$ & $1$ & $-i$ & $-i$ & $i$ &
$-i$ & $-1$ & $-i$ & $1$ & $-i$ & $-i$ & $1$ & $-1$ \\
$14$ & $1$ & $i$ & $-i$ & $i$ & $i$ & $1$ & $-i$ & $1$ &
$1$ & $-1$ & $i$ & $i$ & $i$ & $i$ & $1$ & $i$ &
$-i$ & $1$ & $i$ & $1$ & $i$ & $i$ & $1$ & $-1$ \\
$15$ & $1$ & $-i$ & $-i$ & $-i$ & $1$ & $-i$ & $-1$ & $-i$ &
$-i$ & $i$ & $1$ & $1$ & $i$ & $1$ & $-i$ & $-i$ &
$-i$ & $1$ & $-i$ & $-1$ & $i$ & $-i$ & $1$ & $1$ \\
$16$ & $1$ & $-i$ & $i$ & $i$ & $1$ & $-i$ & $1$ & $i$ &
$1$ & $1$ & $i$ & $-i$ & $-i$ & $i$ & $-i$ & $1$ &
$i$ & $1$ & $-i$ & $1$ & $-i$ & $i$ & $1$ & $1$ \\
$17$ & $1$ & $-1$ & $-i$ & $1$ & $-i$ & $i$ & $-i$ & $-i$ &
$-i$ & $-1$ & $1$ & $i$ & $i$ & $i$ & $i$ & $-i$ &
$1$ & $1$ & $i$ & $-1$ & $i$ & $-i$ & $1$ & $-1$ \\
$18$ & $1$ & $1$ & $i$ & $1$ & $1$ & $1$ & $1$ & $-1$ &
$1$ & $i$ & $i$ & $1$ & $1$ & $-1$ & $1$ & $1$ &
$-1$ & $i$ & $1$ & $1$ & $-1$ & $1$ & $-i$ & $i$ \\
$19$ & $1$ & $1$ & $i$ & $1$ & $i$ & $-i$ & $-i$ & $-i$ &
$1$ & $i$ & $i$ & $1$ & $i$ & $i$ & $-i$ & $i$ &
$-i$ & $1$ & $1$ & $1$ & $-i$ & $-i$ & $1$ & $1$ \\
$20$ & $1$ & $-1$ & $-i$ & $1$ & $1$ & $1$ & $-1$ & $1$ &
$-i$ & $-1$ & $1$ & $i$ & $-1$ & $1$ & $1$ & $1$ &
$1$ & $1$ & $1$ & $-i$ & $1$ & $1$ & $-i$ & $-i$ \\
$21$ & $1$ & $i$ & $-1$ & $1$ & $i$ & $1$ & $-1$ & $-i$ &
$-i$ & $i$ & $-i$ & $i$ & $i$ & $-i$ & $-i$ & $i$ &
$-i$ & $1$ & $i$ & $-1$ & $1$ & $-i$ & $1$ & $-1$ \\
$22$ & $1$ & $-i$ & $1$ & $1$ & $1$ & $-i$ & $-i$ & $1$ &
$-i$ & $-i$ & $i$ & $i$ & $i$ & $i$ & $-i$ & $-i$ &
$i$ & $1$ & $i$ & $1$ & $i$ & $1$ & $1$ & $1$ \\
$23$ & $1$ & $-i$ & $1$ & $1$ & $1$ & $-i$ & $-i$ & $1$ &
$1$ & $1$ & $1$ & $1$ & $-1$ & $1$ & $-1$ & $1$ &
$-1$ & $i$ & $1$ & $-i$ & $-1$ & $1$ & $-i$ & $1$ \\
$24$ & $1$ & $i$ & $-1$ & $1$ & $i$ & $1$ & $-1$ & $-i$ &
$1$ & $-1$ & $-1$ & $1$ & $1$ & $1$ & $1$ & $1$ &
$1$ & $i$ & $1$ & $i$ & $1$ & $1$ & $1$ & $i$
\end{tabular}
\caption{Cocycle for nontrivial element of $H^2(S_4,U(1))$.
The transpositions are numbered, as given in
table~\ref{table:s4-nums}.
\label{table:s4-cocycle} }
\end{center}
\end{sidewaystable}

\begin{sidewaystable}%[h!]
\begin{center}
\begin{tabular}{c|rrrrrrrrrrrrrrrrrrrrrrrr}
 & $1$ & $2$ & $3$ & $4$ & $5$ & $6$ & $7$ & $8$ & $9$ & $10$ & $11$
& $12$ & $13$ & $14$ & $15$ & $16$ & $17$ & $18$ & $19$ & $20$ & $21$ &
$22$ & $23$ & $24$ \\
\hline
$1$ & $1$ & $1$ & $1$ & $1$ & $1$ & $1$ & $1$ & $1$ & $1$ & $1$ & $1$ & $1$ &
$1$ & $1$ & $1$ & $1$ & $1$ & $1$ & $1$ & $1$ & $1$ & $1$ & $1$ & $1$
\\
$2$ & $1$ & $1$ & $-1$ & $-1$ & $0$ & $0$ & $0$ & $0$ & $0$ & $0$ & $0$ &
$0$ & $0$ & $0$ & $0$ & $0$ & $0$ & $0$ & $0$ & $0$
& $-1$ & $-1$ & $1$ & $1$
\\
$3$ & $1$ & $-1$ & $1$ & $-1$ & $0$ & $0$ & $0$ & $0$ & $0$ & $0$ & $0$ & $0$ &
$0$ & $0$ & $0$ & $0$ & $-1$ & $1$ & $-1$ & $1$ & $0$ & $0$ & $0$ & $0$
\\
$4$ & $1$ & $-1$ & $-1$ & $1$ & $0$ & $0$ & $0$ & $0$ &
$0$ & $0$ & $0$ & $0$ & $-1$ & $1$ & $1$ & $-1$ &
$0$ & $0$ & $0$ & $0$ & $0$ & $0$ & $0$ & $0$
\\
$5$ & $1$ & $0$ & $0$ & $0$ & $1$ & $0$ & $0$ & $0$ & $1$ & $0$ & $0$ & $0$ &
$0$ & $0$ & $0$ & $0$ & $0$ & $0$ & $0$ & $0$ & $0$ & $0$ & $0$ & $0$
\\
$6$ & $1$ & $0$ & $0$ & $0$ & $0$ & $1$ & $0$ & $0$ &
$0$ & $0$ & $0$ & $1$ &
$0$ & $0$ & $0$ & $0$ & $0$ & $0$ & $0$ & $0$ & $0$ & $0$ & $0$ & $0$
\\
$7$ & $1$ & $0$ & $0$ & $0$ & $0$ & $0$ & $1$ & $0$ &
$0$ & $1$ & $0$ & $0$ &
$0$ & $0$ & $0$ & $0$ & $0$ & $0$ & $0$ & $0$ & $0$ & $0$ & $0$ & $0$
\\
$8$ & $1$ & $0$ & $0$ & $0$ & $0$ & $0$ & $0$ & $1$ &
$0$ & $0$ & $1$ & $0$ &
$0$ & $0$ & $0$ & $0$ & $0$ & $0$ & $0$ & $0$ & $0$ & $0$ & $0$ & $0$
\\
$9$ & $1$ & $0$ & $0$ & $0$ & $1$ & $0$ & $0$ & $0$ &
$1$ & $0$ & $0$ & $0$ &
$0$ & $0$ & $0$ & $0$ & $0$ & $0$ & $0$ & $0$ & $0$ & $0$ & $0$ & $0$
\\
$10$ & $1$ & $0$ & $0$ & $0$ & $0$ & $0$ & $1$ & $0$ &
$0$ & $1$ & $0$ & $0$ &
$0$ & $0$ & $0$ & $0$ & $0$ & $0$ & $0$ & $0$ & $0$ & $0$ & $0$ & $0$
\\
$11$ & $1$ & $0$ & $0$ & $0$ & $0$ & $0$ & $0$ & $1$ &
$0$ & $0$ & $1$ & $0$ &
$0$ & $0$ & $0$ & $0$ & $0$ & $0$ & $0$ & $0$ & $0$ & $0$ & $0$ & $0$
\\
$12$ & $1$ & $0$ & $0$ & $0$ & $0$ & $1$ & $0$ & $0$ &
$0$ & $0$ & $0$ & $1$ &
$0$ & $0$ & $0$ & $0$ & $0$ & $0$ & $0$ & $0$ & $0$ & $0$ & $0$ & $0$
\\
$13$ & $1$ & $0$ & $0$ & $-1$ &
$0$ & $0$ & $0$ & $0$ & $0$ & $0$ & $0$ & $0$ &
$1$ & $0$ & $0$ & $-1$ &
$0$ & $0$ & $0$ & $0$ & $0$ & $0$ & $0$ & $0$
\\
$14$ & $1$ & $0$ & $0$ & $1$ &
$0$ & $0$ & $0$ & $0$ & $0$ & $0$ & $0$ & $0$ &
$0$ & $1$ & $1$ & $0$ &
$0$ & $0$ & $0$ & $0$ & $0$ & $0$ & $0$ & $0$
\\
$15$ &  $1$ & $0$ & $0$ & $1$ &
$0$ & $0$ & $0$ & $0$ & $0$ & $0$ & $0$ & $0$ &
$0$ & $1$ & $1$ & $0$ &
$0$ & $0$ & $0$ & $0$ & $0$ & $0$ & $0$ & $0$
\\
$16$ & $1$ & $0$ & $0$ & $-1$ &
$0$ & $0$ & $0$ & $0$ & $0$ & $0$ & $0$ & $0$ &
$-1$ & $0$ & $0$ & $1$ &
$0$ & $0$ & $0$ & $0$ & $0$ & $0$ & $0$ & $0$
\\
$17$ & $1$ & $0$ & $-1$ & $0$ &
$0$ & $0$ & $0$ & $0$ & $0$ & $0$ & $0$ & $0$ & $0$ & $0$ & $0$ & $0$ &
$1$ & $0$ & $-1$ & $0$ & $0$ & $0$ & $0$ & $0$
\\
$18$ & $1$ & $0$ & $1$ & $0$ &
$0$ & $0$ & $0$ & $0$ & $0$ & $0$ & $0$ & $0$ & $0$ & $0$ & $0$ & $0$ &
$0$ & $1$ & $0$ & $1$ & $0$ & $0$ & $0$ & $0$
\\
$19$ & $1$ & $0$ & $-1$ & $0$ &
$0$ & $0$ & $0$ & $0$ & $0$ & $0$ & $0$ & $0$ & $0$ & $0$ & $0$ & $0$ &
$-1$ & $0$ & $1$ & $0$ & $0$ & $0$ & $0$ & $0$
\\
$20$ & $1$ & $0$ & $1$ & $0$ &
$0$ & $0$ & $0$ & $0$ & $0$ & $0$ & $0$ & $0$ & $0$ & $0$ & $0$ & $0$ &
$0$ & $1$ & $0$ & $1$ & $0$ & $0$ & $0$ & $0$
\\
$21$ & $1$ & $-1$ & $0$ & $0$ &
$0$ & $0$ & $0$ & $0$ & $0$ & $0$ & $0$ & $0$ & $0$ & $0$ & $0$ & $0$ &
$0$ & $0$ & $0$ & $0$ & $1$ & $-1$ & $0$ & $0$
\\
$22$ & $1$ & $-1$ & $0$ & $0$ &
$0$ & $0$ & $0$ & $0$ & $0$ & $0$ & $0$ & $0$ & $0$ & $0$ & $0$ & $0$ &
$0$ & $0$ & $0$ & $0$ & $-1$ & $1$ & $0$ & $0$
\\
$23$ & $1$ & $1$ & $0$ & $0$ &
$0$ & $0$ & $0$ & $0$ & $0$ & $0$ & $0$ & $0$ & $0$ & $0$ & $0$ & $0$ &
$0$ & $0$ & $0$ & $0$ & $0$ & $0$ & $1$ & $1$
\\
$24$ & $1$ & $1$ & $0$ & $0$ &
$0$ & $0$ & $0$ & $0$ & $0$ & $0$ & $0$ & $0$ & $0$ & $0$ & $0$ & $0$ &
$0$ & $0$ & $0$ & $0$ & $0$ & $0$ & $1$ & $1$
\end{tabular}
\caption{Twisted sector phases for discrete torsion in $S_4$ orbifold,
derived from cocycle in table~\ref{table:s4-cocycle}.  A zero entry
indicates a non-commuting pair of group elements.
Transpositions corresponding to group element numbers are listed in
table~\ref{table:s4-nums}.
\label{table:s4-phases} }
\end{center}
\end{sidewaystable}

\begin{table}[h]
\begin{center}
\begin{tabular}{c|c||c|c||c|c}
$1$ & $1$ & $9$ & $(243)$ & $17$ & $(23)$ \\
$2$ & $(13)(24)$ & $10$ & $(134)$ & $18$ & $(1342)$ \\
$3$ & $(14)(23)$ & $11$ & $(142)$ &  $19$ & $(14)$ \\
$4$ & $(12)(34)$ & $12$ & $(123)$ & $20$ & $(1243)$ \\
$5$ & $(234)$ & $13$ & $(34)$ & $21$ & $(24)$ \\
$6$ & $(132)$ & $14$ & $(1324)$ &  $22$ & $(13)$ \\
$7$ & $(143)$ & $15$ & $(1423)$ &  $23$ & $(1432)$ \\
$8$ & $(124)$ &  $16$ & $(12)$  &  $24$ & $(1234)$
\end{tabular}
\caption{Assignments of transpositions to $S_4$ element numbers in
tables~\ref{table:s4-cocycle}, \ref{table:s4-phases}.
\label{table:s4-nums}
}
\end{center}
\end{table}

The group $S_4$ has five conjugacy classes:
\begin{equation}
\begin{array}{c}
\{ 1 \}, \\
\{ (12), (13), (14), (23), (24), (34) \}, \\
\{ (12)(34), (13)(24), (14)(23) \}, \\
\{ (123), (132), (234), (243), (341), (314), (412), (421) \}, \\
\{ (1234), (1243), (1324), (1342), (1423), (1432) \}.
\end{array}
\end{equation}
Of these, the conjugacy classes whose elements $g$ have the property
that $\omega(g,h) = \omega(h,g)$ for all $h$ that commute with $g$ are
\begin{equation}
\begin{array}{c}
\{ 1 \}, \\
\{ (123), (132), (234), (243), (341), (314), (412), (421) \}, \\
\{ (1234), (1243), (1324), (1342), (1423), (1432) \}.
\end{array}
\end{equation}
Thus, we see that with this discrete torsion, there are precisely
three irreducible projective representations of $S_4$.

\end{document}